\documentclass[english,final]{sdqthesis}
\pdfoutput=1

\usepackage{standalone}
\author{Jonas Schneider}

\title{Analytic Performance Model\\
of a Main-Memory Index Structure}

\thesistype{Bachelor's Thesis}

\nogrouplogo

\reviewerone{Prof.\ Dr.-Ing.\ Klemens Böhm}

\advisorone{Dr.-Ing.\ Martin Schäler}
\advisortwo{David Broneske, M.Sc.}

\editingtime{November 18, 2015}{March 18, 2016}

\usepackage{tabularx}

\usepackage{csquotes}
\usepackage{upquote,mathdots}

\newcommand{\say}[1]{\textquote{#1}}
\newcommand{\lvar}[1]{\ensuremath{\mathit{#1}}}

\usepackage{tikz}
\usetikzlibrary{shapes,arrows,positioning,fit,calc,tikzmark,matrix}
\usetikzlibrary{spy,decorations.fractals,backgrounds,patterns}
\usetikzlibrary{lindenmayersystems}
\usetikzlibrary{shadings}
\pgfdeclarelindenmayersystem{Sierpinski triangle}{
  \rule{F -> G-F-G}
  \rule{G -> F+G+F}}

\usepackage{pgfplots}
\usepackage{pgfplotstable}
\usepackage{tikz-qtree}

\usepackage{subcaption}
\usepackage{float}
\newfloat{listing}{tbhp}{lst}
\floatname{listing}{Listing}

\usepackage[final]{pdfpages}

\usepackage[ruled,vlined,linesnumberedhidden]{algorithm2e}
\DontPrintSemicolon

\newcommand{\NN}{\mathbb{N}}
\newcommand{\RR}{\mathbb{R}}
\newcommand{\PP}{\mathbb{P}}
\newcommand{\EE}{\mathbb{E}}
\newcommand{\Ud}[2]{\mathcal{U}\{#1, #2\}}
\newcommand{\Uc}[2]{\mathcal{U}(#1, #2)}
\newcommand{\Bin}[1]{\model{Bin}(#1)}

\usepackage{amsmath}

\newcommand{\const}{\mathrm{const.}}

\newcommand{\bplustree}{$\mathrm{B}^+$-Tree}
\newcommand{\kdtree}{K-D-Tree}

\newcommand{\naive}{na\"{\i}ve}

\DeclareMathOperator{\R}{\mathrm{R}}
\DeclareMathOperator{\Elf}{\mathbf{Elf}}
\DeclareMathOperator{\visits}{\mathbf{visits}}
\DeclareMathOperator{\fanout}{\mathbf{fanout}}
\DeclareMathOperator{\oplabel}{\mathbf{label}}
\newcommand{\card}[1]{\|#1\|}
\DeclareMathOperator{\mono}{\mathbf{mono}}
\DeclareMathOperator{\avgsize}{\mathbf{expsize}}

\DeclareMathOperator{\buckets}{\mathbf{buckets}}
\DeclareMathOperator{\monobuckets}{\mathbf{monobuckets}}

\usepackage[binary-units]{siunitx}
\usepgfplotslibrary{units}
\pgfplotsset{
  unit code/.code 2 args=
    \expandafter\expandafter\expandafter\expandafter
      \expandafter\expandafter\expandafter\si
        \expandafter\expandafter\expandafter\expandafter
          \expandafter\expandafter\expandafter{#2}%
}

\settitle

\usepackage[citestyle=numeric,style=numeric,backend=biber]{biblatex}
\DefineBibliographyStrings{english}{%
  backrefpage = {page},
  backrefpages = {pages},
}
\usepackage{doi}
\DeclareFieldFormat{doi}{%
  \iffieldundef{url}{%
    \mkbibacro{DOI}\addcolon\space\href{http://dx.doi.org/#1}{\doi{#1}}%
  }{%
  }%
}
\DeclareFieldFormat{urldate}{%
  \iffieldundef{doi}{%
    \mkbibparens{\bibstring{urlseen}\space#1}%
  }{%
  }%
}

\usepackage[noabbrev]{cleveref}
\renewcommand{\autoref}[1]{\Cref{#1}}
\newcommand{\Autoref}[1]{\Cref{#1}}

\newtheorem{hyp}{Hypothesis}
\crefname{hyp}{hypothesis}{hypotheses}
\Crefname{hyp}{Hypothesis}{Hypotheses}

\usepgfplotslibrary{statistics}
\usepackage{environ}
\makeatletter
\newsavebox{\measure@tikzpicture}
\NewEnviron{scaletikzpicturetowidth}[1]{%
  \def\tikz@width{#1}%
  \def\tikzscale{1}\begin{lrbox}{\measure@tikzpicture}%
  \BODY
  \end{lrbox}%
  \pgfmathparse{#1/\wd\measure@tikzpicture}%
  \edef\tikzscale{\pgfmathresult}%
  \BODY
}
\makeatother

\makeatletter
\DeclareRobustCommand{\rvdots}{%
  \vbox{
    \baselineskip4\p@\lineskiplimit\z@
    \kern-\p@
    \hbox{.}\hbox{.}\hbox{.}
  }}
\makeatother

\begin{document}
\setpdf

\maketitle

\frontmatter

\thispagestyle{empty}
\null\vfill
\noindent\hbox to \textwidth{\hrulefill}
I declare that I have developed and written the enclosed
thesis completely by myself, and have not used sources or means without
declaration in the text.
Furthermore, I have observed and complied with the regulations
to ensure good scientific practice at KIT.

\textbf{Karlsruhe, March 18, 2016}
\vspace{1.5cm}

\dotfill\hspace*{8.0cm}\\
\hspace*{2cm}(\theauthor)
\cleardoublepage

\setcounter{page}{1}
\pagenumbering{roman}
\includeabstract
\tableofcontents
\listoffigures
\listoftables

\mainmatter
\chapter{Introduction}
\label{ch:Introduction}

In large-scale data analytics, the availability of increasing amounts of main memory
at affordable cost has led to fundamental changes in the design of database systems.
While traditional systems use secondary storage, such as hard drives, for storing data,
newer systems use \emph{main-memory} based storage, usually implemented
using Random Access Memory~(RAM).
Compared to hard drives, RAM provides significantly shorter latency and higher bandwidth.

The paradigm shift from disk-based to RAM-based storage led to innovations in other areas of data processing systems.
In particular, it influenced the design of \emph{index structures},
i.e.\ data structures that provide ways to phyiscally organise a data set.
Traditionally, index structures suffered from the performance bottleneck of hard disks.
With the advent of the much faster RAM, the design constraints for efficient index structures
have changed.
For disk-based structures, the access latency dominated the other parts of the system.
For main-memory structures, more intricacies and impact factors have to be considered.

At the same time of the shift to main-memory storage, the expectations of database users have
also changed.
One traditionally difficult task is the evaluation of
\emph{Multi-Column Selection Predicates~(MCSPs)},
selection operators that operate on multiple data dimensions simultaneously.
Proposed index structures for efficiently evaluating MCSPs suffer from a variety of
problems, which are commonly called the \say{curse of dimensionality}.
In our case, these problems cause severe performance degradation that
render index structures ineffective.

Recent literature proposed the \say{Elf}, a novel index structure designed
for the efficient evaluation of range queries, a class of MCSPs.
The Elf has an optimised memory layout that exploits the particularities
of a main-memory environment, such as memory alignment and CPU branch prediction.

With its optimised design, the Elf has been shown to outperform other recently proposed
index structures under the right circumstances.
However, its actual run-time cost has only been verified empirically in a limited number of scenarios.
In particular, we find that the Elf's performance is very sensitive to its parameter configuration
and to the queries that are executed.
Choosing a poor configuration for the same query and data
can immediately lead to extraordinarily worsened performance.

With better knowledge of the performance differences between configurations,
database administrators can tune the Elf
configuration towards their specific workloads and data sets.
In addition, advanced applications of the Elf, for example as a join preprocessing step
within a larger query plan, have been proposed as future work.
However, their development requires better knowledge about the performance characteristics of the Elf.

Finally, better knowledge of the index performance can be exploited at run-time
to yield better response times.
This process, known as query planning,
uses information about the query and the dataset
to decide whether to use an index structure or to resort to scanning the dataset.
Good planning decisions can lead to significant performance improvements.
However, good planning decisions require an accurate predictive
model of query performance on the given data structure.
Therefore, unpredictable performance directly inhibits the viability of the Elf.

Consequently, we want to initiate a first step towards predicting the Elf's performance
ahead-of-time, before the costly construction of the structure itself.

\section{Goal of this Thesis}

The goal of this thesis is to develop an improved understanding of the Elf data structure,
and to model its performance characteristics in a wide range of scenarios.
To this end, we provide the following contributions.

\begin{enumerate}
  \item
    Based on a literature review of proposed multi-dimensional index structures,
    we describe how the Elf compares to them and how it uniquely tackles the problems
    encountered for high dimensional data sets.
  \item
    We develop a formal model that predicts the runtime cost of the Elf search, without building the Elf first.
    Our model is based on the size and shape of the structure, and how the query interacts with it.
  \item
    We provide a theoretical underpinning of the impact of skewed and correlated data distributions
    on the performance of the Elf.
    In particular, we describe how it behaves in scenarios that lead to degenerate performance for other indexing methods.
  \item
    We provide empirical evidence that the Elf response time is indeed robust and follows our predictions.
    As such, we argue that the Elf design is appropriate for applying the metrics for disk-based index tructures
    to a main-memory structure.
\end{enumerate}

\section{Outline}

The remainder of this thesis is structured as follows.
In \Autoref{chp:background},
we introduce the range query problem and describe proposed approaches to evaluate this kind of query.
In particular, we introduce the Elf data structure and highlight its unique properties.
Additionally, we give an overview on how the performance of classical disk-based index structures is modelled.

In \Autoref{chp:model}, we analyse the Elf search algorithm and propose a model of its execution time.
We describe the relationship between the size of the Elf and the query execution time.
Finally, we extend our model to account for query selectivites, skewed data distributions, and correlations.

In \Autoref{chp:results}, we show the results of our empirical evaluation of the accuracy of our model.
We discuss our methods and results in \Autoref{chp:discussion}, where we also describe potential shortcomings of our model.
We conclude this thesis in \Autoref{chp:conclusion} with a summary and present opportunities for future work.

\chapter{Background}\label{chp:background}
In our work, we consider the evaluation of range queries on a data set representing a set of points in Euclidean space.
In this chapter, we describe the range query problem and refer to related work that introduced a variety of approaches for optimising the evaluation of range queries.
In particular, we describe multi-dimensional tree-based methods.
These methods form the foundation for the Elf index structure, which we describe in detail.
Finally, we show the common method for developing a cost model for an index structure like the Elf.

\section{The range query search problem}
A \emph{range query} \cite{DBLP:books/daglib/0098215,thesis:howat2012,harangsri1998query} is a specific type of multi-column selection predicate (MCSP),
which is a logical predicate that defines a subset of a multi-dimensional data set.
In particular, a range query selects exactly the data points that are contained within an axis-aligned hypercuboid intersecting the data space.

\begin{figure}[ht]
  \centering
  \begin{tikzpicture}\begin{axis}[
      width=0.8\textwidth,
      xmin=0, xmax=12, ymin=0, ymax=5,
      axis lines=middle,
      axis equal image,
      xtick=\empty, ytick=\empty,
      xlabel=$X_1$, ylabel=$X_2$,
      mark size=1.5,
      every axis x label/.style={
          at={(ticklabel* cs:1.05)},
          anchor=west,
      },
      every axis y label/.style={
          at={(ticklabel* cs:1.05)},
          anchor=south,
      },
      extra x ticks={2,6},
      extra y ticks={1,3},
      extra tick style={grid=major},
      extra x tick labels={$l_1$,$u_1$},
      extra y tick labels={$l_2$,$u_2$},
  ]
  \addplot [only marks, draw=red!75,fill=red!75, restrict x to domain=2:6,restrict y to domain=1:3] table {\datasettable};
  \addplot [only marks, draw=black, restrict x to domain=0:2] table {\datasettable};
  \addplot [only marks, draw=black, restrict x to domain=6:100] table {\datasettable};
  \addplot [only marks, draw=black, restrict y to domain=0:1] table {\datasettable};
  \addplot [only marks, draw=black, restrict y to domain=3:100] table {\datasettable};
  \draw [red!75, thick] (axis cs:2,1) rectangle (axis cs:6,3);

  \end{axis}
  \end{tikzpicture}
  \caption[Example two-dimensional data set with a range query rectangle]%
  {An example two-dimensional data set together with a range query rectangle (a two-dimensional hypercuboid).
  The points marked in red match the query.}
  \label{fig:rangeq}
\end{figure}
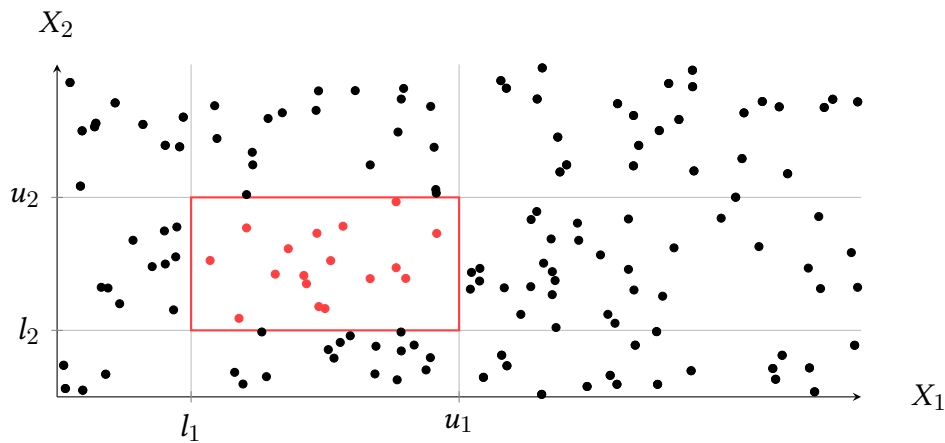

Let $R$ be a $k$-dimensional relation with columns $X_1, \dots, X_k$, i.e.\
$R \subset X_1 \times \dots \times X_k$.
\Autoref{fig:rangeq} shows a simple example range query over $R$ for $k=2$,
where a hypercuboid is simply a rectangle.
\pgfmathsetseed{1138} 
\pgfplotstableset{ 
    create on use/x/.style={create col/expr={abs(rand*12)}},
    create on use/y/.style={create col/expr={abs(rand*5)}}
}
\pgfplotstablenew[columns={x,y}]{150}\datasettable

The parameters of the range query are
the lower and upper bound values $l_i$ and $u_i$ for each dimension $i=1,\dots,k$.
These \emph{query windows} uniquely determine the hypercuboid.
Many commonly used predicates, such as $=,<,\leq,>,\geq$, and $\texttt{BETWEEN}$,
can be represented in this definition, as shown in \autoref{tbl:predmap}.
The range query is the combination of one such subpredicate for each attribute.
\begin{table}[htb]\centering
\begin{tabular}{ll} \toprule
  SQL predicate & Query window \\ \midrule
  \texttt{x = c} & $[c,c]$ \\
  \texttt{x < c} & $[\operatorname{min}, c-1]$ \\
  \texttt{x <= c} & $[\operatorname{min}, c]$ \\
  \texttt{x > c} & $[c+1, \operatorname{max}]$ \\
  \texttt{x >= c} & $[c, \operatorname{max}]$ \\
  \texttt{x BETWEEN a AND b} & $[a,b]$ \\
  \texttt{1} & $[\operatorname{min},\operatorname{max}]$ \\ \bottomrule
\end{tabular}
\caption[Mapping from SQL predicates to query windows]{Mapping from SQL predicates to query windows.
  $\operatorname{min}$ and $\operatorname{max}$
  denote the minimum and maximum of the attribute domain, respectively.
  }\label{tbl:predmap}
\end{table}

One characteristic property of a range query is its \say{specificity}.
The overall fraction of points of matched by the
query is called the \emph{(total) selectivity} $\sigma$ of the query.
For each dimension $i$, the fraction of points matched by the subpredicate
in dimension $i$ is called is called the \emph{partial selectivity}
$\sigma_i$ of the query in attribute $i$.

As seen in \autoref{tbl:predmap}, the predicate \say{any value in column $i$ matches} is
not treated as an exceptional case.
Range queries that contain this predicate
are sometimes also called \emph{partial match queries}.

\section{Introduction to indexing}\label{sec:others}
A variety of methods has been proposed to efficiently evaluate range queries, both in single- and multi-dimensional data sets.
In this section, we give an overview over the basic concepts of indexing.
First, we describe the simplest evaluation method: the linear scan over the data set.
Then, we introduce the {\bplustree} as an effective single-dimensional index.
Finally, we illustrate the complications arising from multiple dimensions using the inverted list approach to multi-dimensional indexing.

Together with tree-based multi-dimensional index structures, which we discuss in the following section,
the methods we present contain import concepts for the Elf.
Later, we will compare the Elf to each one of them and describe similarities and differences to the known methods.

\subsection{Linear scans}
Perhaps the simplest method of evaluating any kind of MCSP, including a range query, is to scan through the list of data points sequentially.
For each data point, it is determined whether the point matches the query, and it is added to the result set if so.

This so-called \emph{linear scan} takes $\mathcal{O}(nk)$ time, where $n$ is the number of data points, and $k$ is the number of dimensions.
In particular, the runtime does not depend on e.g.\ the selectivity of the query.

Despite this high asymptotic complexity, optimised variants of the linear scan remain competitive in many scenarios.
Different memory layouts lead to \emph{row stores}, where the attributes of each row are laid out sequentially, and \emph{column stores} \cite{stonebraker2005c}, where the values of a single column are laid out sequentially.

Both modern hard disk drives and CPUs are often tuned to quickly detect and accelerate the processing of sequential data accesses.
Recent work on main-memory sequential scans \cite{bitweaving,Boncz:1999:DAO:645925.671364,Zhou:2002:IDO:564691.564709} exploits instruction-level parallelism and other features of modern CPUs to achieve competitive performance.

However, even these accomplishments cannot improve the linear asymptotic complexity of the linear scan.
For large data sets, especially in cases where only a very small fraction of the data is requested (e.g.\ range queries with low total selectivity), the scanning overhead becomes prohibitively expensive.

\subsection{{\bplustree}s and generalised variants}\label{sec:btree}
\begin{figure}[tb]
  \centering
  \begin{minipage}[t][][b]{.25\linewidth}\centering
  \begin{tabular}{ll} \toprule
    Key & Value \\ \midrule
    1 & A \\
    2 & B \\
    3 & C \\
    4 & D \\
    10 & E \\ \bottomrule
  \end{tabular}
\end{minipage}%
\begin{minipage}[t][][b]{.7\linewidth}
\centering
\pgfdeclarelayer{background}
\pgfdeclarelayer{foreground}
\pgfsetlayers{background,main,foreground}
\begin{tikzpicture}[scale=0.8]
\tikzstyle{n} = [draw,shape=circle,minimum size=2.8em,inner sep=0pt,fill=red!20]
\tikzstyle{bign} = [draw,shape=circle,minimum size=2em,inner sep=0pt,fill=green!20]




\node (A) at (33.895bp,18.0bp) [bign] {$A$};
  \node (C) at (177.89bp,18.0bp) [bign] {$C$};
  \node (B) at (105.89bp,18.0bp) [bign] {$B$};
  \node (E) at (321.89bp,18.0bp) [bign] {$E$};
  \node (D) at (249.89bp,18.0bp) [bign] {$D$};
  \node (A_1) at (197.89bp,263.0bp) [n] {$$[1,10]$$};
  \node (C_1) at (35.895bp,91.0bp) [n] {$$[1,1]$$};
  \node (C_3) at (249.89bp,91.0bp) [n] {$$[4,10]$$};
  \node (C_2) at (141.89bp,91.0bp) [n] {$$[2,3]$$};
  \node (B_1) at (141.89bp,177.0bp) [n] {$$[1,3]$$};
  \node (B_2) at (237.89bp,177.0bp) [n] {$$[4,10]$$};
  \draw [->] (B_1) -- (C_1);
  \definecolor{strokecol}{rgb}{0.0,0.0,0.0};
  \pgfsetstrokecolor{strokecol}
  \draw (99.395bp,134.0bp) node {$1$};
  \draw [->] (C_3) -- (D);
  \draw [->] (B_1) -- (C_2);
  \draw (145.39bp,134.0bp) node {$3$};
  \draw [->] (C_2) -- (C);
  \draw [->] (C_3) -- (E);
  \draw [->] (A_1) -- (B_1);
  \draw (177.39bp,220.0bp) node {$3$};
  \draw [->] (C_1) -- (A);
  \draw [->] (C_2) -- (B);
  \draw [->] (A_1) -- (B_2);
  \draw (227.89bp,220.0bp) node {$10$};
  \draw [->] (B_2) -- (C_3);
  \draw (251.89bp,134.0bp) node {$10$};

\begin{pgfonlayer}{background}
  \draw[rounded corners=2em,line width=3.5em,blue!20,cap=round]
  (A_1.center) -- (B_1.center) -- (C_1.center) -- (A.center);
  \draw[rounded corners=2em,line width=3.5em,blue!20,cap=round]
                  (B_1.center) -- (C_2.center) -- (B.center);
  \draw[rounded corners=2em,line width=3.5em,blue!20,cap=round]
                  (B_1.center) -- (C_2.center) -- (C.center);
\end{pgfonlayer}
\end{tikzpicture}
\end{minipage}
\caption[Example {\bplustree} for a map from keys to values]{
  The {\bplustree} for an example map from keys to values.
  The interior nodes are labelled with the interval they represent.
  The edges are labelled with the pivot values,
  i.e.\ an edge with label $l$ is traversed if $x \leq l$.
  Highlighted in blue are the nodes that are visited when evaluating the range query $[1;2]$.
}
  \label{fig:btree}
\end{figure}
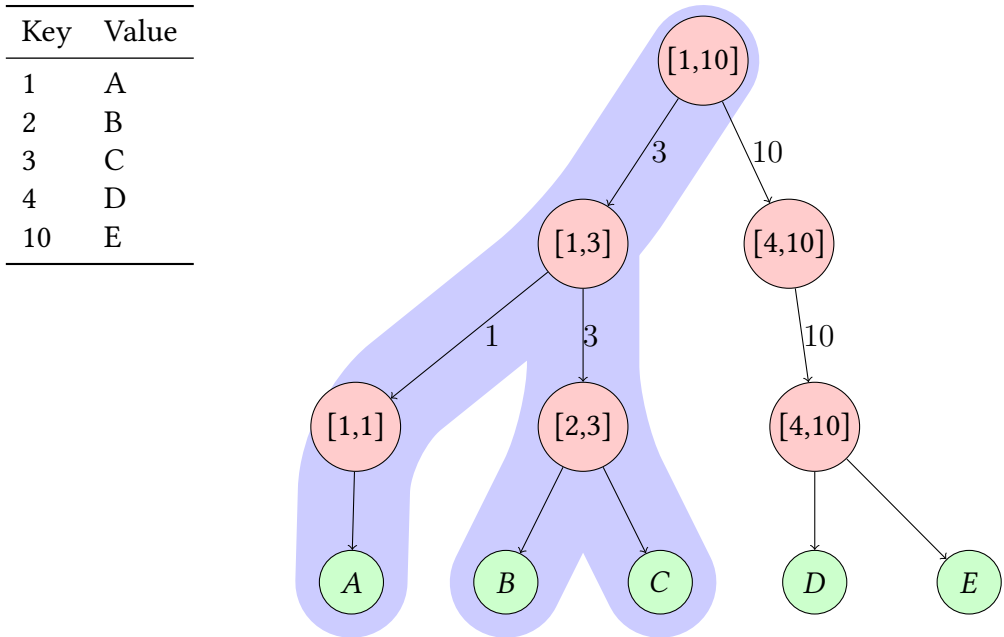

In the single-dimensional case, search trees such as the {\bplustree} \cite{Comer:1979:UB:356770.356776}, derived from the original B-Tree \cite{bayer1972organization}, have become the method of choice for evaluating selection predicates.
The {\bplustree} is a self-balancing search tree that is used to store values identified by a numerical key.
\Autoref{fig:btree} shows the {\bplustree} for an example data set.

The {\bplustree} is designed such that during search, no items that are \say{far away} from the intended position have to be accessed.
This is achieved by storing, for a group of values, a lower and upper bound for the keys of the group.
If the query interval does not intersect this bounding interval, the items in the group cannot possibly match the query, and therefore do not have to be checked.
Skipping the individual comparisons for the items contained in the group reduces the overall execution time.

Exploiting this property of quickly checking a coarse bound and pruning the search space
is called \emph{branch-and-bound} \cite{mehlhorn2008algorithms}, an algorithm design pattern.
\Autoref{fig:btree} shows how the pruning behaviour allows the {\bplustree} search algorithm to skip accessing some of the data points:
For the range query $[1;2]$, the region $[4;10]$ does not intersect the query bounds.
Therefore, the search can skip the descent into the right half of the tree.

The concept of the {\bplustree} has received great attention and has been implemented as the Generalised Search Tree~(GiST) framework \cite{aoki1998generalizing}.
Users of the framework can implement specialised index structures while relying on the generic rebalancing and query functions provided by the framework.
For example, the multi-dimensional R-Tree and {\kdtree} structures, which we discuss in \autoref{sec:trees}, can be implemented using GiST \cite{eltabakh2006space}.

The {\bplustree} itself, however, is inapplicable for multi-dimensional queries.
In theory, it is possible to build a {\bplustree} using the values of multiple columns as keys.
This is the approach taken by popular DBMSes such as PostgreSQL \cite{postgresmultiindex} for multi-dimensional indices.

However, the fundamental principle exploited by the B-Tree is the fact that for one-dimensional data sets, there exists a total ordering that preserves spatial locality \cite{908985}.
What this means is that values that are \say{close together} also have keys that are close together.
However, in higher-dimensional space, such an ordering does not exist anymore.

Therefore, the {\bplustree} is only applicable in a multi-dimensional scenario if the total selectivity is dominated by the selection on a single attribute.
In this case, simply indexing the dominant attribute suffices to achieve acceptable performance.
This assumption, which we further discuss in the following subsection, does not hold for many OLAP query workloads, including the popular TPC-H benchmark \cite{council2008tpc}.

\subsection{Inverted lists and the problem of high partial selectivity}\label{sec:inverted}
For single-dimensional queries, structures like the {\bplustree} are generally sufficient.
However, multi-dimensional indexing requires a different approach.
We will demonstrate this by describing a scenario in which {\naive} approaches to
multi-dimensional indexing perform poorly.

First, consider any multi-dimensional query $Q$
that can be represented as the logical conjunction (AND)
of several single-dimensional subpredicates $Q_i$:
$$Q = Q_1 \wedge Q_2 \dots \wedge Q_k$$
The result set of $Q$ therefore consists of exactly the tuples that invidiually satisfy each of the subpredicates.

A rather simple approach to evaluating $Q$ is the so-called \emph{inverted list} method \cite{DBLP:books/daglib/0098215}.
The inverted list method evaluates each subpredicate individually,
and then merges the results together to obtain the result of the original query.
In particular, the evaluation is performed in two steps:
\begin{enumerate}
  \item For each column $i=1,\dots,k$, evaluate the single-dimensional $Q_i$.
  This results in the set $S_i$ of tuples that match $Q_i$.
  To evaluate the subpredicate,
  any single-dimensional index structure, such as the {\bplustree}, may be used.

  \item
  Compute the result $S$ by intersecting the intermediate result sets:
  $$S = S_1 \cap S_2 \cap \dots \cap S_k$$
\end{enumerate}

The performance of this approach is, perhaps unsurprisingly, questionable.
To highlight the case of particularly poor performance, we will assume the following:
while $Q$ has low selectivity (i.e.\ $\card{S}$ is small),
each $Q_i$ has a high selectivity.
This is equivalent to stating that $Q$ has low \emph{total} selectivity and high
\emph{partial} selectivities.

Queries that satisfy this criterion are commonplace, and they are present in popular benchmarks
like the TPC-H decision support benchmark \cite{council2008tpc}.
\Autoref{tbl:largeintermediate} shows the total and partial selectivities of Q19 from this benchmark.
While the total selectivity is low, each of the partial selectivities is comparatively high.
\begin{table}[ht]\centering
  \begin{tabular}{ll} \toprule
    Selectivity & Predicate \\ \midrule
    $\sigma_1 \approx \num{0.22}$ & $Q19_1 = \say{\texttt{l\_quantity BETWEEN 5 AND 15}}$ \\
    $\sigma_2 \approx \num{0.14}$ & $Q19_2 =
      \say{\texttt{l\_shipmode in
      (`AIR', `AIR REG')}}$ \\
    $\sigma_3 \approx \num{0.25}$ & $Q19_3 = \say{\texttt{l\_shipinstruct = `DELIVER IN PERSON'}}$ \\ \addlinespace
    $\sigma \approx \num{0.008}$ & $Q19 = Q19_1 \wedge Q19_2 \wedge Q19_3$ \\ \bottomrule
  \end{tabular}
  \caption[Total selectivity and partial selectivities of TPC-H query Q19]{
    Total selectivity and partial selectivities of TPC-H query Q19.
    The total selectivity is significantly lower than each of the partial selectivities.
  }\label{tbl:largeintermediate}
\end{table}


Since every partial selectivity $\sigma_i$ is high, the following holds:
$$\card{S} = \sigma\card{R} \ll \sigma_i \card{R} = \card{S_i}$$
Therefore, the intermediate data sets $S_i$ are very large in comparison to the overall result set $S$,
resulting in high asymptotic complexity.
Either the large set has to be explicity stored in memory, leading to high space complexity,
or it has to be implicitly iterated over, leading to high time complexity.

The discrepancy between partial and total selectivites is possible because the total selectivity is the product of all partial selectivities.
The magnitude of the discrepancy increases with the dimensionality of the data set.
To illustrate this, a query $Q$ on a 20-dimensional data set might
have a partial selectivity of $0.9$ in every column.
This means that each $S_i$ will have size $\card{S_i} = 0.9\card{R}$.
However, the total selectivity of $Q$ is only $0.9^{20} \approx 0.12 \ll 0.9$.


Consequently, the space complexity of the inverted list approach is at least
$$\Omega\left(\card{R} \cdot \min_i \sigma_i \right)$$
where the $\sigma_i$ denote the partial selectivities of the query.
Therefore, if \emph{all} of the dimensions are queried with high partial selectivity,
even the best strategy for computing the intersection requires computing a large intermediate set.

In conclusion, this class of queries is problematic for methods based on the inverted index approach.
The asymptotic complexity is not rooted within the individual single-dimensional indexes, though;
it is present regardless of the exact type of single-dimensional index that is used.
Therefore, the entire concept of composing multiple single-dimensional index structures in this way
is suboptimal for efficiently processing multi-column queries with low total selectivity.

Later, we will see how the Elf builds on this method by essentially building one inverted list
\emph{for each value} in the first attribute of the relation,
instead of one inverted list for each dimension.

\section{Multi-dimensional index structures}\label{sec:trees}
The drawbacks of the inverted list method
suggest that the multi-dimensional indexing problem cannot simply be solved by
composing multiple single-dimensional index structures.
Instead, specialised multi-dimensional index structures are needed.

In this section, we will introduce a variety of \emph{tree-based} multi-dimensional index structures.
Other proposed approaches to multi-dimensional indexing, such as space-filling curves \cite{908985},
grid files~\cite{Nievergelt:1984:GFA:348.318586}
and vector approximation methods~\cite{Ferhatosmanoglu:2000:VAB:354756.354820}
are, while interesting, not relevant to our discussion of the Elf.
Therefore, we will not describe them further.


The main property exploited by tree-based methods,
both single-dimensional and multi-dimensional,
is that the query shape for a range query is \emph{convex}.
What this means is that
the desired data points form a cluster in the data
space that never contains any undesired data points \enquote{in between} the desired points.

Two corollaries of this property lead to the two main classes of multi-dimensional tree-based index structures \cite{DBLP:books/daglib/0098215}:
\begin{enumerate}
  \item
    \emph{Two points that are close together are likely either both included or not included in the result set.}
    Therefore, it is desirable to group data points that are close together in space and treat them as a single object.
    This group of data points is more efficient to handle than the set of data points itself, thereby reducing search time.
    This approach leads to the \emph{data-partitioning} class of index structures.

  \item
    \emph{Two points that are far away from each other are likely not both included in the result set.}
    (This is true if the total selectivity of the query is assumed to be low.)
    Therefore, it is desirable to partition the data space, ahead of time, into a number of regions.
    The search algorithm then only has to examine a subset of the regions to answer the query, thereby reducing search time.
    This approach leads to the \emph{space-partitioning} class of index structures.
\end{enumerate}
In the following, we will describe the R-Tree, a data-partitioning index structure, and the \kdtree, a space-partitioning index structure.
They each exploit the principles described above.
Later, we will see that the Elf structure borrows concepts from both structures,
essentially forming a hybrid between the two classes.

\subsection{R-Tree}\label{sec:rtree}
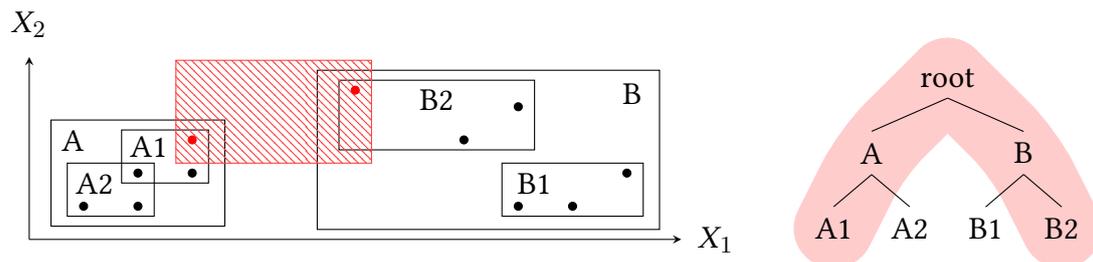
\begin{figure}[htb]
  \centering
\pgfdeclarelayer{background}
\pgfdeclarelayer{foreground}
\pgfsetlayers{background,main,foreground}
\begin{tikzpicture}\begin{axis}[name=space,
    width=0.8\textwidth-2cm,
    height=4cm,
    xmin=0, xmax=12, ymin=0, ymax=5.5,
    axis lines=middle,
    xtick=\empty, ytick=\empty,
    xlabel=$X_1$, ylabel=$X_2$,
    mark size=1.5,
    every axis x label/.style={
        at={(ticklabel* cs:1.01)},
        anchor=west,
    },
    every axis y label/.style={
        at={(ticklabel* cs:1.05)},
        anchor=south,
    },
]
\draw (axis cs:0.4,0.4) rectangle (axis cs:3.6,3.6);
\draw (axis cs:5.3,0.3) rectangle (axis cs:11.6,5.1);
\draw (axis cs:0.7,0.7) rectangle (axis cs:2.3,2.3);
\draw (axis cs:1.7,1.7) rectangle (axis cs:3.3,3.3);
\draw (axis cs:5.7,4.8) rectangle (axis cs:9.3,2.7);
\draw (axis cs:8.7,0.7) rectangle (axis cs:11.3,2.3);
\draw [red!75, pattern=north west lines, pattern color=red!75] (axis cs:2.7,2.3) rectangle (axis cs:6.3,5.4);

\addplot [only marks, draw=black] coordinates {
  (1,1) (2,1)
  (3,2) (2,2) 
  (9,4) (11,2) (8,3)
  (9,1) (10,1)
};
\addplot [only marks, fill=red,red] coordinates { (3,3)  (6,4.5) };

\node at (axis cs:0.8,3) {A};
\node at (axis cs:11.1,4.4) {B};
\node at (axis cs:1.2,1.7) {A2};
\node at (axis cs:2.2,2.8) {A1};
\node at (axis cs:7.5,4.2) {B2};
\node at (axis cs:9.3,1.7) {B1};

\end{axis}
\tikzset{level/.style={sibling distance=2cm/#1,level distance=1cm}}
\node [right=3cm of space.north east,
  anchor=north west] (mem) {root}
  child {node {A}
    child {node {A1}}
    child {node {A2}}
    }
  child {node {B}
    child {node {B1}}
    child {node {B2}}
    }
;

\begin{pgfonlayer}{background}
  \draw[rounded corners=1em,line width=2.5em,red!20,cap=round]
  (mem.center) -- (mem-1.center) -- (mem-1-1.center);
  \draw[rounded corners=1em,line width=2.5em,red!20,cap=round]
  (mem.center) -- (mem-2.center) -- (mem-2-2.center);
\end{pgfonlayer}

\end{tikzpicture}
  \caption[R-Tree for an example data set]{
  Left:
  Example data set with R-Tree bounding rectangles and an example range query (red).
  Right:
  In-memory layout of the R-Tree.
  Only the highlighted nodes are visited during search.}
  \label{fig:rtree}
\end{figure}

The R-Tree \cite{Guttman:1984:RDI:602259.602266} finds groups of data points that are close to each other and represents each group by its minimum bounding rectangle (MBR).
The MBRs themselves are grouped again to form a tree with the original tuples as leaves.
\Autoref{fig:rtree} shows the R-Tree for an example data set.
Groups of points are represented by their minimal bounding rectangles.

The search algorithm starts at the root and checks, for each group, if its bounding rectangle intersects the query rectangle.
If they intersect, the algorithm recurses into the group.
If they do not intersect, any points in the group cannot possibly match the query, and the descent is skipped.

While simple in concept, building an efficient R-Tree involves a number of nontrivial decisions, such as how to choose which points to group.
Poor decisions can lead to degenerated trees with undesired search performance.
Additionally, the original version of the R-Tree allows the bounding rectangles
of different groups to overlap to achieve balanced sizes for the groups.

Multiple variants of the R-Tree, such as the $\R^+$-Tree, the $\R^*$-Tree and the X-Tree have been proposed to try and address these problems using more sophisticated splitting and grouping heuristics
\cites{Arge:2008:PRP:1328911.1328920,Beckmann:1990:RER:93605.98741,Sellis:1987aa,Xia:2005:IRO:1097064.1097083,berchtold2001index}.

\subsection{\kdtree}
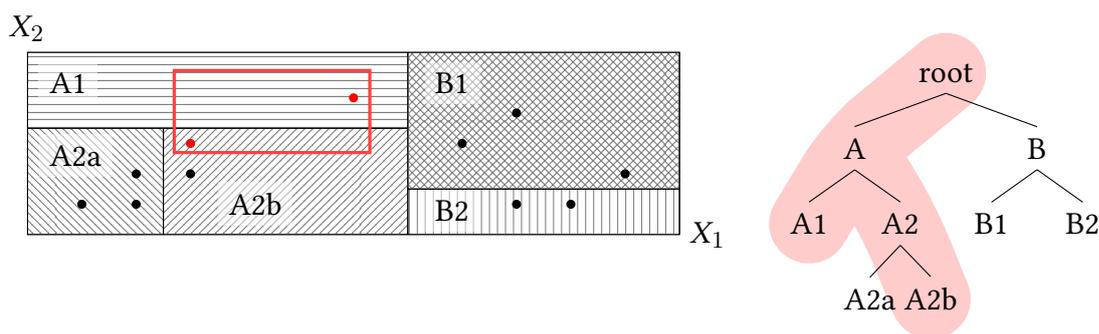
\begin{figure}[htb]
  \centering
\pgfdeclarelayer{background}
\pgfdeclarelayer{foreground}
\pgfsetlayers{background,main,foreground}
\begin{tikzpicture}
[
  desc/.style={
    fill=white,
    fill opacity=0.5,
    text opacity=1,
    anchor=west,
  },
]
\begin{axis}[name=space,
    width=0.8\textwidth-2cm,
    height=4cm,
    xmin=0, xmax=12, ymin=0, ymax=6,
    xtick=\empty, ytick=\empty,
    xlabel=$X_1$, ylabel=$X_2$,
    mark size=1.5,
    every axis x label/.style={
        at={(ticklabel* cs:1)},
        anchor=west,
    },
    every axis y label/.style={
        at={(ticklabel* cs:1)},
        anchor=south,
    },
]
\addplot [only marks, draw=black] coordinates {
  (1,1) (2,1)
  (3,2) (2,2) 
  (9,4) (11,2) (8,3)
  (9,1) (10,1)
};
\draw [pattern=north west lines, pattern color=black!50] (axis cs:0,0) rectangle (axis cs:2.5,3.5);
\draw [pattern=north east lines, pattern color=black!50] (axis cs:2.5,0) rectangle (axis cs:7,3.5);
\draw [pattern=horizontal lines, pattern color=black!50] (axis cs:0,3.5) rectangle (axis cs:7,6);
\draw [pattern=vertical lines, pattern color=black!50] (axis cs:7,0) rectangle (axis cs:12,1.5);
\draw [pattern=crosshatch, pattern color=black!50] (axis cs:7,1.5) rectangle (axis cs:12,6);
\addplot [only marks, fill=red,red] coordinates { (3,3)  (6,4.5) };

\draw [red!75, very thick] (axis cs:2.7,2.7) rectangle (axis cs:6.3,5.4);

\node [desc] at (axis cs:0.2,5) {A1};
\node [desc] at (axis cs:0.2,2.6) {A2a};
\node [desc] at (axis cs:3.5,1) {A2b};
\node [desc] at (axis cs:7.3,5) {B1};
\node [desc] at (axis cs:7.3,0.8) {B2};

\end{axis}

\tikzset{level/.style={sibling distance=2.4cm/#1,level distance=1cm}}
\node [right=3cm of space.north east,
  anchor=north west] (mem) {root}
  child {node {A}
    child {node {A1}}
    child {node {A2}
        child {node {A2a}}
        child {node {A2b}}
      }
    }
  child {node {B}
    child {node {B1}}
    child {node {B2}}
    }
;

\begin{pgfonlayer}{background}
  \draw[rounded corners=1em,line width=2.5em,red!20,cap=round]
  (mem.center) -- (mem-1.center) -- (mem-1-1.center);
  \draw[rounded corners=1em,line width=2.5em,red!20,cap=round]
  (mem.center) -- (mem-1.center) -- (mem-1-2.center) -- (mem-1-2-2.center);
\end{pgfonlayer}

\end{tikzpicture}
  \caption[{\kdtree} for an example data set]{
  Left:
  Example data set with {\kdtree} partitions and an example range query (red).
  Each partition is drawn in an individual hatching.
  Right: In-memory layout of the {\kdtree}.
  Only the highlighted nodes are visited during search.}
  \label{fig:kdtree}
\end{figure}

The {\kdtree} and its variants
are examples of \emph{space-partitioning} index structures
\cite{Bentley:1975:MBS:361002.361007,robinson1981kdb}.
The initial data space is split in one of the dimensions along a hyperplane.
Both resulting halves of the data space are recursively split again,
alternating between the dimensions.
This process continues until every partition contains a sufficiently small number of points.
The partitions are then stored as a tree, with the inner nodes containing the \emph{pivot values} of the split.
\Autoref{fig:kdtree} shows the {\kdtree} for an example data set.
In the example, the data space is recursively split
until each partition contains at most three points.
The partitions are made alternately along the $X_1$ and $X_2$ axes.
During the search, only partitions that intersect the query rectangle (red)
have to be examined further; the rest of the space is be efficiently \emph{pruned}.

The performance of the {\kdtree} primarily depends on the size of the query shape,
and the number of dimensions \cite{Freidman1977}.
For point queries, only one side of each partition has to be examined, since the query point
cannot intersect \emph{both} partitions.
A larger query shape increases the likelihood that both partitions intersect.
Another impact factor is the positioning of the points;
if the pivot values are chosen poorly,
the data points will be unevenly distributed between the partitions.
In this case, the search will have to visit many more branches than with a better choice of pivots.
As we will see in the following section, the \say{curse of dimensionality}
brings a similar effect if the dimensionality of the data set increases.
This leads to deteriorating performance of the {\kdtree} for high-dimensional data sets.

\section{Challenges of high-dimensional data sets}\label{sec:curse}
\subsection{The curse of dimensionality}
Tree-based multi-dimensional index structures perform well for low-selectivity
queries in a small number of dimensions.
However, as the number of dimensions increases,
there is evidence for a deterioration
in the efficiency of tree-based structures.
This is known as the \emph{curse of dimensionality}.
Around ten to sixteen dimensions are usually cited as as the point
at which proposed multi-dimensional index structures again perform worse
than sequential scans \cites{weber1998quantitative,berchtold2001index}.
We will therefore call data sets with dimension 1 \emph{single-dimensional},
dimensions 2--16 \emph{multi-dimensional},
and dimensions greater than 16 \emph{high-dimensional}.
We will focus on multi-dimensional data sets,
which are commonly found in business data analytics.
In contrast, high-dimensional data sets are  commonly found in image and multimedia processing.

The deterioration of index performance on the transition from multi-dimensional to high-dimensional data
can be explained by the exponential increase in \emph{volume} of a high-dimensional data space.
For example, a 16-dimensional data space with ten possible values in each dimension
allows $10^{16}$ different combinations of values.
Therefore, $10^{16}$ unique data points would be required to fully occupy the data space.
In practice, the size of data sets does not increase exponentially with the number of dimensions.
Therefore, the data space is occupied very sparsely, containing large parts of \say{empty space}.

The rapid expansion in volume also causes distance metrics to become ill-defined \cite{aggarwal2001surprising}.
For example, consider the family of $L_p$ norms for $\RR^k$, defined as follows for
$x,y\in\RR^k, p\geq 1$:
$$
\|x-y\|_p =
\|(x_1,\dots,x_k)-(y_1,\dots,y_k)\|_p =
\left(
  \sum_{i=1}^k (x_i-y_i)^p
\right)^{1/p}
$$
A distance metric like the $L_p$ norm is usually employed e.g.\ by the R-Tree
to find groups of tuples that are close together,
and by the {\kdtree} to find a pivot value that evenly partition the data space.

For a set of points $R$, large increases in $k$ leads to a confusing situation:
the computed distance between a point $x\in R$ and its nearest neighbour approaches
the distance between $x$ and the point farthest away from $x$, leading to:
$$\min_{y\in R\setminus\{x\}} \|x-y\|_p \approx \max_{y\in R\setminus\{x\}} \|x-y\|_p$$
Therefore, the distance between points becomes useless as a metric for measuring
the \say{similarity} of points.
In practice, this leads to degenerate tree structures where the evaluation of most queries
deteriorates to poorly-optimised linear scans.

\subsection{Overcoming the curse}
Due to the curse of dimensionality, high-dimensional data sets lead to poor performance
for proposed multi-dimensional index structures.
Commonly, attempts are made to reduce the dimensionality of the data set, using
Principal Component Analysis~(PCA) or other techniques
to obtain a data set with a reduced number of dimensions \cites{Pearson1901,Sarveniazi2014}.
Through the reduction in dimensionality, the adverse affects of the curse of dimensionality can be reduced as well.
In fact,
\say{there is a consensus in the high-dimensional data analysis community that the only
reason any methods work in very high dimensions is that, in fact, the data are not
truly high-dimensional}
\cite{levina2004maximum}.
However, for our case of range queries on multi-dimensional
(not high-dimensional) data sets,
the dimension reduction only shifts the problems from the index structure
to the dimension reduction method,
since the reduced dimensions cannot be queried.

Another interesting approach to overcome the curse of dimensionality is
through the use of \emph{space-dividing methods}.
One common property of traditional structures like the R-Tree and the {\kdtree}
is that the partitions of the data space do not reduce the dimension of the data space.
For example, the R-Tree might partition a two-dimensional rectangular space into several smaller rectangles, each still two-dimensional.
The \emph{tree striping} method~\cite{DBLP:books/daglib/0098215} instead attempts to \emph{divide} the data space into two or more subspaces of lower dimension than the original space.
These subspaces are then independently searched, and the results merged to obtain the final result.
This approach is interesting, because the space division also reduces the adverse effects of the curse of dimensionality.

In its proposed form, though, the tree striping approach is impractical.
In particular, it assumes that the data space can be decomposed into disjoint subspaces such that the original data set is the Cartesian product of the of the two subspaces.
This is not a property exhibited by real-world or benchmark data sets.
Later, we will see that the Elf performs a space division similar to the tree striping method,
without exhibiting this limitation.


\section{The Elf}
The Elf data structure~\cite{elf} was recently proposed to overcome the issues of traditional high-dimensional index structures.
In this section, we will introduce the Elf through an example, compare the Elf to the other index structures described above,
and describe the algorithms involved in manipulating the Elf.

\subsection{Introduction}
We will illustrate the design of the Elf data structure using an example relation shown in \autoref{fig:elfintrodata}.
The corresponding Elf is shown in \autoref{fig:elfintroelf}.
The basic structure of the Elf is a tree of all combinations of attribute values present in the data set.
For each tuple, the labels on the edges from the root to the leaf correspond to the row of values in the table.
For example, the path from the root to C has the edge labels $(1,3,3)$, equal to the attribute values of C.
This results in a fixed-depth tree, with exactly one level for each attribute $X_1,\dots,X_k$.
Note that the Elf assumes an ordering of the attributes within the tree.
This ordering is an important parameter which we will discuss later.
\begin{figure}[tb]
  \centering
\pgfdeclarelayer{background}
\pgfdeclarelayer{foreground}
\pgfsetlayers{background,main,foreground}

\begin{tikzpicture}[
  n/.style={
    draw,
    shape=circle,
    minimum size=3em,
    fill=red!20,
  },
  l/.style={
    draw=none,
    fill=none,
    auto=right,
  },
  level distance=3cm,
  sibling distance=.6cm,
]


\node [align=left,text width=4cm] at (-7cm,0cm) {
 \begin{tabular}{llll} \toprule
    $X_1$ & $X_2$ & $X_3$ & $\mathit{Ref}$ \\ \midrule
    1     & 3     & 5     & A \\
    2     & 4     & 5     & B \\
    1     & 3     & 3     & C \\
    2     & 2     & 1     & D \\
    3     & 3     & 3     & E \\ \bottomrule
  \end{tabular}
  \smallskip
  {\setcapindent{0pt}
  \captionof{table}{Example relation with $k$=3}\label{fig:elfintrodata}}
};

\tikzstyle{bign} = [draw,shape=circle,minimum size=2em,inner sep=0pt,fill=green!20]
\begin{scope}[nodes=n]
\Tree [.\node(t){$(~)$};
    \edge node[l] {1};
    [.\node(a1){$(1)$};
      \edge node[l] {3};
      [.\node(a2){$(1,3)$};
        \edge node[l] {3};
        [.\node[bign]{C}; ]
        \edge node[l,auto=left](a3) {5};
        [.\node[bign](A){A}; ]
        ]
    ]
    \edge node[l] {2};
    [.\node[xshift=-1.7mm]{$(2)$};
      \edge node[l] {2};
      [.{$(2,2)$}
        \edge node[l] {1};
        [.\node[bign]{D}; ]
      ]
      \edge node[l,auto=left] {4};
      [.{$(2,4)$}
        \edge node[l,auto=left] {5};
        [.\node[bign]{B}; ]
      ]
    ]
    \edge node[l,auto=left] {3};
    [.{$(3)$}
      \edge node[l] {3};
      [.{$(3,3)$}
        \edge node[l] {3};
        [.\node[bign]{E}; ]
      ]
    ]
  ]
\end{scope}

\begin{pgfonlayer}{background}
  \draw[rounded corners=2em,line width=5em,blue!15,cap=round]
  (t.center) -- (a1.center) -- (a2.center) -- (A.center);
\end{pgfonlayer}
\end{tikzpicture}
  \caption[An example Elf]{
    An example relation together with the corresponding Elf in prefix tree layout.
    Interior nodes are marked with their prefix and shown in red.
    Leaf nodes are shown in green.
    The blue highlight marks the sequence of values for tuple A.
  }\label{fig:elfintroelf}
\end{figure}

Every node of the Elf is either an interior node, or a leaf.
The root is an interior node.
The nodes connected to an interior node are called its \emph{subnodes}.
The subnodes that are themselves interior nodes are called \emph{sub-Elfs},
since we will later see that they always represent another \say{smaller} Elf structure.
For a given interior node, the set of leafs reachable from it is called the set of tuples \emph{below} the node.
Similarly, the first dimension $X_1$ is called the \emph{topmost} or \emph{first} dimension.

\paragraph{Exploiting prefix redundancies.}
Each interior node corresponds to a unique \emph{prefix} of the data set.
For example, node $(1)$ contains all tuples that have $X_1=1$, and arbitrary values in the other columns.
A \emph{prefix redundancy} of length $l$ is the occurence of multiple data points that share the same values in the first $l$ attributes of the relation.
For example, the tuples A and C share a prefix redundancy of length 2, while B and D share a prefix redundancy of length 1.
Within the tree, multiple occurences of a prefix redundancy are always stored only once.
The tree branches out at every point where a prefix redundancy ends.
For data sets that have a large number of prefix redundancies, this leads to a compact structure.
The fact that each prefix is uniquely \say{owned} by a single node is also desirable when searching, as we will see later.

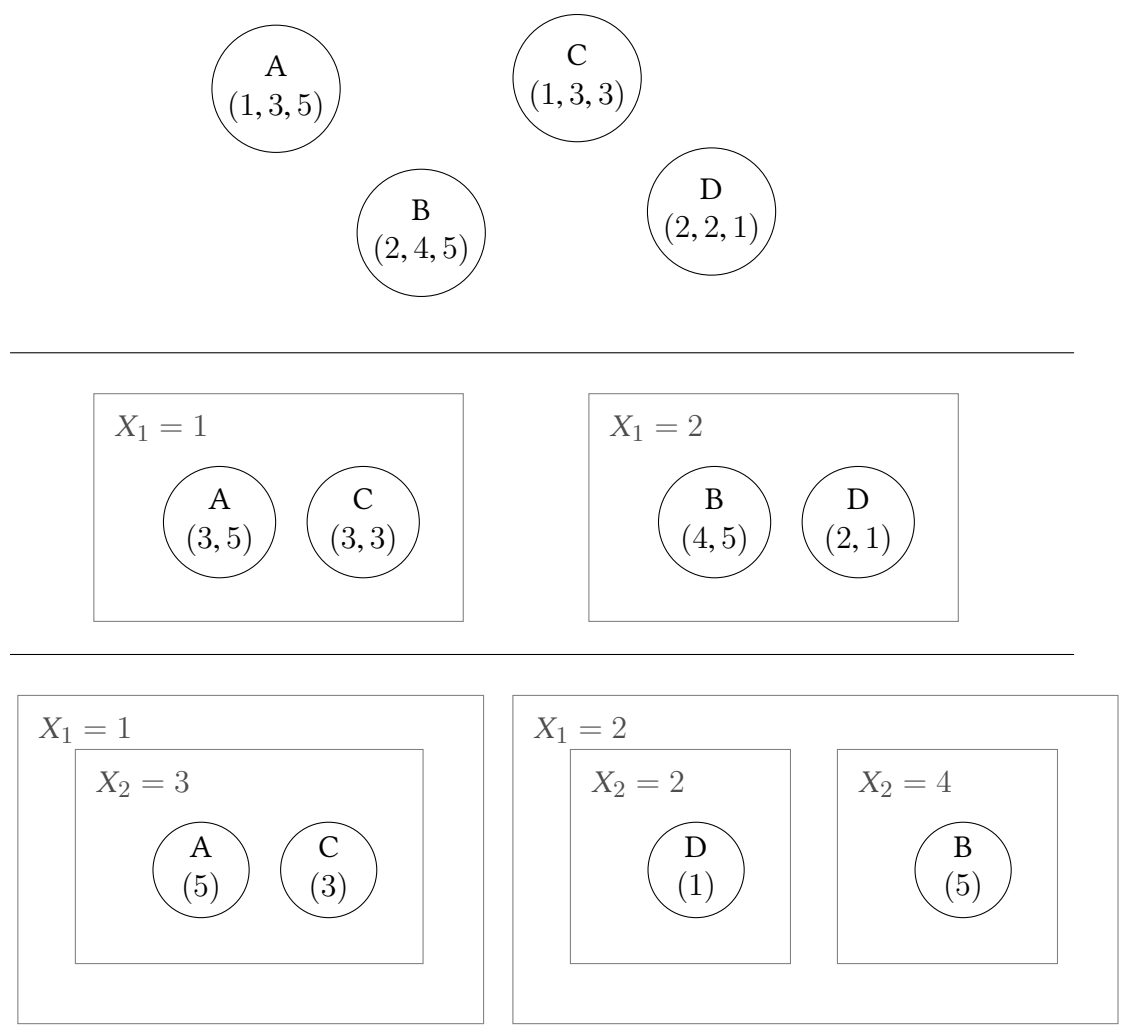
\begin{figure}[tb]
  \centering
\begin{tikzpicture}[
  grouplabel/.style={font=\color{black!70}},
  node distance= 4mm and 4mm,
]
\tikzstyle{n1} = [draw,shape=circle,minimum size=4em,inner sep=0pt,align=center]
\tikzstyle{n2} = [draw,shape=circle,minimum size=3.5em,inner sep=0pt,align=center]
\tikzstyle{n3} = [draw,shape=circle,minimum size=3em,inner sep=0pt,align=center]

\begin{scope}[xshift=2.5cm]
  \node[n1] (A1) {A\\$(1,3,5)$};
  \node[n1, below right=1cm of A1] (B1) {B\\$(2,4,5)$};
  \node[n1, above right=1.2cm of B1] (C1) {C\\$(1,3,3)$};
  \node[n1, below right=0.8cm of C1] (D1) {D\\$(2,2,1)$};
\end{scope}
\draw (-1cm,-3.5cm) -- (13cm,-3.5cm);

\begin{scope}[yshift=-4.5cm,xshift=1cm]
  \node (x11) [grouplabel] { $X_1=1$ };
  \node[n2,below=0.5cm of x11.east] (A2) {A\\$(3,5)$};
  \node[n2,right=of A2] (C2) {C\\$(3,3)$};
  \node[draw=none,below right=0.5cm of C2] (d11) {};
  \node (c11) [draw=black!50, fit={(x11) (A2) (C2) (d11)}] {};

  \node (x12) [grouplabel,right=5cm of x11] { $X_1=2$ };
  \node[n2,below=0.5cm of x12.east] (B2) {B\\$(4,5)$};
  \node[n2,right=of B2] (D2) {D\\$(2,1)$};
  \node[draw=none,below right=0.5cm of D2] (d12) {};
  \node (c12) [draw=black!50, fit={(x12) (B2) (D2) (d12)}] {};
\end{scope}
\draw (-1cm,-7.5cm) -- (13cm,-7.5cm);

\begin{scope}[yshift=-8.5cm]
  \node (x21) [grouplabel] { $X_1=1$ };
  \node (x211) [grouplabel,below=of x21.east] { $X_2=3$ };
    \node[n3,below=0.5cm of x211.east] (A3) {A\\$(5)$};
    \node[n3,right=of A3] (C3) {C\\$(3)$};
    \node[draw=none,below right=0.5cm of C3] (d21) {};
  \node (c211) [draw=black!50, fit={(x211) (A3) (C3) (d21)}] {};
  \node[draw=none,below right=0.5cm of c211] (d22) {};
  \node (c21) [draw=black!50, fit={(x21) (c211) (d22)}] {};

  \node (x22) [grouplabel,right=5cm of x21] { $X_1=2$ };
    \node (x221) [grouplabel,below=of x22.east] { $X_2=2$ };
      \node[n3,below=0.5cm of x221.east] (D3) {D\\$(1)$};
      \node[draw=none,below right=0.5cm of D3] (d221) {};
    \node (c221) [draw=black!50, fit={(x221) (D3) (d221)}] {};

    \node (x222) [grouplabel,right=2cm of x221.east] { $X_2=4$ };
      \node[n3,below=0.5cm of x222.east] (B3) {B\\$(5)$};
      \node[draw=none,below right=0.5cm of B3] (d222) {};
    \node (c222) [draw=black!50, fit={(x222) (B3) (d222)}] {};

  \node[draw=none,below right=0.5cm of c222] (d2223) {};
  \node (c21) [draw=black!50, fit={(x22) (c221) (c222) (d2223)}] {};
\end{scope}
\end{tikzpicture}
  \caption[Intermediate grouping stages of the Elf construction]{
    Intermediate grouping stages of the Elf construction.
    Tuple E is omitted for visual clarity.
  }\label{fig:elfgrouping}
\end{figure}

\paragraph{Building the tree through recursive grouping.}
The iterative elimination of prefix redundancies is equivalent to recursively grouping the data set.
These grouping steps are illustrated in \autoref{fig:elfgrouping}, and we will describe them here.
First, the data points are grouped by their value of the first attribute $X_1$.
This leads to a number of groups, one for every distinct value in $X_1$.
Within each group, all points share the same value in the first attribute.
This value uniquely identifies the group and is called the group's \emph{label}.
Storing the value for each item in a group is therefore redundant and can be omitted.

Now, the points within each group are grouped \emph{again}; this time using the values of $X_2$.
This creates, within each group, a set of sub-groups.
The recursion continues until the data is grouped for all attributes.

Now, the groups can be efficiently stored as a tree of the group labels.
Each edge is marked with the label of the next group that the edge leads to.
This forms the Elf data structure as shown previously in \autoref{fig:elfintroelf}.

\paragraph{Depth-first search with pruning.}
A range query can be evaluated on the Elf with a depth-first search through the tree.
The search starts at the root and recursively descends into the tree.
The main optimisation afforded by the structure of the tree is that
each interior node represents a unique prefix (the \emph{pruning property}).
The prefix values can be compared with the query range to determine whether
the prefix matches the query.
If it does not match, descending into the subtree can be skipped entirely.




\subsection{Recursive definition}\label{sec:elfformal}
In this section, we show the \emph{self-similarity} of the Elf by describing its construction through a recursive formula.
In particular, we will show that every subtree of the Elf is again another Elf;
in fact, it is the Elf of a lower-dimensional subset of the original data set.

Consider a $r$-dimensional relation $R$ with attributes $(X_1, \dots, X_r)$.
Building the Elf for $R$ is accomplished by computing the recursive function $\Elf_r(R)$.

$\Elf_k(S)$ is built from a set $S$ of points in a $k$-dimensional Euclidean space.
We assume that each point $s \in S$ corresponds to exactly one tuple in $R$, identified by a unique tuple ID.

\paragraph{Base case.}
If $k=1$, then $\Elf_1(S)$ is simply a map from every attribute value present in $S$ to the ID of the tuple having that value.

\paragraph{Reductive step.}
Otherwise, let $P$ be the unique set of $(k-1)$-dimensional parallel hyperplanes such that:
\begin{enumerate}
  \item All planes $p \in P$ contain at least one point. (No empty planes are created.)
  \item Every point in $S$ is contained by exactly one plane in $P$.
  \item
    All planes $p \in P$ are normal to the $X_{r-k+1}$ axis.
    (In the first iteration, $r=k$, and therefore the planes are normal to the $X_1$ axis.)
\end{enumerate}
\begin{figure}[tb]\centering
  \begin{tikzpicture}
  \begin{axis}[
    3d box=background,
    samples=5,
    domain=-4:4,
    xticklabels={},
    yticklabels={},
    zticklabels={},
    xlabel=$X_2$,ylabel=$X_3$,zlabel=$X_1$,
  ]
    \addplot3[surf,fill=blue!60] coordinates {
        (0,0,0) (.4,0,0) (2.4,0,0) (3,0,0)

        (0,3,0) (.4,3,0) (2.4,3,0) (3,3,0)
    };

    \addplot3[surf,fill=yellow!60] coordinates {
        (0,0,2) (1.5,0,2) (3,0,2)

        (0,3,2) (1.5,3,2) (3,3,2)
    };

    \addplot3[surf,fill=red!60] coordinates {
        (0,0,4) (1,0,4) (2,0,4) (3,0,4)

        (0,3,4) (1,3,4) (2,3,4) (3,3,4)
    };

    \addplot3[only marks,mark=*,mark size=2pt] coordinates {
        (1,2.7,4)
        (2,1.7,4)

        (1.5,1,2)

        (.4,1,0)
        (2.4,1.8,0)
    };

  \end{axis}
\end{tikzpicture}
  \caption[Elf planes partitioning a three-dimensional data set]
    {Three two-dimensional Elf planes partitioning a three-dimensional example data set along the $X_1$ axis, i.e.\ for the initial case where $k=r$.
    Every plane contains at least one point, and all planes are parallel to each other.
    These partitions represent the first level of the final Elf.
    }\label{fig:lasagna}
\end{figure}
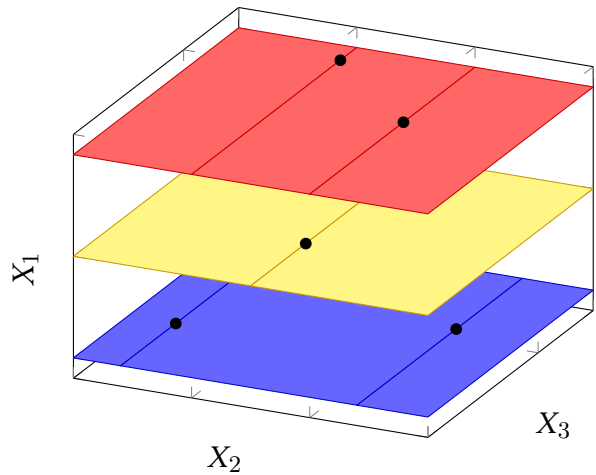
\Autoref{fig:lasagna} illustrates the position of the planes and how they \say{slice} up the data space.
From the properties of $P$ follows that:
\begin{enumerate}
  \item
    For each plane $p\in P$, all points $x_1, x_2 \in p$ have the same value in the attribute $X_{r-k+1}$.
    This value is called $\oplabel(p)$.
  \item
    The labels of two different planes in $P$ are different.
\end{enumerate}
Therefore, we can store each plane $p \in P$ more efficiently.
Instead of storing the set of points within $p$ as a set of $k$-dimensional points,
we only store $p$'s label once,
and then remove the $X_{r-k+1}$ coordinate from every point in $p$, obtaining $p'$.

Now $p'$ is just a set of $(k-1)$-dimensional points.
We recursively apply this technique and do not store $p'$ directly, but instead compute $\Elf_{k-1}(p')$.

This leads us to the following equation:
let $p_1, p_2, \dots, p_l$ denote the planes in $P$, and let $p'_1, p'_2, \dots, p'_l$ denote the sets of reduced points as described above.
Then:
\begin{equation}
  \Elf_k(S) = \left\{ \, (\oplabel(p_1), \Elf_{k-1}(p'_1)), \,\,\dots,\,\, (\oplabel(p_l), \Elf_{k-1}(p'_l)) \,\right\}
\end{equation}

At each level, $k$ decreases by 1.
In the $i$'th level of recursion, the planes will be normal to the $X_i$ axis, starting at $X_1$.
The base case where $k=1$ (see above) stops the recursion.
Therefore, the recursive application leads to a tree of fixed depth $k$.

In summary, we have reduced the problem of indexing a $k$-dimensional data set to indexing multiple $(k-1)$-dimensional data sets.
The $k$-dimensional data set is partitioned by creating a number of parallel $(k-1)$-dimensional hyperplanes normal to an axis so that every data point lies in exactly one of the planes.
For each plane, we store its characteristic label,
and then construct the $(k-1)$-dimensional Elf for the points contained within.
This creates a tree of fixed depth, with the tuple IDs as leaves.

\subsection{In-memory layout}\label{sec:hashmap}
So far, we have described the Elf in its \emph{prefix tree layout}.
However, this layout this very inefficient for storing in memory directly.
Instead, the Elf is brought into an optimised \emph{linearised layout},
through an intermediate step called the \emph{dimension list} layout.
Additionally, two types of deteriorations are addressed, which we will describe in this section.

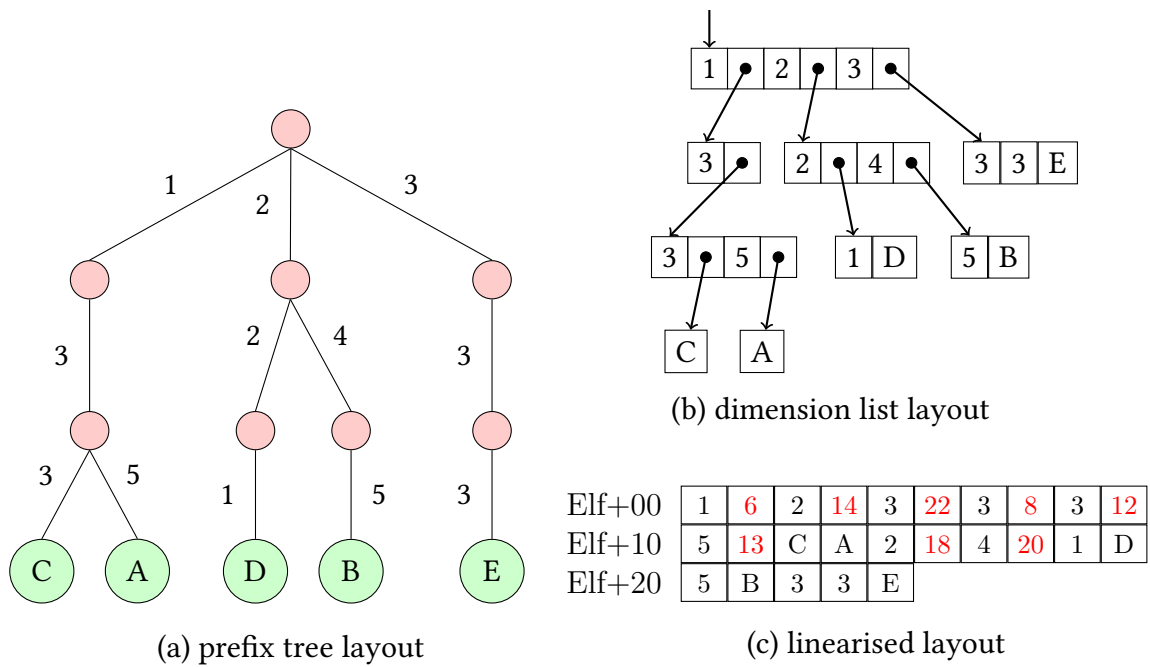
\begin{figure}[tb]
  \centering
\begin{tikzpicture}[
  nodes={draw},
  n/.style={
    draw,
    shape=circle,
    minimum size=1.2em,
    fill=red!20,
  },
  l/.style={
    draw=none,
    fill=none,
    auto=right,
  },
  c/.style={
    font={\fontsize{10}{10}\selectfont},
    inner sep=0,
    minimum height=5mm,
    minimum width=6mm,
    align=center,
  },
  level distance=2cm,
  sibling distance=.4cm,
]
\tikzstyle{bign} = [draw,shape=circle,minimum size=2em,inner sep=0pt,fill=green!20]
\begin{scope}[xshift=-7.2cm,nodes=n,yshift=-8mm]
\Tree [.\node(t){};
    \edge node[l] {1};
    [.{}
      \edge node[l] {3};
      [.{}
        \edge node[l] {3};
        [.\node[bign]{C}; ]
        \edge node[l,auto=left] {5};
        [.\node[bign]{A}; ]
        ]
    ]
    \edge node[l] {2};
    [.\node[xshift=-1.7mm]{};
      \edge node[l] {2};
      [.{}
        \edge node[l] {1};
        [.\node[bign]{D}; ]
      ]
      \edge node[l,auto=left] {4};
      [.{}
        \edge node[l,auto=left] {5};
        [.\node[bign]{B}; ]
      ]
    ]
    \edge node[l,auto=left] {3};
    [.{}
      \edge node[l] {3};
      [.{}
        \edge node[l] {3};
        [.\node[bign]{E}; ]
      ]
    ]
  ]
\end{scope}
\node [below=6.3cm of t,l] {(a) prefix tree layout};

\begin{scope}[xshift=-0.5cm]
\matrix (root) [draw=none,column sep=-\pgflinewidth]
{
\node[rounded corners=0pt] (root1) {1};
& \node[rounded corners=0pt] (root1p) {\phantom{1}};
& \node[rounded corners=0pt] (root2) {2};
& \node[rounded corners=0pt] (root2p) {\phantom{1}};
& \node[rounded corners=0pt] (root3) {3};
& \node[rounded corners=0pt] (root3p) {\phantom{1}};
\\};
\draw[thick,<-] (root1.north) -- ++(0,5mm);
\end{scope}
\begin{scope}[yshift=-1.25cm]
  \begin{scope}[xshift=-1.5cm]
  \matrix (rootto1) [draw=none,column sep=-\pgflinewidth,rounded corners=3pt]
  {
  \node[rounded corners=0pt] (rootto1to3) {3};
  & \node[rounded corners=0pt] (rootto1to3p) {\phantom{1}};
  \\};
  \end{scope}

  \begin{scope}[xshift=0.25cm]
  \matrix (rootto2) [draw=none,column sep=-\pgflinewidth,rounded corners=3pt]
  {
    \node[rounded corners=0pt] (rootto21) {2};
  & \node[rounded corners=0pt] (rootto21p) {\phantom{1}};
  & \node[rounded corners=0pt] (rootto22) {4};
  & \node[rounded corners=0pt] (rootto22p) {\phantom{1}};
  \\};
  \end{scope}

  \begin{scope}[xshift=2.4cm]
  \matrix [draw=none,column sep=0]
  {
  \node[rounded corners=0pt] (Emono) {\vphantom{D}3};
  &\node[rounded corners=0pt] {\vphantom{D}3};
  & \node[rounded corners=0pt] {E};\\
  };
  \end{scope}
\end{scope}
\draw[fill=black] (root1p.center) circle (2pt);
\draw[fill=black] (root2p.center) circle (2pt);
\draw[fill=black] (root3p.center) circle (2pt);
\draw[thick,->] (root1p.center) -- (rootto1to3.north);
\draw[thick,->] (root2p.center) -- (rootto21.north);
\draw[thick,->] (root3p.center) -- (Emono.north);

\begin{scope}[yshift=-2.5cm]
  \begin{scope}[xshift=-1.5cm]
  \matrix (rootto1to3) [draw=none,column sep=-\pgflinewidth,rounded corners=3pt]
  {
    \node[rounded corners=0pt] (rootto1to31) {3};
  & \node[rounded corners=0pt] (rootto1to31p) {\phantom{1}};
  & \node[rounded corners=0pt] (rootto1to32) {5};
  & \node[rounded corners=0pt] (rootto1to32p) {\phantom{1}};
  \\};
  \end{scope}
  \draw[fill=black] (rootto1to3p.center) circle (2pt);
  \draw[->,thick] (rootto1to3p.center) -- (rootto1to31.north);

  \begin{scope}[xshift=0.5cm]
  \matrix [draw=none,column sep=0]
  {
  \node[rounded corners=0pt] (Dmono) {\vphantom{D}1};
  & \node[rounded corners=0pt] {D};\\
  };
  \end{scope}

  \begin{scope}[xshift=2cm]
  \matrix [draw=none,column sep=0]
  {
  \node[rounded corners=0pt] (Bmono) {\vphantom{B}5};
  & \node[rounded corners=0pt] {B};\\
  };
  \end{scope}

  \draw[fill=black] (rootto21p.center) circle (2pt);
  \draw[fill=black] (rootto22p.center) circle (2pt);
  \draw[->,thick] (rootto21p.center) -- (Dmono.north);
  \draw[->,thick] (rootto22p.center) -- (Bmono.north);

\end{scope}

\begin{scope}[yshift=-3.75cm]
  \begin{scope}[xshift=-2cm]
  \matrix [draw=none,column sep=0]
  {
  \node[rounded corners=0pt] (Cmono) {C};\\
  };
  \end{scope}

  \begin{scope}[xshift=-1cm]
  \matrix [draw=none,column sep=0]
  {
  \node[rounded corners=0pt] (Amono) {A};\\
  };
  \end{scope}
\end{scope}

\draw[fill=black] (rootto1to31p.center) circle (2pt);
\draw[fill=black] (rootto1to32p.center) circle (2pt);
\draw[->,thick] (rootto1to31p.center) -- (Cmono);
\draw[->,thick] (rootto1to32p.center) -- (Amono);

\node [below=4.2cm of root.center,xshift=4mm,l] {(b) dimension list layout};

\begin{scope}[yshift=-6.3cm,xshift=1cm]
  \newcommand{\jonasc}[1]{\node[c,rounded corners=0pt] {\ttfamily\vphantom{A}#1};}
  \newcommand{\jonascx}[2]{\node (#2) [c,rounded corners=0pt] {\ttfamily\vphantom{A}#1};}
  \newcommand{\jonasp}[1]{\node[c,rounded corners=0pt,text=red] {\ttfamily\vphantom{A}#1};}
  \newcommand{\jonaspx}[2]{\node (#2) [c,rounded corners=0pt,text=red] {\ttfamily\vphantom{A}#1};}
  \matrix (mem) [draw=none,column sep=0]
  {
  \jonascx{1}{rowA} &
  \jonasp{6} &
  \jonasc{2} &
  \jonasp{14} &
  \jonasc{3} &
  \jonasp{22} &

  \jonasc{3} &
  \jonasp{8} &

  \jonasc{3} &
  \jonasp{12} \\
  \jonascx{5}{rowB} &
  \jonasp{13} &
  \jonasc{C} &
  \jonasc{A} &

  \jonasc{2} &
  \jonasp{18} &
  \jonasc{4} &
  \jonasp{20} &
  \jonasc{1} &
  \jonasc{D} \\
  \jonascx{5}{rowC} &
  \jonasc{B} &

  \jonasc{3} &
  \jonasc{3} &
  \jonasc{E} \\
  };
  \node[left=1mm of rowA,draw=none] {\texttt{Elf+00}};
  \node[left=1mm of rowB,draw=none] {\texttt{Elf+10}};
  \node[left=1mm of rowC,draw=none] {\texttt{Elf+20}};
\end{scope}

\node [below=1cm of mem.center,xshift=-5mm,l] {(c) linearised layout};

\end{tikzpicture}
  \caption[Comparison between different layouts of the Elf]{
  Comparison of prefix tree layout, dimension list layout, and linearised layout of the Elf.
  In the linearised layout, pointer values are shown in red.
  }
  \label{fig:elfmemlayout}
\end{figure}

Interior nodes themselves only store redundant information.
The prefix they own is given by the labels on the path from the root to the node.
Therefore, the node labels are dropped.
To obtain an efficient in-memory representation, each node is represented by the set of its outgoing edges.
This set is stored in the form of an array that contains the edge labels and pointers to the nodes on the other side of the edge.
This array is called a \emph{dimension list}.
The \emph{linearised layout} is given by concatenating all dimension lists in a depth-first manner
from left to right.
\Autoref{fig:elfmemlayout} shows an example Elf in tree layout, dimension list layout,
and linearised memory layout.

After prefix redundancies end, the tree layout of the Elf resembles a linked list.
For example, in \autoref{fig:elfmemlayout}, the interior nodes leading up to tuple E
never branch out to other nodes.
Traversing these linked lists is inefficient.
Since we assumed that no tuples share the same value in all dimensions,
each tuple has at least $l+1$ unique values near its leaf, where $l\in\{0,\dots,k-1\}$.
These unique last values are stored sequentially in memory, in the form of an \emph{$l$-monolist}.
This elimininates the slow linked list traversal.
The figure depicts the 2-monolist for tuple E, two 1-monolists for tuples D and B,
and two 0-monolists for tuples C and A.

Finally, the root dimension list of the Elf could be stored in an optimised fashion.
Observe that all tuples are below the root dimension list, and all possible values of $X_1$
are present in it; otherwise, they wouldn't be possible values.
Therefore, storing the values $(1,2,3)$ could be omitted, leaving only the pointers.
This \say{hash map} method is proposed by the original Elf authors for efficiently storing a large-cardinality attribute at the top of the Elf.
However, for simplicity,
we will not consider this special property in our further discussion of the Elf.
Instead, we assume that the root attribute has cardinality 1,
i.e.\ all tuples share the same value.
This assumption essentially disables the hash map technique.
When we consider the \say{first} attribute of the Elf, we will from here on
refer to the first attribute below the hash map.

\subsection{Comparison to related approaches}
The Elf borrows concepts from all the index structures that were introduced in \cref{sec:others,sec:trees}.
In the following, we will briefly describe how the Elf is similar to and different from each of them.

\paragraph{Data grouping through similarity.}
Like the R-Tree, the Elf groups similar data points into a group that is efficient to handle.
The R-Tree uses the \emph{distance} between points to identify which points to group:
a number of points whose pairwise distances are below a certain threshold are considered \say{similar}.
These similar points are then grouped together and represented by their MBR.
The Elf, however, does not use the distance between points as the criterion for determining similarity.
Instead, two tuples are similar iff they share a specific attribute value.
This makes the Elf a \emph{feature-based} method instead of a \emph{distance-based method} \cite{754960}.

\paragraph{Recursive application of inverted lists.}
On the topmost level, the Elf works like the inverted list approach.
For the first attribute, the Elf determines all the unique values contained in the dataset, and creates a map using these attribute values as keys.
However, the inverted list approach maps these keys directly to the list of data points that share the key attribute value.
Instead, the Elf recursively applies the inverted list concept, now determining all the unique values in the second attribute.
This process is recursively applied until the last attribute is reached, where an ordinary inverted index is used.

\paragraph{Hybrid between row and column stores.}
In a \emph{row store}, the attribute values for a single tuple are stored adjacent to each other.
In constrast to that, in a \emph{column store} all the values for a single column are stored adjacent to each other.
The Elf is a hybrid between these structures, as illustrated in \autoref{fig:beyond_row_column}.
\begin{figure}[htb]
  \begin{minipage}[b]{.33\linewidth}\centering
  \begin{tikzpicture}[overlay, remember picture]
    \draw[rounded corners=1mm,yshift=3mm,line width=4mm,blue!20,cap=round] (pic cs:cl) -- (pic cs:cr);
  \end{tikzpicture}
  \begin{tabular}{llll}
    $X_1$ & $X_2$ & $X_3$ & $\mathit{Ref}$ \\ \midrule
    1     & 3     & 5     & A \\
    2     & 4     & 5     & B \\
    \tikzmark{cl}1     & 3     & 3     & C\tikzmark{cr} \\
    2     & 2     & 1     & D \\
  \end{tabular}
  \subcaption{row store}
\end{minipage}\begin{minipage}[b]{.33\linewidth}\centering
  \begin{tikzpicture}[overlay, remember picture]
    \draw[rounded corners=1mm,yshift=3mm,xshift=3mm,line width=4mm,blue!20,cap=round] (pic cs:ct) -- (pic cs:cb);
  \end{tikzpicture}
  \begin{tabular}{llll}
    1 & 2 & \tikzmark{ct}1 & 2 \\
    3 & 4 & 3 & 2 \\
    5 & 5 & 3 & 1 \\
    A & B & \tikzmark{cb}C & D
  \end{tabular}
  \subcaption{column store}
\end{minipage}\begin{minipage}[b]{.33\linewidth}\centering
  \begin{tikzpicture}
  \pgfdeclarelayer{background}
  \pgfdeclarelayer{foreground}
  \pgfsetlayers{background,main,foreground}

    \matrix (table) {
      \node (a) {1}; & & & &\node (b) {2}; & & \node {}; & \node {};  \\
      \node (a1) {3}; & & & &\node (b1) {2}; & & \node (b2) {4}; & \node {}; \\
      \node (a11) {3}; & \node (a11m) {C}; & \node (a12) {5}; & \node (a12m) {A};
      & \node (b11) {1}; & \node (b11m) {D};
      & \node (b21) {5}; & \node (b21m) {B}; \\
    };

    \draw[->] (a1) -- (a); 
    \draw[->] (b1) -- (b); 
    \draw[->] (b) -- (b2);
    \draw[->] (a11) -- (a1); 
    \draw[->] (a11m) -- (a11); 

    \draw[->] (a1) -- (a12);
    \draw[->] (a12m) -- (a12); 

    \draw[->] (b11) -- (b1); 
    \draw[->] (b21) -- (b2); 

    \draw[->] (b11m) -- (b11); 
    \draw[->] (b21m) -- (b21); 

\begin{pgfonlayer}{background}
  \draw[rounded corners=1mm,line width=4mm,blue!20,cap=round]
  (a.center) -- (a1.center) -- (a11.center) -- (a11m.center);
\end{pgfonlayer}
\end{tikzpicture}
  \subcaption{Elf}
\end{minipage}
  \caption[Comparison between Elf and traditional row stores and column stores]{
  Comparison between Elf and traditional row stores and column stores.
  The attribute values for tuple \say{C} are highlighted.
  While the row store is completely horizontal and the column store completely vertical,
  the Elf is a mixture of both.
  }\label{fig:beyond_row_column}
\end{figure}
On the first levels, the Elf resembles a deduplicated column store.
Each node points to a different subtree at the next level of the structure.
However, the deeper levels of the Elf start to resemble a row store.
This is especially apparent when considering monolists, which are by definition a row store.
Monolists occur more frequently in deeper levels of the Elf,
and the final level is always a monolist.

\paragraph{Dimensionality reduction through space division.}
Finally, the Elf can be regarded as a practical implementation of the space division approach.
The \say{tree striping} technique divided the $k$-dimensional data space
into two lower-dimensional subspaces of dimensions $k=k_1+k_2$,
whose cartesian product again formed the original data space.
The Elf essentially applies this technique with \say{$k_1=0$ and $k_2=k-1$}.
However, the Elf does not partition the entire data space,
but only the \emph{data set}, i.e.\ the non-empty part of the data space.
The partitions are created using a number of planes that assign a unique region to every data point
(see \autoref{fig:lasagna}).
The label of the Elf plane becomes the first subspace of the tree striping technique,
containing only a single value and therefore having dimension $k_1=0$.
The Elf plane becomes the second subspace;
the points contained in it do not vary in one of the original dimensions,
and therefore
$k_2=k-1$.
The \enquote{missing dimension} is restored by building a multitude of these structures,
and pointing to each of them from the dimension list.

In summary, the Elf shares important properties with a number of known index structures.
However, it is hard to classify the Elf in conventional terms as a space-partitioning or data-partitioning structure, as a distance-based or feature-based structure, or as a row-store or column-store structure.

\subsection{Performance of the Elf search}
While we will discuss the performance of the Elf search in great detail in \autoref{chp:model},
we will briefly describe here how the design of the Elf structure allows for improved search
performance compared to linear scanning.

The Elf tree is searched using a depth-first search starting from the root.
However, for low-selectivity queries, only a small fraction of nodes is visited.
Using the Elf structure's pruning property,
comparing an entire prefix to the query bounds requires only a single comparison
for all tuples sharing that prefix.
Consequently,
the total number of comparisons during the Elf search is greatly reduced compared to a linear scan,
where at least one attribute of every tuple is always compared.

While the pruning property exists in similar fashion
for other tree-based multi-dimensional index structures,
for the Elf, it does not lead to issues arising from the curse of dimensionality.
The Elf tackles the problem of the sparsity of data in high  dimensions
by ignoring the empty space around the data points.
In essence, this space is \say{already pruned} for every query.
Therefore, unlike traditional multidimensional index structures,
the Elf does not suffer from the sparseness of data in high-dimensional spaces.

The effectiveness of the pruning during the Elf search highly depends on the order of dimensions within the tree.
For example, consider a two-dimensional relation $R=\{X_1,X_2\}$, each containing $\vee_1=\vee_2=1000$ different attribute values.
Now consider the partial match query $X_1=1$.
If the Elf is constructed in the order $(X_1,X_2)$, only the interior node $(1)$ is accessed;
the query result is immediately given by the set of all tuples below this node.
However, if the Elf is constructed in the reversed order $(X_2,X_1)$, \emph{every}
interior node $(j), j \in [0;\vee_2]$ is accessed, since each query result is stored
below one of the interior nodes $(j,1)$.
One heuristic for the ordering is therefore that the attributes with low partial selectivity should be stored close to the top of the Elf.
Since the search time of the Elf varies so drastically with different configurations,
we will later develop a prediction model for the search time in a given configuration.




\section{Index structure performance prediction}
In addition to good performance (or \emph{low cost}), predictable performance is
an important property for the suitability of an index structure.
In this section, we motivate why accurately estimating the
performance of an index structure can actually increase performance itself.
To this end, we will give an introduction to query planning and optimisation.
We describe the approaches used for proposed prediction methods for
existing index structures,
and the challenges encountered when modelling the cost
of main-memory index structures, such as the Elf.

\subsection{Query planning and optimisation}
Usually, an index structure provides a performance improvement only for certain kinds  of queries.
For example, linear scans already perform well for selection predicates with large selectivity.
Using nearly any proposed sophisticated tree-based structure for this kind of query would lead to poor performance \cite{qin2007towards}.
Therefore, the database system has to dynamically analyse the query to determine which of
a set of available strategies for executing the query is the most cost-effective.

Choosing between different execution strategies for a given query
is the problem tackled by \emph{query optimisers} in modern database systems \cites{agrawal2000automated,harangsri1998query}.
The run-time choices for evaluating a given query are called \emph{plans}.
For example, we might have a choice between the two plans (1) \say{Scan for requested attribute values} and (2) \say{Look up requested values in a \bplustree}.
For large-selectivity queries, plan (1) might be preferrable to plan (2), while plan (2) yields superior performance for low-selectivity queries up to some selectivity $\sigma_\lvar{eq}$, where both plans perform equally well.


In commercial database systems, the responsibilities of the query optimiser go even further,
e.g.\ when deciding on the evaluation order of joins.
This leads to a much larger number of query plans available to the optimiser, requiring sophisticated planning methods.
However, in this work, we will focus solely on the evaluation of selection predicates.
For comparing two plans, the optimiser will generally predict the cost (i.e.\ execution time)
of a plan using a \emph{cost model} that estimates the complexity of executing the given plan.
The optimiser then chooses the plan with the lowest cost estimate.

It has been shown that the optimiser cost models
of popular database systems like DB2~\cite{Reiss:2003:CSQ:872757.872804}
and PostgreSQL~\cite{qin2007towards}
often produce inaccuracies of several orders of magnitude,
even when optimising simple selection predicates.
This is problematic, because an inaccurate cost estimate can lead to the optimiser
choosing a poor execution plan instead of a faster one.

One common metric for measuring the performance impact of a poor query plan
is the \emph{global relative cost~(GRC)} of the plan \cite{Reiss:2003:CSQ:872757.872804}.
The GRC of a plan $p$ is defined as
$$\operatorname{GRC}(p) = \frac{\operatorname{Cost}(p)}{\operatorname{Cost}(p^*)}$$
where $\operatorname{Cost}(x)$ is the actual cost of plan $x$,
and $p^*$ is the cheapest plan that evaluates the same query as $p$.
Therefore, GRC models the relative increase in cost incurred by choosing a suboptimal plan
$p$ over the optimal, but likely unknown plan $p^*$, since $\operatorname{GRC}(p^*)=1$.

Now again consider the above scenario, where one query plan involves a linear scan,
and the other involves searching an index structure.
Ideally, the plan with the lowest actual cost would be chosen.
However, assume that the cost model overestimates the cost of using the index structure by a factor of $\epsilon$.
It can be shown that in this case,
the optimiser chooses a plan $p$ which has suboptimal cost of up to
$\operatorname{GRC}(p) = \epsilon$ \cite{qin2007towards}.
Therefore, increased accuracy of the cost model is likely
to lead to perfomance improvements in query processing.

\subsection{Approaches to cost estimation}
The methods for estimating the performance of proposed index structures can generally be divided into two categories \cite{6544899}: (1) \emph{simulation-based} or sampling-based and (2) \emph{model-based} or parametric.

Simulation-based approaches
\cite{Duggan:2011:PPC:1989323.1989359,Ahmad:2011:ISR:2035111.2035115}
are usually dynamic algorithms that execute queries or parts of queries before deciding on the query plan for future queries.
This approach is considered appropriate when modelling the performance of a complex software system whose behaviour is only predictable empirically.

On the other hand, \emph{model-based} approaches try to quantify the performance of a smaller system,
like an index structure, using mathematical analysis.
Some work has been done to model the performance of range queries regardless of the index structure used
\cite{pagel1993towards,bohm2000cost}.
However, their assumptions on (1) data uniformity and independence,
(2) query uniformity and selectivity, and (3) storage layout and performance are generally
too strict to lead to meaningful real-world results.
In general, the performance of multi-dimensional range queries depends highly on the design of the chosen index structure.
Usually, a simpler design of the index structure leads to more readily quantifiable performance.
Performance models for the popular R-Tree \cite{theodoridis1996model} and {\kdtree}
\cite{Freidman1977} methods have been developed.
These models estimate the performance of each index structure
under specific assumptions about the query and data distribution,
and are used by query optimisers to estimate the cost of the search.

\subsection{The page count metric}\label{sec:pagecount}

Index structures for large databases generally assume a \emph{disk-based} storage of the index.
Disk storage systems provide a block device interface to the database system.
The database can request a set of \emph{pages} to be fetched from the disk.
Each page fetch, however, incurs a significant delay (the \emph{seek time}) since the disk controller has to mechanically move a
magnetic head to the specific physical position of the requested page on the disk.

The delays incurred by the the mechanical action make the hard disk a very slow form of memory
compared to the computer's main memory.
In algorithm design, the so-called \emph{external memory model} is used
to differentiate between the access costs of different forms of memory \cite{mehlhorn2008algorithms}.
However, the disk is in fact so slow
(up to tens of milliseconds compared to \SIrange{50}{150}{\nano\second} for a RAM access)
that it becomes the the bottleneck for index operations.
Therefore, the performance impact of other parts of the database system,
such as the CPU caches and main memory,
is negligible in comparison to the hard disk access times.
A complex analysis of these other impacts is therefore not required.
Following these considerations, the \emph{number of accessed disk pages}
has been established as the dominant cost factor,
which is predicted by the cost model of a query optimiser.

However, a significant reduction in the cost of Random Access Memory~(RAM)
has recently led to the advance of in-memory \cite{boehm2011efficient,bitweaving}
or combined disk-memory \cite{grund2010hyrise} index structures.
Instead of on a disk, the index is now stored in the much faster RAM.

Fast access to the stored index data eliminates
the disk seek time as the dominant factor for the performance of index operations.
In turn, this raises the relative importance of other impact factors,
such as the in-memory layout of the index structure.
The memory layout, in turn, determines the effectiveness of the CPU cache.
Memory accesses that can be answered from the CPU cache can be hundreds of times faster than uncached accesses \cite{mehlhorn2008algorithms}.
A cache-conscious memory layout can reduce the overall execution time of already
highly-optimised structures, e.g.\ the {\bplustree}, by up to \SI{30}{\percent} \cite{rao2000making}.

Two basic principles of a cache-conscious algorithm design are to
(1) store locally-used data inside a single cache block, and
(2) align data structures to cache block boundaries.
However, the exact prediction and eviction policies of the CPU cache
are generally implemented in hardware and depend on the used CPU microarchitecture.
Therefore, achieving the highest levels of cache consciousness requires
considerable efforts spent for architecture-specific tuning \cite{Kim:2011:DFA:2043652.2043655}.
On the other hand, generic algorithms that are not optimised towards a specific architecture
incur different levels of efficiency on different architectures.

With the influence of CPU caches, memory access times are not guaranteed to be uniform.
Therefore, it is not obvious what the preferrable performance metric for a main-memory index structure is.
The direct equivalent to the number of accessed disk pages would be the number of accessed memory cells; however, the behaviour of the CPU cache could make this number meaningless, even if correctly predicted.

The results for our performance model for the Elf, described in \autoref{chp:results},
suggest that in our scenario, the number of accessed memory locations is still an accurate metric
for the execution time.
This means that the Elf's performance can be predicted by estimating the number of accessed memory
locations ahead of time.
This is the foundation on which we will build our performance model for the Elf in the following chapter.

\section{Combinatorics}\label{sec:combinatorics}
Our performance model for the Elf is only a stochastical approximation of the expected behaviour.
Therefore, we require a number of definitions for working with uncertain quantities, which we will introduce in this section.

First, let $\Ud{a}{b}$ denote the the uniform distribution over a discrete interval $\{a,a+1,\dots,b\} \subset \NN$,
and let $\Uc{a}{b}$ denote the the uniform distribution over a continuous interval $[a,b] \subset \RR$.
Secondly, in our analysis of the Elf, we will encounter two different types of \emph{occupancy problems},
stochastic problems which are commonly called \say{balls and urns} problems \cite{Beeler2015}.
These problems concern the probabilities (denoted $\PP$) and expected values (denoted $\EE$) that arise in situations where each of a number of balls is distributed randomly into exactly one of a number of urns.

In the remainder of this section, we formulate two occupancy problems and derive solutions to each.
We initially motivate the problem by briefly explaining the role of each solution in our cost model.
Finally, we define two functions, $\buckets$ and $\monobuckets$, that we will later use
to refer to the solutions to these particular problems.

\subsection{Estimating distinct values in a sample}
The first problem considers the expected size of an individual dimension list of the Elf.
The length of a dimension list is equal to the number of unique attribute values of the tuples contained within the list.
If we assume that the attribute values are uniformly distributed, and that enough tuples are present, the problem can be stated as follows:
\begin{quote}
Let $X$ be a random variable with a uniform distribution over $\{1,2,\dots,c\}$.
Now let $(x_1, x_2, \dots, x_k)$ be a list of $k$ independent samples of $X$.
What is the expected cardinality of $\{x_1, x_2, \dots, x_k\}$, i.e.\ how many \emph{distinct} values were drawn?
\end{quote}
The problem is equivalent to this balls-and-urns problem:
\begin{quote}
  Assume $c$ urns, each containing an infinite supply of indistinguishable balls of a particular unique color.
  Repeat $k$ times: choose an urn at random, and draw a ball from it.
  How many different colors are present among the drawn balls?
\end{quote}
The solution to this problem can be derived as follows \cite{72229}:
Consider the random number $N_k$ of different values chosen after $k$ picks.
Naturally, $N_0 = 0$ almost surely.
Knowing $N_k$, the probability of picking a new item with the $k+1$'th pick is $\frac{c-N_k}{c}$,
because out of the $c$ possible values, the $N_k$ already picked values are disallowed.
Therefore, with the $k+1$'th pick included, the expected value increases to
$\EE (N_{k+1} | N_k) = N_k + \frac{c - N_k}{c}.$

Finally, the expected number of different values after $k$ picks $n_k = \EE N_k$ is:
\begin{align*}
  n_0 &= 0 \\
  n_{k+1} &= n_k + \frac{c - n_k}{c} \\
          &= 1 + a n_k, \quad \mathrm{with} \, \, a = \frac{c - 1}{c}.
\end{align*}
Solving this recursion leads to
\begin{align*}
  n_k = \frac{1-a^k}{1-a} = c \frac{c^k - (c - 1)^k}{c^k}.
\end{align*}
We define a function $\buckets$ that computes this solution:
\begin{equation}
  \label{eq:buckets}
  \buckets\colon \NN \times \NN \to \RR, \,\, (c, k) \, \mapsto \, c \frac{c^k - (c-1)^k}{c^k}
\end{equation}

\subsection{Estimating single-occurrent values in a sample}
Our second problem considers the occurence of monolists within the Elf.
A monolist is created when only one tuple remains for some fixed attribute values.
Given the number $k$ of subnodes for some Elf, we want to estimate how many of these subnodes
will contain only one tuple out of the $n$ total tuples.
Conceptually, this represents the following occupancy problem:
\begin{quote}
Uniformly distribute $n$ balls into $k$ urns.
Assuming that each urn now contains at least one ball, how many urns contain exactly one ball?
\end{quote}
The solution to this problem is equal to the solution of the following simpler problem: \say{Given $n-k$ balls are uniformly distributed into $k$ buckets, how many buckets stay empty?}

For any fixed item, the probability of a fixed urn receiving it is $\frac{1}{k}$.
The probability of a fixed urn \emph{not} receiving this item is $1 - \frac{1}{k} = \frac{k-1}{k}$.
Therefore, the probability of a fixed urn $a$ not receiving any items out of $n-k$ draws is:
$$\PP[\textnormal{urn $a$ is empty}] = \left(\frac{k-1}{k}\right)^{n-k}.$$
Finally, the expected total number of empty urns can be computed as:
\begin{align*}
  \EE[\textnormal{number of empty urns}] &= \sum_{i=1}^k \PP[\textnormal{urn $i$ is empty}] \\
  &= k \, \PP[\textnormal{urn $a$ is empty}] \quad \quad \textrm{(for some $a$)} \\
  &= k \left(\frac{k-1}{k}\right)^{n-k}
\end{align*}
We write this solution as a function $\monobuckets$:
\begin{equation}
  \label{eq:monobuckets}
  \monobuckets\colon \NN \times \NN \to \RR, \,\, (k, n) \, \mapsto \, k \left(\frac{k-1}{k}\right)^{n-k}
\end{equation}

\chapter{Predicting the Elf search time}\label{chp:model}

In this chapter, we describe our method to predict the time taken for searching the Elf index structure.
First, we motivate our approach by finding the dominant operations of the search algorithm.
We explain our design that uses the \emph{size} of the Elf to estimate the search time.
We then describe our model and algorithm to predict this size.
Finally, we show how to apply our model to non-uniform and correlated data sets.

\section{Analysing the search algorithm}\label{sec:analysis}
\begin{algorithm}[tb]
  \caption{Searching the Elf\label{alg:search}}
  \SetKwFunction{Search}{Search}
  \SetKwFunction{SearchMonolist}{SearchMonolist}
  \SetKwData{Results}{results}
  \SetKwProg{Proc}{Procedure}{ is}{end}
  \SetKwInOut{Input}{Input}\SetKwInOut{Output}{Output}
  \Input{$\Elf_k(R)$\\Query: lower bounds $l_1,\dots,l_k$, upper bounds $u_1,\dots,u_k$}
  \Output{query result $= \{ \operatorname{TID}(x) \, \, \vert \, \, x = (x_1,\dots,x_k) \in R: \,\, \forall_i:\, l_i \leq x_i \leq u_i \}$}
  \Begin{
    \Results $\gets \emptyset$\;
    \Search{root of $\Elf_k(R)$, 1, \Results}\;
    \Return{\Results}\;
  }
  \BlankLine
  \SetKwData{Node}{node}
  \SetKwData{Dim}{dim}
  \SetKwData{Subnode}{subnode}
  \SetKwData{E}{e}
  \Proc{\Search{\Node, \Dim, \Results}}{
    \While{\Node has unvisited outgoing edges}{
      \ShowLn\label{line:nextsibling}$\E \gets$ next unvisited outgoing edge of \Node (\E is now visited)\;
      \If{$l_{\Dim} \leq \operatorname{label}(\E) \leq l_{\Dim}$}{
        \ShowLn\label{line:traverse}$\Subnode \gets$ traverse \E\;
        \If{\Subnode is a monolist}{
          \tcc{$\Dim \leq k$}
          \SearchMonolist{\Subnode, $\Dim+1$, \Results}\;
        }
        \Else{
          \tcc{$\Dim < k$}
          \Search{\Subnode, $\Dim+1$, \Results}\;
        }
      }
    }
  }

  \SetKwData{i}{i}
  \Proc{\SearchMonolist{\Node, \Dim, \Results}}{
    \For{$\i \gets \Dim$ to $k$}{
      \If{$!(l_i \leq \Node[i] \leq u_i)$}{
        \Return\;
      }
    }
    $\Results \gets \Results \cup \{ \operatorname{TID}(\Node) \}$\;
  }
\end{algorithm}
As described in the previous chapter, the search time of main-memory index structures is
affected by
\say{strange} effects of e.g.\ the CPU cache,
which can cause chaotic and nondeterministic performance.
Therefore, finding a dominant operation in the algorithm is expected to be difficult.
However, we argue that the design of the Elf ensures that one specific factor,
the \emph{number of visited Elf nodes}, is the dominant factor for the execution time.
In this section, we present our motivation for this hypothesis.
Starting from a description of the algorithm,
we discuss the impact of each individual operation on the total execution time.
Our goal is to disregard as many factors as possible to obtain a simple and testable model of the algorithm performance.

\Autoref{alg:search} shows the method used by Köppen~et.~al.~\cite{elf}
to evaluate a range query using a previously built Elf.
The main procedure \model{Search} is, first and foremost,
a depth-first tree traversal starting at the top of the Elf.
A depth-first search is characterised by two operations:
\begin{enumerate}[label=(\arabic*)]
  \item Retrieve the next outgoing edge of a node (line~\ref{line:nextsibling}).
  \item Traverse an edge, descending to its child node (line~\ref{line:traverse}).
\end{enumerate}

In the following, we will argue that all the other operations only take place exactly once, and in constant time, for one of the above operations (1) and (2).

\paragraph{Query bounds comparison.}
First, the comparison between the edge labels and the query bounds
is executed exactly once for each enumerated edge, and therefore for each occurence of (1).
Since the query bounds $l_1,\dots,l_k$,$u_1,\dots,u_k$ are used throughout the entire execution,
they are likely stored in cache.
The labels were loaded in the preceding (1) and are cached as well.
Therefore, the comparison is cheap.

\paragraph{Monolist check.}
This check is executed for each traversed edge, and therefore for each occurence of (2).
Due to the in-memory layout of the Elf, the presence of a monolist is signalled by the most significant bit of the pointer in the dimension list.
Therefore, this is a cheap comparison between a constant and a value in the CPU cache.

\paragraph{\model{Search} and \model{SearchMonolist}.}
For each occurence of (2), exactly one of \model{Search} or \model{SearchMonolist} is called.
\model{SearchMonolist} has approximately constant execution time, since it does not recurse
and the preceding (2) already loaded the monolist.
On the other hand, \model{Search} can lead to further recursions.
However, every further recursion is also preceded by exactly one further occurence of (2).

Using these arguments, we reduce the list of possible impact factors to the operations (1) and (2).
Each occurence of either operation results in a constant delay of the execution.
From here on, we will consider the cost of the associated computation to be included in the respective operation.
For (1), the associated computation is only performed on cached data, and therefore considered negligible.
For (2), the same argument applies, except for the cost of the function call, which has to be explicitly considered.

What remains is to estimate the costs of (1) and (2) themselves.
In principle, they both access an equal amount of \say{new} data in memory,
data that is only accessed at most once since a dimension list is never visited twice.

The total number of occurences of (1) is linear to the length of the dimension lists.
However, the memory layout of the Elf is highly optimised for (1).
The elements of a single dimension list are placed sequentially in memory,
and traversed sequentially as well.
This leads to high spatial locality: eight 64-bit dimension list entries fit into
a 64-byte x86 cache line.
Additionally, hardware prefetching will likely predict further sequential accesses.

While (1) is cache-optimised, the situation is very different for (2).
In fact, (2) likely jumps to an uncached location, since the
subnodes are placed in memory behind the entirety of the current dimension list.
Furthermore, (2) also includes write accesses to the program stack, since it includes a function call to either \model{Search} or \model{SearchMonolist}.
Therefore, we expect (2) to be at least an order of magnitude slower than (1).

Summarising the last two paragraphs, (2) is the dominant operation, unless dimension lists are excessively long.
However, at each level of the Elf, the data set is divided into a number of partitions.
These partitions are at the next level partitioned further.
Therefore, the expected length of each individual dimension list decreases exponentially with level.
As we will see later, the exact value of the exponent depends on the variance of the data set,
however the exponential decrease is always present regardless of distribution.
Therefore, most dimension lists are short.

In conclusion, we find that (2) is the dominant operation.
The number of occurences of (2) is equal to the number of calls
to \model{Search} and \model{SearchMonolist} combined.
We say that a call to one of these methods \say{visits} the given node.
This line of argument leads us to the central hypothesis for estimating the execution time:

\begin{hyp}\label{hyp:linearvisit}
  The execution time of \Autoref{alg:search} is approximately linear to the number of Elf nodes visited.
\end{hyp}

This means that we can treat each Elf node similar to how
cost models for disk-based index structures (see \autoref{sec:pagecount}) treat disk pages;
each page access incurs a cost that is nonzero, and constant on average.
To the best of our knowledge, this is a unique feature of the Elf
that distinguishes it from other main-memory index structures.

In our evaluation of the model, presented in \autoref{chp:results}, we train a linear regression model
on the relationship between the number of visited nodes and the actual search time.
This model is used to finally predict the search time of the Elf.

In the remainder of this chapter, we will describe how we predict the number of visited nodes.

\section{Taking measure of the Elf}
In the previous section, we established that the number of visited nodes during the Elf search has a major influence on the execution time of the search.
Therefore, we now focus on estimating this visit count.
In this section, we will make a number of assumptions about the usage scenario of the Elf.
As we will see, these assumptions ensure that all the nodes at a given depth of the Elf are similar to each other.
This allows us to formulate the \emph{Elf metrics}, which quantify the shape of the Elf using
a small set of numerical values that determine the shape.

First, we observe that the number of visited nodes depends on a variety of impact factors, including:
\begin{enumerate}[label={(\arabic*)}]
  \item The cardinalities of the columns,
  \item partial selectivities of the query, and
  \item data distribution and correlations.
\end{enumerate}
In this section, we will focus on (1) by making fixed assumptions about the other impact factors.
First, we assume that the Elf is searched with selectivity 1, i.e.\ every node is visited.
Then, we make two major assumptions about the data set.
Let $X_i$ be the random variable that values for column $i$ are sampled from.
\begin{description}
  \item[independence assumption]
    We assume that all the $X_i$'s are stochastically independent.
  \item[uniformity assumption]
    We assume that each $X_i$ is uniformly distributed, i.e.\ $X_i \sim \Ud{0}{\vee_i}$,
    where $\vee_i$ denotes the \emph{cardinality} of column $i$.
\end{description}
We will relax these assumptions in \autoref{sec:correlation} and \autoref{sec:nonuniform}, respectively.

In the following, we describe our approach to \say{taking the Elf's measurements},
and how the shape of the Elf varies when the cardinalities of the columns change.
This leads us to a number of definitions of quantities, which we call the \emph{Elf metrics}.

\subsection{Motivation}\label{sec:visitmotivation}
The recursive definition of the Elf, described in \autoref{sec:elfformal},
suggests that the Elf has a kind of \say{fractal} topology.
This is to say that that the sub-Elfs for deeper dimensions
are similar in structure to the entirety of the Elf.

\begin{figure}[htb]\centering
 \pgfplotscreateplotcyclelist{my black white}{%
solid, every mark/.append style={solid, fill=black}, mark=*\\%
dotted, every mark/.append style={solid, fill=black}, mark=square*\\%
densely dotted, every mark/.append style={solid, fill=black}, mark=otimes*\\%
loosely dotted, every mark/.append style={solid, fill=black}, mark=triangle*\\%
dashed, every mark/.append style={solid, fill=black},mark=diamond*\\%
loosely dashed, every mark/.append style={solid, fill=black},mark=*\\%
densely dashed, every mark/.append style={solid, fill=black},mark=square*\\%
dashdotted, every mark/.append style={solid, fill=black},mark=otimes*\\%
dasdotdotted, every mark/.append style={solid},mark=star\\%
densely dashdotted,every mark/.append style={solid, fill=black},mark=diamond*\\%
}
\centering
 \pgfplotstableread{data/visit_counts.dat}{\visitcounts}
  \begin{tikzpicture}
  \begin{loglogaxis}[enlarge x limits=false,ymin=1,width=\textwidth-4cm,legend pos=north west,
  xlabel={$\vee_1$},ylabel={Number of visited nodes},scaled y ticks={real:0.002}, 
  legend entries={$\visits_1$,$\visits_2$,$\visits_3$,$\visits_4$,$\visits_5$,$\visits_6$},
  legend pos=outer north east,
  cycle list name=my black white,
]
  \tikzset{every pin edge/.style={solid, thin}}
  \addplot +[mark=none,unbounded coords=discard,very thick] table [x=gamma1, y=visited1] {\visitcounts}
      node [pos=0.4,pin={175:$\visits_1$},inner sep=0pt] {};
  \addplot +[mark=none,unbounded coords=discard,very thick] table [x=gamma1, y=visited2] {\visitcounts} node [pos=0.9,pin={170:$\visits_2$},inner sep=0pt] {};
  \addplot +[mark=none,unbounded coords=discard,very thick] table [x=gamma1, y=visited3] {\visitcounts} node [pos=0.7,pin={170:$\visits_3$},inner sep=0pt] {};
  \addplot +[mark=none,unbounded coords=discard,very thick] table [x=gamma1, y=visited4] {\visitcounts} node [pos=0.45,pin={170:$\visits_4$},inner sep=0pt] {};
  \addplot +[mark=none,unbounded coords=discard,very thick] table [x=gamma1, y=visited5] {\visitcounts} node [pos=0.3,pin={170:$\visits_5$},inner sep=0pt] {};
  \addplot +[mark=none,unbounded coords=discard,very thick] table [x=gamma1, y=visited6] {\visitcounts} node [pos=0.65,pin={-170:$\visits_6$},inner sep=0pt] {};
  \end{loglogaxis}
  \end{tikzpicture}
  \caption[Number of visited nodes at each Elf level]{
    Number of visited nodes at each Elf level,
    as the cardinality of the first column increases.
  }\label{fig:visits_pattern}
\end{figure}
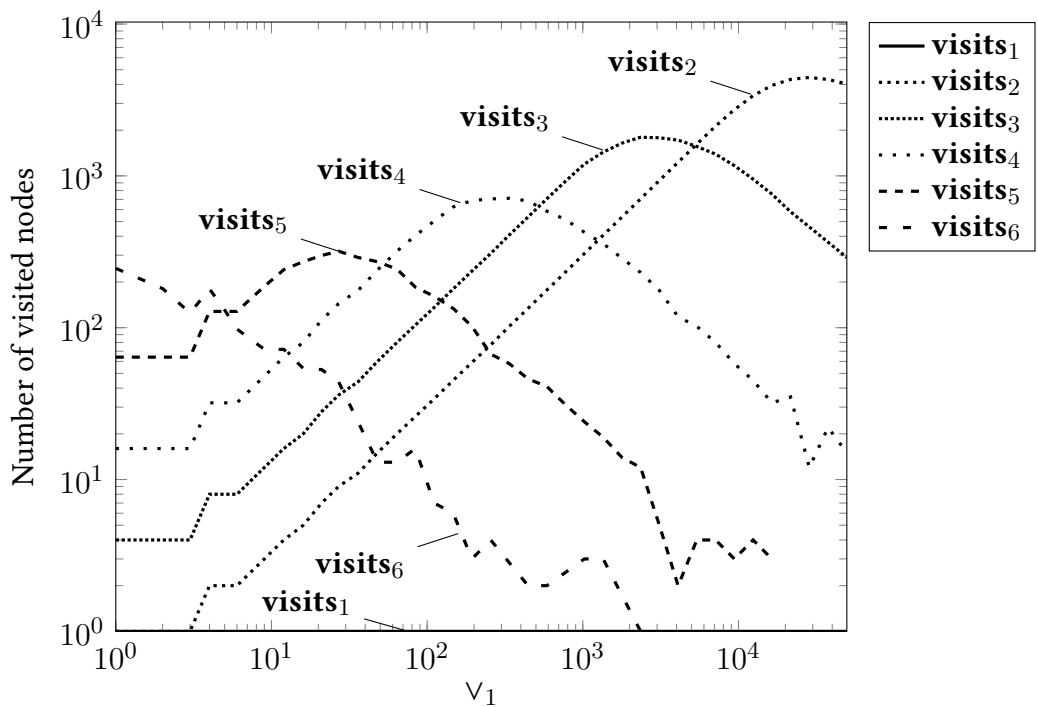

In an initial series of experiments, we therefore
focussed on the cardinality of the first dimension only.
We varied this cardinality while leaving all other parameters equal.
Then, we measured the total number of nodes of the Elf.
Additionally, we tracked at which \emph{depth} each node is stored.
This gives the values $\visits_i$, which denote the number of visited nodes at depth $i$ of the Elf, where $i=1,\dots,k$.
Shown in \autoref{fig:visits_pattern}, these numbers follow a very peculiar pattern.
\pgfplotscreateplotcyclelist{my black white}{%
solid, every mark/.append style={solid, fill=black}, mark=*\\%
dotted, every mark/.append style={solid, fill=black}, mark=square*\\%
densely dotted, every mark/.append style={solid, fill=black}, mark=otimes*\\%
loosely dotted, every mark/.append style={solid, fill=black}, mark=triangle*\\%
dashed, every mark/.append style={solid, fill=black},mark=diamond*\\%
loosely dashed, every mark/.append style={solid, fill=black},mark=*\\%
densely dashed, every mark/.append style={solid, fill=black},mark=square*\\%
dashdotted, every mark/.append style={solid, fill=black},mark=otimes*\\%
dasdotdotted, every mark/.append style={solid},mark=star\\%
densely dashdotted,every mark/.append style={solid, fill=black},mark=diamond*\\%
}

The number of nodes at each dimension follows a pattern that is defined
by a period of gradual increase, followed by a peak, followed by a period of gradual decrease.
We interpret this pattern as the consequence of two main factors:
\begin{enumerate}[label={(\arabic*)}]
  \item
  \emph{Higher cardinality leads to higher fanout.}
  If more distinct values are present for the first attribute,
  the first dimension list has more entries.
  However, every entry gets its own unique node in the next dimension.
  Therefore, the number of nodes in the next dimension increases.
  While we will introduce a formal definition of \say{fanout} later,
  for now we use it to describe that adding an element to a dimension list
  leads to an entirely new dimension list at the next level, which is much
  larger than the single element added to the first list.

  \item
  \emph{Lower number of tuples per dimension list leads to more monolists.}
  As the fanout of a dimension increases due to (1), the tuples are
  distributed between an increasing number of sub-lists.
  The uniformity assumption implies that an approximately equal
  number of tuples is assigned to each sub-list.
  Therefore, the number of tuples stored per sub-list decreases.
  This leads to a higher probability of a sub-list only storing one tuple.
  Consequently, the expected number of monolists increases.
  However, each monolist forms a premature leaf of the tree, i.e.\ a leaf
  that is not at depth $k$.
  This means that for each monolist that occurs, one node is \say{stolen}
  from the \emph{next} dimension.
  Therefore, one less node is visited in the next dimension.
\end{enumerate}
In the first dimension, only a single node is visited. Therefore $\visits_1=1$ constantly.

Starting from the second dimension, (1) and (2) start to counteract each other:
at first, the number of nodes rises constantly due to (1).
This rise in turn increases the impact of (2).
The rise continues until the fanout becomes
so large that the effects of (2) overpower the effects of (1).
Then, gradually less nodes are present in the deeper dimensions.

These observations motivate us to model the number of visited nodes analytically,
for each layer of the Elf.
Afterwards, we simply take the sum of the $\visits_i$ to obtain the total number of visited nodes.

\subsection{Elf metrics}
We will now introduce the \emph{Elf metrics}, a small set of quantities
that capture the impact factors for the number of visited nodes during the evaluation of a range query $Q$.

\begin{figure}[!htb]\centering\begin{tikzpicture}
[
    mycell/.style={draw, minimum size=0.7cm},
    dot/.style={mycell,
        append after command={\pgfextra
        \fill[black] (\tikzlastnode) circle[radius=3pt];
        \endpgfextra}},
    gauge/.style={
      {Bar[]}-{Bar[]},
    },
    halfgauge/.style={
      -{Bar[]},
    },
    dimlist/.style={
      matrix of nodes, row sep=-\pgflinewidth, column sep=-\pgflinewidth,
      nodes={mycell}, nodes in empty cells
    },
    monolist/.style={
      dimlist,
    },
    tid/.style={
      rounded corners=3pt,
    },
    fulldimlist/.style={
      dimlist,nodes={dot},
    },
    dimleg/.style={
      align=left,anchor=west,
      fill=white,
    },
    p/.style={
      ->,
    },
    partp/.style={
      draw,
    },
    dimsep/.style={
      draw=black!30,
      very thick,
      dashed
    },
    fade/.style={
    },
    match/.style={
      append after command={\pgfextra
        \fill[orange] (-.35cm,-.35cm) rectangle (.35cm,.35cm);
        \fill[black] (\tikzlastnode) circle[radius=3pt];
      \endpgfextra},
    },
    node distance=5mm,
]
\newcommand{\gaugedist}{0.7cm}
\newcommand{\gaugelabeldist}{-2mm}


\matrix (root) [fulldimlist] at (0,2cm) {\\};
\draw [p] ([yshift=0.4cm]root-1-1.north) -- (root-1-1.north);

\matrix (dim1) [fulldimlist] {&|[match]|&|[match]|&|[match]|&|[match]|&&\\};
\draw[gauge] ([yshift=\gaugedist]dim1.west) -- ([yshift=\gaugedist]dim1.east);
\node at ([yshift=\gaugedist+\gaugelabeldist]dim1.north) {$\fanout_1$};

\draw [p] (root-1-1.center) -- ++(0,-.5cm) -| (dim1-1-1.north);

\begin{scope}[yshift=-5cm]
  \begin{scope}[fade]
    \matrix (dim2left) [fulldimlist] at (-6.08cm,0) {&\\};
    \draw (dim2left-1-1.north west) -- ++(-3mm,0);
    \draw (dim2left-1-1.south west) -- ++(-3mm,0);
    \node at ([xshift=-3mm]dim2left-1-1.west) {\dots};
  \end{scope}

  \matrix (dim21) [fulldimlist,right=of dim2left] {&&&\\};
  \draw[gauge] ([yshift=\gaugedist]dim21.west) -- ([yshift=\gaugedist]dim21.east);
  \node at ([yshift=\gaugedist+\gaugelabeldist]dim21.north) {$\fanout_2$};

  \matrix (dim22) [fulldimlist,right=of dim21] {&&&\\};

  \matrix (dim23) [monolist,right=of dim22] {$1$\\\rvdots\\|[tid]|$E$\\};
  \matrix (dim24) [monolist,right=of dim23] {$3$\\\rvdots\\|[tid]|$F$\\};
  \matrix (dim25) [monolist,right=of dim24] {$5$\\\rvdots\\|[tid]|$G$\\};
  \node [right=3mm of dim25] {\dots};

\end{scope}
\newcommand{\shiftbase}{5mm}
\newcommand{\shiftspread}{5mm}
\draw[partp]  (dim1-1-1.center)
    -- ++(0,-\shiftbase-3*\shiftspread)
    -- ++(-4cm,0) node[anchor=east] {\dots};

\draw[p]  (dim1-1-2.center)
    -- ++(0,-\shiftbase-4*\shiftspread)
    -| (dim21-1-1.north);

\draw[p]  (dim1-1-3.center)
    -- ++(0,-\shiftbase-4*\shiftspread)
    -| (dim22-1-1.north);

\draw[p]  (dim1-1-4.center)
    -- ++(0,-\shiftbase-3*\shiftspread)
    -| (dim23-1-1.north);

\draw[p]  (dim1-1-5.center)
    -- ++(0,-\shiftbase-2*\shiftspread)
    -| (dim24-1-1.north);

\draw[p]  (dim1-1-6.center)
    -- ++(0,-\shiftbase-1*\shiftspread)
    -| (dim25-1-1.north);

\draw[partp]  (dim1-1-7.center)
    -- ++(0,-\shiftbase-0cm)
    -- ++(4.5cm,0) node[anchor=west] {\dots};

\draw[gauge]
     ([yshift=-1.4cm]dim21.west)
  -- ([yshift=-1.4cm]dim24.east)
  node [yshift=-3mm,midway] {$\visits_2$};

\draw[gauge]
     ([yshift=-1.8cm]dim23.west)
  -- ([yshift=-1.8cm]dim24.east)
  node [yshift=-3mm,midway] {$\mono_2$};


\draw[dimsep] (-7cm,1.3cm) -- ++(14cm,0);
\draw[dimsep] (-7cm,-3cm) -- ++(14cm,0);
\draw[dimsep] (-7cm,-7.5cm) -- ++(14cm,0);
\draw[dimsep] (-7cm,-9.9cm) -- ++(14cm,0);
\begin{scope}[xshift=-7cm]
  \node (dim0) [dimleg] at (0,1.8cm) {Depth 0 (hash map)};
  \node (dim1) [dimleg] at (0,.8cm) {Depth 1};
  \node (dim2) [dimleg] at (0,-3.5cm) {Depth 2};
  \node (dim3) [dimleg] at (0,-8cm) {Depths $3,\dots,k$};
  \node (dimk1) [dimleg] at (0,-10.4cm) {Depth $k+1$};
\end{scope}

\draw[partp] (dim21-1-4.south)
  -- ++(0,-3cm)
  ++(0,-2.5mm)
  node {\rvdots};
\draw[p]
  (dim21-1-4.south) ++(0,-3.5cm)
  -- ++(0,-1.6cm)
  node (tid2) [anchor=north,monolist] {|[tid]|$C$\\};

\draw[partp] (dim21-1-3.south)
  -- ++(0,-3cm)
  ++(0,-2.5mm)
  node {\rvdots};
\draw[p]
  (dim21-1-3.south) ++(0,-3.5cm)
  -- ++(-1cm,-1.6cm)
  node (tidx) [anchor=north,monolist] {|[tid]|$A$\\};
\draw[partp]
  (dim21-1-3.south) ++(0,-3.5cm)
  -- ++(-.4cm,-.4cm)
  node (tidx) [anchor=north east,yshift=3mm,xshift=1mm,rotate=-5] {$\iddots$};

\node (tid1) [left=-1mm of tid2,monolist] {|[tid]|$B$\\};
\draw[p]
  (dim21-1-4.south) ++(0,-3.4cm) ++(-1mm,-1mm)
  -- (tid1.north);

\draw[p] (dim22-1-1.south)
  -- ++(0,-2.8cm)
  node (tid3) [anchor=north,monolist] {\rvdots\\|[tid]|$D$\\}
   ;

\draw[partp] (dim21-1-1.south) -- ++(0,-.3cm) node[anchor=north] {\rvdots};
\draw[partp] (dim21-1-2.south) -- ++(0,-.3cm) node[anchor=north] {\rvdots};

\draw[partp] (dim22-1-2.south) -- ++(0,-.3cm) node[anchor=north] {\rvdots};
\draw[partp] (dim22-1-3.south) -- ++(0,-.3cm) node[anchor=north] {\rvdots};
\draw[partp] (dim22-1-4.south) -- ++(0,-.3cm) node[anchor=north] {\rvdots};

\draw[halfgauge]
     ([yshift=-3mm,xshift=-2cm]tid1.south west) node {\dots}
     ++(4mm,0)
  -- ([yshift=-3mm]tid2.south east)
  node [yshift=-3mm,midway] {$\avgsize_2$};

\end{tikzpicture}%
  \caption[Overview of the Elf metrics]{Overview of the Elf metrics, focussing on the metrics
  for the second dimension $X_2$, which is stored at depth 2.
  Non-pointer entries of dimension lists have been omitted.
  Highlighted is the query window for $X_1$.
  }\label{fig:elfmetrics}%
\end{figure}
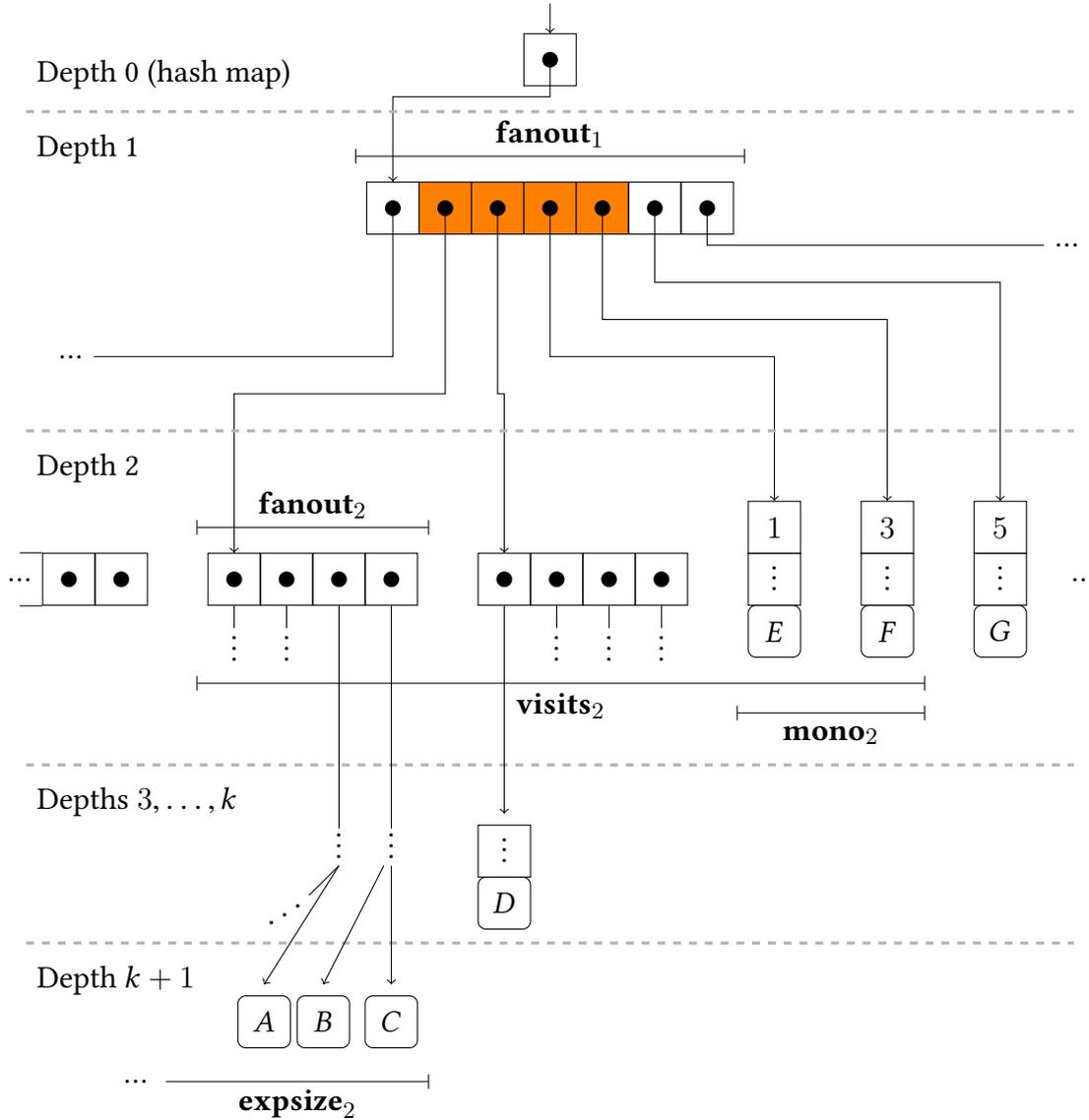

It follows from the independence and uniformity assumptions that all the non-monolist nodes at a specific depth are approximately equal to each other.
Therefore, for each level of the Elf, we only need to consider the \say{mean} or \emph{expected} node at that level.
Figure~\ref{fig:elfmetrics} gives an overview of the Elf metrics, which we describe below.
In this example, tuples $\{A,B,C,D,E,F\} \subset Q$, i.e.\ they are contained in the result set for query $Q$,
while $G$ is not.

\begin{description}
  \item[$\visits_i$] The number of nodes at depth $i$ that are visited during the search.
    This includes both dimension lists and monolists.
  \item[$\mono_i$]
    The number of monolists \emph{encountered} at depth $i$,
    i.e.\ the number of visited nodes at depth $i$ that are monolists.
    In such a monolist, the first value stored in the monolist is the value of column $i$.
    The number $\mono_i$ is equal to the number of visited $(k-i)$-monolists.
  \item[$\avgsize_i$]
    The expected number of tuples that are \emph{below} any dimension list at depth $i$.
    This number is equal to the expected number of tuples
    that share a prefix redundancy of length $i-1$.
  \item[$\fanout_i$]
    The expected length of any dimension list at depth $i$.
    This is equal to the expected number of different attribute values for each list.
\end{description}
In the following sections, we will describe how to estimate these Elf metrics ahead of time
and later show that this model of the Elf structure
allows an accurate prediction of the Elf size.

\section{Computing the Elf metrics}\label{sec:basicmodel}
In order to predict the size of the Elf, we have introduced a number of metrics for the structure.
In this section, we describe our technique to compute these metrics,
which include the number of nodes visited during the Elf search.
Our technique is an iterative algorithm that works from the top of the Elf to the bottom.
We will first give the metrics for the top level of the Elf,
and then describe a set of equations for deriving by induction
the metrics for a deeper level $i+1$ given the metrics for level $i$.
Finally, we give a compact representation of this prediction algorithm.

\subsection{Base case}
First, note that we ignore the hash map property, and assume that the first dimension list only contains a single entry.
Therefore, computing the Elf metrics starts at subscript 1 instead of 0.
At the top of the Elf, we visit only one list that contains the entirety of $R$.
Since we can assume that $\card{R}>1$, the first dimension list is not a monolist. Therefore $\visits_1=1$, $\mono_1 = 0$, and $\avgsize_1=\card{R}$.

Finally, $\fanout_1$ is determined by examining the distribution of the data set. Since $X_1$ is uniformly distributed over $\vee_1$ different values, $\fanout_1$ is given by the answer to the question:
\enquote{Given $\vee_1$ urns containing distinctly colored balls.
Now $\card{R}$ times, an urn is chosen independently, and one ball is drawn from it.
How many different colors of balls are drawn?}
Equation \ref{eq:buckets} gave the solution to this combinatorial problem as a function $\buckets$, that maps the cardinality of the distribution and the number of draws to the expected number of unique values.
Therefore we get $\fanout_1 = \buckets(\vee_1, \card{R})$.

\subsection{Inductive step}
Starting from the base case values, we will compute the number of visited nodes in deeper levels of the Elf iteratively.
Therefore, let $i \in \{1,\dots,k-1\}.$

\paragraph{Number of visited nodes.}
The number of nodes visited at depth $i+1$ is determined as follows.
We visited $\visits_i$ nodes in layer $i$ during our search.
$\mono_i$ of these nodes were monolists, and monolists don't lead to any further nodes in layer $i+1$.
However, for each of the non-monolist nodes, we visit $\fanout_i$ nodes at depth $i+1$. Therefore:
\begin{equation}\label{eq:searchvisits}
  \visits_{i+1} = (\visits_i-\mono_i) \cdot \fanout_i.
\end{equation}

\paragraph{Number of monolist encounters.}
On average, $\avgsize_i$ tuples below every node at depth $i$ will be distributed into $\fanout_i$ nodes each.
For large $\avgsize_i$, modelling this process is unproblematic, as the uniform distribution of values simply leads to an equal partitioning of
\begin{equation}\label{eq:naive_avgsize}
  \frac{\avgsize_i}{\fanout_i}
\end{equation}
tuples stored per subnode.
However, in deeper levels of the Elf, $\avgsize_i$ will rapidly decrease.
This increases the probability of a subnode storing just one tuple.
In this case, the subnode will be the start of a monolist.
We can estimate the statistically expected number of these \emph{monolist encounters}.
The expected number is the solution of:
\enquote{Given $\avgsize_i$ balls are uniformly distributed into $\fanout_i$ urn.
Assuming each urn gets at least one ball, how many urns get exactly one ball?}
Equation \ref{eq:monobuckets} answered this combinatorial problem as $\monobuckets(\avgsize_i, \fanout_i)$.
This expected number of monolists is encountered for every non-monolist node visited at depth $i.$
The total expected number of monolist encounters at level $i+1$ is therefore:
\begin{equation}\label{eq:searchmono}
  \mono_{i+1} = (\visits_i-\mono_i) \cdot \monobuckets(\avgsize_i, \fanout_i).
\end{equation}

\paragraph{Number of remaining tuples.}
We now know that $\fanout_i$ subnodes are below each of the nodes at dimension $i$, and $\monobuckets(\fanout_i, \avgsize_i)$ of those are monolists.
Therefore, we can now compute the average number of tuples that are stored in each non-monolist node at depth $i+1$ as follows:
\begin{equation}\label{eq:searchavgsize}
  \avgsize_{i+1} = \frac{\avgsize_i-\monobuckets(\avgsize_i, \fanout_i)}{\fanout_i-\monobuckets(\avgsize_i, \fanout_i)}
\end{equation}
This value differs from the value in Equation \ref{eq:naive_avgsize} in that it incorporates the number of monolists: each monolist decreases the number of tuples to be distributed, but also decreases the number of buckets the tuples are distributed into.
In essence, seen from the perspective of the non-monolist subnodes, the tuples stored in monolists simply vanish.

\paragraph{Number of subnodes.}
Finally, for each of the subnodes at depth $i+1$, the $\avgsize_{i+1}$ tuples are stored in an Elf dimension list, grouped by their value of $X_{i+1}$.
From the uniformity assumption, it follows that $X_{i+1}$ is uniformly distributed over $\vee_{i+1}$ values.
The expected length of these dimension lists is therefore equivalent to the solution of this combinatorical problem:
\enquote{Given $\vee_{i+1}$ urns containing distinctly colored balls.
Now $\avgsize_{i+1}$ times, an urn is chosen independently, and one ball is drawn from it.
How many different colors of balls are drawn?}
Like in the base case, the solution is given by the cardinality estimation function $\buckets$:
\begin{equation}\label{eq:searchfanout}
  \fanout_{i+1} = \buckets(\vee_{i+1}, \avgsize_{i+1})
\end{equation}

This closes the loop: now we have computed $\visits_{i+1}$, and also all the values necessary to compute $\visits_{i+2}$. By induction, we can now compute all the $\visits_j, j = 1, \dots, k$.

The final addition we will make before presenting the
overall algorithm is in the incorporation of the partial selectivities $\sigma_i$.
The partial selectivity $\sigma_i$ indicates that an expected number of $\sigma_i\cdot\vee_i$
attribute values of $X_i$ match the query bounds.
Therefore, out of the $\visits_{i+1}$ possible visits that could be made from dimension list entries
at dimension $i$, only a fraction of $\sigma_i$ of those is actually taken.
Consequently, we simply have to scale $\visits_{i+1}$ by a factor of $\sigma_i$ to correctly account for the selectivity of the query.

\Autoref{alg:searchtime} shows our method that combines the  considerations described in this section
into a compact procedure for predicting the number of visited nodes during the Elf search.

\begin{algorithm}[tb]
  \caption{Compute the expected number of visited nodes during Elf search\label{alg:searchtime}}
  \SetKwInOut{Input}{Input}\SetKwInOut{Output}{Output}
  \Input{$\card{R}, (\vee_1, \dots, \vee_k), (\sigma_1, \dots, \sigma_k)$}
  \Output{$\visits_1, \dots, \visits_k$}
  \BlankLine
  $\visits_1 \gets 1$\;
  $\mono_1 \gets 0$\;
  $\avgsize_1 \gets \card{R}$\;
  $\fanout_1 \gets \buckets(\vee_i, \card{R})$\;
  \BlankLine
  \For{$i$ from $1$ to $k-1$}{
    $\visits_{i+1} \gets \sigma_i \cdot (\visits_i-\mono_i) \cdot \fanout_i$\;
    $\operatorname{mono\_per\_list} \gets \monobuckets(\avgsize_i, \fanout_i)$\;
    $\mono_{i+1} \gets \sigma_i \cdot (\visits_i-\mono_i) \cdot \operatorname{mono\_per\_list}$\;
    $\avgsize_{i+1} \gets (\avgsize_i-\operatorname {mono\_per\_list})/(\fanout_i-\operatorname {mono\_per\_list})$\;
    $\fanout_{i+1} \gets \buckets(\vee_{i+1}, \avgsize_{i+1})$\;
  }
  $\visits_{k+1} \gets (\visits_{k}-\mono_{k}) \cdot \fanout_{k}$\;
\end{algorithm}

\section{Modelling correlations as fractal dimensions}\label{sec:correlation}
In the previous sections, we described the expected performance of the Elf under the \emph{independence assumption}, which states that the values in each column do not correlate.
This greatly simplified the analysis of the structure of the Elf in deeper dimension.

However, real-world data sets often contain correlations between at least some of the attributes.
For example, consider the TPC-H \model{lineitem} relation, shown in Table~\ref{tbl:tpchcorrelated}.
\begin{table}[ht]
  \centering
  \begin{tabular}{l c l l}
    \toprule
    \model{order\_id} & \dots & \model{shipdate} & \model{receiptdate} \\
    \midrule
    999  & \dots & 1998-01-10 & 1998-01-14 \\
    1000 & \dots & 1997-04-09 & 1997-04-21 \\
    1001 & \dots & 1994-10-16 & 1994-11-08 \\
    \bottomrule
  \end{tabular}
  \caption[Correlation between  \model{shipdate} and \model{receiptdate} in the TPC-H \model{lineitem} table]{An excerpt from the TPC-H \model{lineitem} table. The dates in the \model{shipdate} and \model{receiptdate} columns are always close to each other.\label{tbl:tpchcorrelated}}
\end{table}
Each entry in the \model{lineitem} table represents an item of a purchase order handled by a retail business.
At some point in time, the item is shipped to the customer.
The shipping date is stored in the attribute \model{shipdate}.
Later, the item arrives and the delivery is acknowledged by the customer.
The date of the delivery is stored in the attribute \model{receiptdate}.

Observe that the values of \model{shipdate} and \model{receiptdate} are related to each other.
For example, it's impossible that an item is received before it is shipped.
It's also quite unlikely that an item is received more than a couple of weeks after it was shipped.
In fact, the TPC-H specification defines $\model{receiptdate} = \model{shipdate} + N$, where $N \sim \Ud{1}{21}$.

In cases like these, we say that the values in these columns \emph{correlate}.
Correlations impair the accuracy of our performance model of the Elf.
In particular, the problem caused by correlations can be illustrated with a simple example:

Consider $\Elf_2$ for a simplified \model{lineitem} relation, containing only \model{shipdate} and \model{receiptdate}
Assume \model{shipdate} is stored at depth 1, with \model{receiptdate} below it at depth 2.
The correlation between the two columns causes the apparent cardinality of the deeper column within the Elf to shrink.
Even though the marginal distribution of each attribute is uniform (and therefore satisfies the uniformity assumption),
the $\fanout_2$ value is vastly overestimated.

The reason for this misprediction can be explained as follows.
Suppose we observe \num{1000} distinct values for both \model{shipdate} and \model{receiptdate} within the table.
When building the Elf, we group the tuples by \model{shipdate},
creating $\fanout_1 = \num{1000}$ dimension lists on the second level.
Now consider the shape of these dimension lists, assuming that the table is sufficiently large.
Eventually, every value of \model{receiptdate} is predicted to appear in every list.
Therefore, \autoref{alg:searchtime} will estimate $\fanout_2 \approx \num{1000}$.

However, as described above, for any fixed value of \model{shipdate},
$\model{receiptdate} = \model{shipdate}\, +\, N$, where $N \sim \Ud{1}{21}$.
This means that
$$\model{receiptdate} \sim \Ud{\model{shipdate}}{\model{shipdate}+21}.$$
Therefore, for sufficiently many tuples, $\fanout_2 = 21 \ll \num{1000}$.
The error can be made arbitrarily large by further increasing the number of distinct values in the first column.

\subsection{The correlation dimension}
In statistical computation, this ``cardinality shrinkage'' phenomenon
is studied as the \emph{correlation dimension} \cites{doi:10.1080/00949659108811357}.
The correlation dimension is a type of fractal dimension of a set of points.
For a fractal object, a fractal dimension measures the amount of ``coverage'' of Euclidean space by the object.

As an example of a classic fractal object,
consider the \emph{Sierpinski triangle}, which is shown in \autoref{fig:sierpinski}.
The Sierpinski triangle is neither one-dimensional (because it has nonzero perimeter)
nor two-dimensional (because it has zero area).
Other fractals that cover ``less'' or ``more'' of the two-dimensional plane have lower or higher fractal dimensions, respectively.
In fact, the Sierpinski triangle has a \emph{fractal dimension}
\footnote{
  This assumes the definition of the \emph{box-counting fractal dimension}.
  Other types of fractal dimensions exist, that we will not discuss further.}
of $\log 3 / \log 2 \approx \num{1.585}$,
just between 1 and 2.
\def\trianglewidth{2cm}%
  \pgfdeclarelindenmayersystem{Sierpinski triangle2}{
      \symbol{X}{\pgflsystemdrawforward}
      \symbol{Y}{\pgflsystemdrawforward}
      \rule{X -> X-Y+X+Y-X}
      \rule{Y -> YY}
  }%
\begin{figure}\centering
  \foreach \level in {1,3,5}{%
  \tikzset{
      l-system={step=\trianglewidth/(2^\level), order=\level, angle=-120}
  }%
  \begin{minipage}[b]{.30\linewidth}\centering
  \begin{scaletikzpicturetowidth}{0.9\textwidth}
  \begin{tikzpicture}[scale=\tikzscale]
      \fill [black!85] (0,0) -- ++(0:\trianglewidth) -- ++(120:\trianglewidth) -- cycle;
      \draw [draw=none] (0,0) l-system
      [l-system={Sierpinski triangle2, axiom=X},fill=white];
  \end{tikzpicture}
  \end{scaletikzpicturetowidth}
  \subcaption{iteration \level}
  \end{minipage}}%
  \caption[The Sierpinski triangle after different numbers of iterations]{
    The Sierpinski triangle after different numbers of iterations.
    The initial shape is an equilateral triangle.
    At each iteration, every triangle is replaced by three smaller triangles, with the triangle-shaped middle portion left out.
    As the number of iterations approaches infinity, the figure tends to infinite perimeter and zero area.
    Therefore, it is neither a one-dimensional or two-dimensional Euclidean shape.
  }\label{fig:sierpinski}
\end{figure}
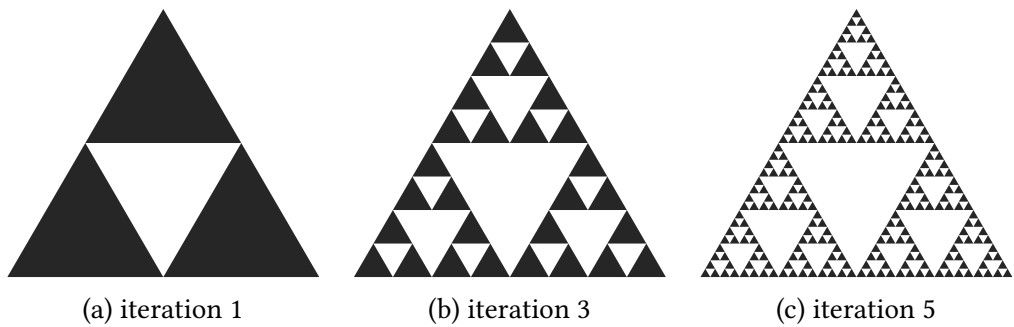

A similar behaviour can be observed for datasets that exhibit correlations, like the \model{lineitem} relation described above \cite{Belussi:1995:ESS:645921.673166}.
Since the fractal dimension can also be explained through the entropy of the data set,
it is also sometimes called the \say{information dimension} or \say{intrinsic dimension}.
Correlations cause a decrease in information-theoretic entropy -- analogously,
a correlation \say{shrinks} the dimensionality of the data set.
This effect can drastically impact the performance of multi-dimensional index structures \cite{faloutsos1994beyond}.
It is one of the factors causing the curse of dimensionality (see \autoref{sec:curse}),
which leads to performance deteriorations in multi-dimensional index structures.

For the Elf, however, the fractal dimension has a very specific and well-defined impact.
In particular, it changes the length of the dimension lists in the correlated dimension that is stored deeper within the Elf.
Consider these scenarios, which are illustrated in figure \ref{fig:fractalcorrelation}:
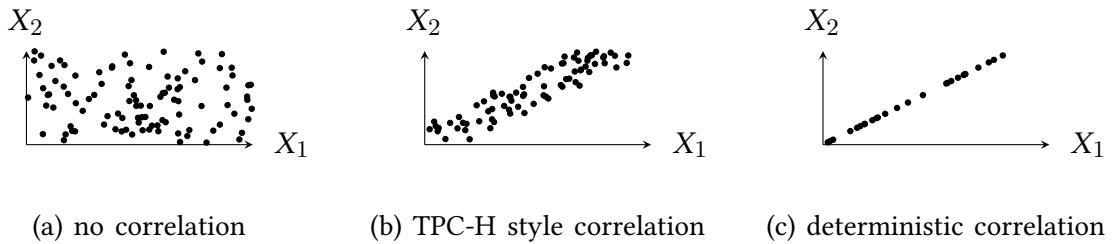
\begin{figure}[tb]
  \centering
  \makebox[\textwidth][c]{\pgfmathsetseed{1234} 
\pgfplotstableset{ 
    create on use/x1/.style={create col/expr={abs(rand*12)}},
    create on use/y1/.style={create col/expr={abs(rand*5)}}
}
\pgfplotstablenew[columns={x1,y1}]{100}\corrtablea
\pgfplotstableset{ 
    create on use/x2/.style={create col/expr={abs(rand*12)}},
}
\pgfplotstablenew[columns={x2}]{100}\corrtableb
\pgfplotstablecreatecol
    [expr={0.5*\thisrow{x2} + rand}]
    {y2}{\corrtableb}
\pgfplotstableset{ 
    create on use/x3/.style={create col/expr={abs(rand*12)}},
}
\pgfplotstablenew[columns={x3}]{30}\corrtablec
\pgfplotstablecreatecol
    [expr={0.5*\thisrow{x3}}]
    {y3}{\corrtablec}
\centering
\pgfsetlayers{background,main}
\begin{tikzpicture}
\begin{axis}[
  name=a,
    xshift=-1.5cm,
    width=0.3\textwidth,
    xmin=0, xmax=12, ymin=0, ymax=5,
    axis lines=middle,
    axis equal image,
    height=5cm,
    xtick=\empty, ytick=\empty,
    xlabel=$X_1$, ylabel=$X_2$,
    mark size=1,
    every axis x label/.style={
        at={(ticklabel* cs:1.05)},
        anchor=west,
    },
    every axis y label/.style={
        at={(ticklabel* cs:1.05)},
        anchor=south,
    },
    ticks=none,
    enlargelimits=false,
]
\addplot [only marks, draw=black] table {\corrtablea};
\end{axis}

\begin{axis}[
  name=b,
  at=(a.right of north east), anchor=left of north west,
  xshift=1cm,
    width=0.3\textwidth,
    xmin=0, xmax=12, ymin=0, ymax=5,
    axis lines=middle,
    axis equal image,
    height=5cm,
    xtick=\empty, ytick=\empty,
    xlabel=$X_1$, ylabel=$X_2$,
    mark size=1,
    every axis x label/.style={
        at={(ticklabel* cs:1.05)},
        anchor=west,
    },
    every axis y label/.style={
        at={(ticklabel* cs:1.05)},
        anchor=south,
    },
    ticks=none,
    enlargelimits=false,
]
\addplot [only marks, draw=black] table {\corrtableb};
\end{axis}

\begin{axis}[
  name=c,
  at=(b.right of north east), anchor=left of north west,
    xshift=1cm,
    width=0.3\textwidth,
    xmin=0, xmax=12, ymin=0, ymax=5,
    axis lines=middle,
    axis equal image,
    height=5cm,
    xtick=\empty, ytick=\empty,
    xlabel=$X_1$, ylabel=$X_2$,
    mark size=1,
    every axis x label/.style={
        at={(ticklabel* cs:1.05)},
        anchor=west,
    },
    ticks=none,
    every axis y label/.style={
        at={(ticklabel* cs:1.05)},
        anchor=south,
    },
    enlargelimits=false,
]
\addplot [only marks, draw=black] table {\corrtablec};
\end{axis}

\node[text width=4cm,align=center,anchor=north] at ([yshift=-5mm]a.south)
  {\subcaption{no correlation}};

\node[text width=6cm,align=center,anchor=north] at ([yshift=-5mm]b.south)
  {\subcaption{TPC-H style correlation}};

\node[text width=5cm,align=center,anchor=north] at ([yshift=-5mm]c.south)
  {\subcaption{deterministic correlation}};
\end{tikzpicture}

}
  \caption[Data sets with different amounts of correlation]{
    Three two-dimensional data sets, each showing a different amount of correlation.}
  \label{fig:fractalcorrelation}
\end{figure}
\begin{enumerate}
  \item No correlation. If correlations are absent, the data points are randomly distributed across the two-dimensional (\model{shipdate},\model{receiptdate}) data space.
  As the number of points approaches infinity, the set of points becomes a two-dimensional plane.

  \item Medium correlation. In the actual TPC-H data, the points always appear in the neighbourhood of the trend line, but randomly deviate from it slightly.
  As the number of points approaches infinity, the set of points forms a \say{band} around the trend line.

  \item Full correlation. If the correlation is absolute, one attribute value implies one specific value in the other attribute.
  In this case, all data points fall on a line.
  As the number of points approaches infinity, the set of points becomes this one-dimensional line.
\end{enumerate}
The actual TPC-H correlation exhibits the fractal dimension property;
while it does not ``flatten'' the \model{receiptdate} dimension entirely,
it significantly reduces the number of unique values encountered.

Note that the loss of dimensionality in the \say{full correlation} scenario would be directly visible in the corresponding Elf:
every value of \model{shipdate} has only one corresponding value of \model{receiptdate}.
Therefore, all dimension lists at depth 2 would only have length 1;
the second dimension would \say{disappear}.

\subsection{Eliminating the effect of correlations on cardinality}
Using the insight from the previous section, we correct the size of the attribute domains
to account for this type of correlation, which we call a \emph{uniform correlation}.

In real-world data sets, we might not know about correlations beforehand.
However, methods have been proposed that use statistical measures
such as the covariance of the attributes or more sophisticated
techniques to automatically find correlations and quantify their strength \cite{ilyas2004cords}.
Therefore, we assume that correlations are known or computed at Elf build time.

To compensate for the effects of the correlation, we reduce the domain size of the correlated attribute.
For example, in the \model{lineitem} data set described above, we would simply assume that
\model{receiptdate} has cardinality \num{21} instead of \num{1000}.
The new corrected cardinality is called then \emph{correlated cardinality}.
The domain is always reduced on the attribute that is ordered deeper within the Elf.
Now, we can treat the data set as if no correlations were present to begin with,
and we can again use \autoref{alg:searchtime} to predict the search performance.



\section{Modelling non-uniform data distributions}\label{sec:nonuniform}
Our model, as described thus far, relies on the \emph{uniformity assumption}.
The uniformity assumption states that that attribute values for a given column $i$ are drawn with equal probability from a discrete uniform distribution $\Ud{0}{\vee_i}$.

Real-world data sets usually do not exhibit this property.
Instead, they often follow non-uniform discrete distributions.
Therefore, assuming a uniform distribution is dangerous;
inevitably, it will lead to errors in the predictions made by the model.

In this section, we describe our histogram-based method to relax the uniformity assumption, instead allowing arbitrary distributions for the data contained in the Elf.
We will first introduce the theory of our approach, and then show how we incorporate it into our model.

\subsection{Histogram theory}
\begin{figure}\centering
  \begin{tikzpicture}
  [
    declare function={
      binom(\n,\p) = \n!/(x!*(\n-x)!)*\p^x*(1-\p)^(\n-x);
      normal(\m,\s) = 1/(\s*sqrt(2*pi))*exp(-((x-\m)^2)/(2*\s^2));
    },
  ]
      \begin{axis}[
        ybar interval,
        ymin=0,
        xmin=0,xmax=6,
        width=\textwidth,
        height=6cm,
        ticks=none,
        ]
        \addplot [samples=7,domain=0:6] {binom(6,0.5)*20};
      \end{axis}
  \begin{axis}
    [
      width=\textwidth,
      height=6cm,
      xmin=0,xmax=20,
      xlabel={$c$},ylabel={$\PP[X=c]$},
      ymajorticks=false,
      xtick=\empty,
      extra x ticks={12},
      extra x tick labels={$\mu$},
    ]
    \addplot [smooth,red,thick,domain=0:20] {normal(12,3)};
  \end{axis}
  \end{tikzpicture}
  \caption{Probability densitity of a normal distribution and a six-bucket histogram approximation}\label{fig:histogram}
\end{figure}
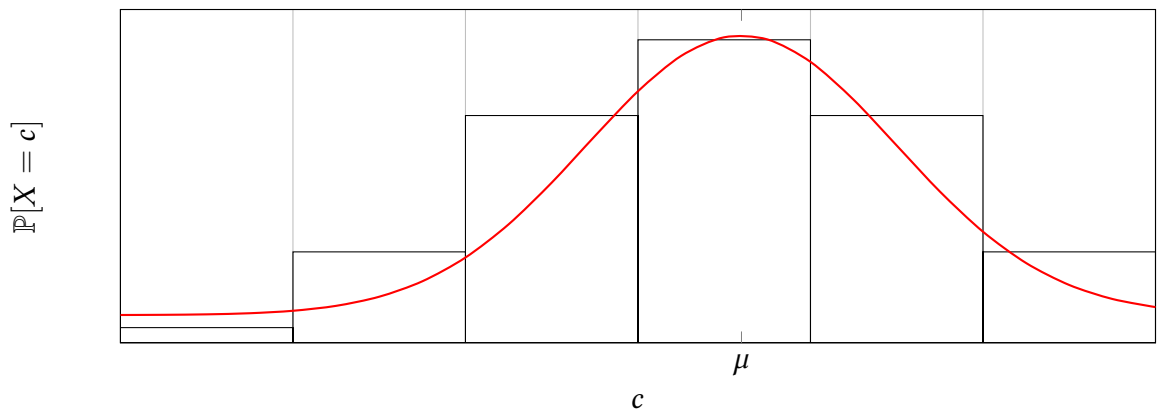
The use of histograms is a widely used approach to approximate non-uniform data sets
in situations where they would otherwise be difficult to handle
\cite{Ioannidis:2003:HH:1315451.1315455}.
A histogram partitions the \emph{attribute domain} into so-called \say{buckets}.
Each value within the domain is assigned to a unique bucket.
Therefore, the histogram also partitions the relation itself.
\footnote{In fact, the partitioning concept of histograms is very similar to the Elf itself: the Elf is essentially a fine-grained histogram with a unique bucket for each unique prefix key in the data set.}
\Autoref{fig:histogram} shows an example data distribution together with a histogram approximation with 5 buckets.
The special case of a single-bucket histogram represents the uniformity assumption.
As the number of buckets tends to infinity, the shape of the histogram converges to the original data distribution.

The main advantage afforded by histograms is that within a bucket, the data distribution is again approximately uniform.
The details on how to optimally choose bucket sizes are still a topic of ongoing research.
In the following, we will only focus on the simplest and oldest type of histogram, the equi-width histogram.
In an \emph{equi-width $k$-histogram}, every bucket has the constant width $n/k$, where $n$ is the size of the data domain, and $k$ is the number of buckets.

We will employ the histogram technique to approximate the size of the Elf
for a non-uniform data set by instead measuring one sub-Elf for each bucket of the histogram.
These sub-Elfs again have approximately uniform distribution, so we can apply our basic model described in \autoref{sec:basicmodel}.
Consider the following example.
Assume that the values $X_1$ in the first column are distributed as follows:
\begin{enumerate}
  \item $1 \leq X_1 \leq 100$
  \item $\PP(X_1 \leq 50) = 0.75$
  \item $\PP(X_1 = i \,\,\vert\,\, X_1 \leq 50) = \PP(X_1 = j \,\,\vert\,\, X_1 > 50) = \const$
\end{enumerate}
This means that the data distribution is a piecewise uniform distribution that is \say{skewed} towards smaller values.
Building a 2-histogram in this case is very simple: the Elf is simply split into two halves, one containing the values $1,\dots,50$, and one containing $51,\dots,100$
Now, within each sub-Elf, $X_1$ is again uniformly distributed.
The final missing piece is now how to merge the sizes of the sub-Elfs back together to obtain the size of the original Elf.

However, due to the design of the Elf, the merging process is very simple: it simply consists of summing the sizes of the sub-Elfs.
Searching through Elf node is accomplished by scanning the dimension list, and appropriately
recursing into the tree where needed.
If the list were to be divided into two parts, each part could be searched independently,
without examining the other part.
Therefore, we can assume that the total search time
is equal to the sum of the search times of the two halves.

This observation is similar to the concept of sub-vector space dimension in algebra:
let $V$ be a vector space and $U,W$ two sub-vector spaces.
Then the \emph{dimension formula for sub-vector spaces} holds:
$$\dim(U+W) = \dim(U) + \dim(W) - \dim(U \cap W).$$
In particular, if the sum is direct, i.e.\ if $\dim(U \cap W) = 0$,
we arrive at the principle described above.
This leads us to the following hypothesis:


\begin{hyp}\label{thm:homomorph}
  Let $A, B \subset R$, and let $f(X)$ be the number of visited nodes
  while searching through $X \subset R$.
  Then
    $$f(A \cup B) \approx f(A) + f(B) - f(A \cap B).$$
  Therefore, if $A \cap B = \emptyset$, then $f(A \cup B) \approx f(A) + f(B)$.
  In particular, if $R=A \oplus B$, then $f(R) \approx f(A) + f(B)$.
\end{hyp}
This hypothesis allows us to predict the visit count for the Elf of a non-uniform data set,
by summing the visit counts for a number of sub-Elfs that have approximately uniform distribution.

\subsection{The histogram algorithm}
Using the results from the previous section,
we will now extend \autoref{alg:searchtime} to handle a non-uniform data distribution in the first column of the Elf, i.e.\ the top-most column.
The technique we present here can be applied multiple times to handle non-uniform distributions in more than one column.

\Autoref{alg:nonuniform} shows our method that uses the histogram technique to estimate the Elf size.
The algorithm is best explained through an example.
Therefore, let $X_1 \sim \operatorname{Bin}(10,0.5)$,
i.e.\ $X_1$ follows a binomial distribution with $n=10$ and $p=0.5$.
This means that $X_1 \in \{0,\dots,10\}$, i.e.\ $\vee_1=10$.

First, we compute the probabilities $\PP[X_1 = x]$ for each $x\in \{0,\dots,\vee_1\}$.
This is the probability of each unique value occuring in a single row of the relation.

Then, we first partition this list of probabilities into $b$ histogram buckets of equal width.
If $b=3$, the three buckets will contain the probabilities of the values
$\{0,1,2,3\}$, $\{4,5,6,7\}$, and $\{8,9,10\}$, respectively.
These ranges are produced by the function \model{Chunks},
which splits a list of values into $b$ disjoint lists
of values.
Each of these sub-lists is contiguous and of approximately equal length.

We iterate over each \model{chunk}, which is one of the sub-lists containing
the probabilities for an interval of attribute values.
We start at the left end of the distribution and iterate to the right.

First, we determine the position of \model{chunk} within the normalised $[0;1]$ query interval.
To this end, we keep track of the number \model{seen}, which represents the upper bound of the previously visited chunk within the normalised interval.
This previous upper bound then becomes the lower bound for the current chunk.
For the first bucket, $\model{seen}=\model{left}=0$.
The upper bound for the current chunk is computed by adding to the lower bound the length of the chunk,
again normalised to $\vee_1$, which is the sum of the length of all chunks.
Therefore, for the last chunk, $\model{right}=1$.
This means that $[\model{left};\model{right}]$ now represents the position of the bucket within the
normalised query interval.

If the query interval and the bucket range do not intersect, i.e.\ if $u_1<\model{left}$,
we will never visit the Elf contained in this bucket, and can therefore skip the next step.
Otherwise, we compute the attributes of the sub-Elf represented by the current bucket.
Instead of the entirety of $R$, the sub-Elf only contains the attribute values that
are contained in the chunk.
Therefore, the sub-Elf differs from the original Elf in three important characteristics:
\begin{itemize}
  \item The total number of tuples is not $\card{R}$, but $\alpha \card{R}$, where $\alpha$
  is the fraction of tuples contained within the bucket.
  \item The domain size of the first column is not $\vee_1$, but instead $\card{\model{chunk}}$, i.e.\ the width of the bucket.
  \item Finally, the query interval in the first column is not $[l_1=0,u_1]$,
  but instead the fraction $\beta$ of coverage of $[\model{left};\model{right}]$ by $[0,u_1]$.
\end{itemize}
The adjusted values for these attributes are the parameters for a call to \model{Predict},
which represents the prediction for a uniform Elf described previously in \autoref{alg:searchtime}.
The sum of the visit counts for the sub-Elf is added to a running total of the visit counts of all sub-Elfs.

Finally, after the size of all chunks is predicted, the sum of visited nodes in each of the sub-Elfs
is returned as the prediction for the visit count of the original Elf, per \autoref{thm:homomorph}.

\begin{algorithm}[tb]
  \caption{Approximate non-uniform distribution using histogram\label{alg:nonuniform}}
  \SetKwInOut{Input}{Input}\SetKwInOut{Output}{Output}
  \SetKwFunction{Predict}{Predict}
  \SetKwFunction{Chunks}{Chunks}
  \Input{$R$ with non-uniform distribution in $X_1$, $\dim(R)=k$\\
      Cardinalities $\vee_1, \dots, \vee_k$\\
      Query windows $[0; u_1], [l_2, u_2], \dots, [l_k; u_k]$; normalised to $[0;1]$\\
      Number of buckets $b \leq \vee_1$}
  \Output{$\sum_i \visits_i$}
  \SetKwData{x}{sum}
  \SetKwData{vleft}{left}
  \SetKwData{vright}{right}
  \SetKwData{probabilities}{probabilities}
  \SetKwData{chunk}{chunk}
  \SetKwData{seen}{seen}
  \SetKwFor{ForEachMap}{foreach}{take}{endfch}
  \BlankLine
  $\x \gets 0 : \NN_0 $\;
  $\seen \gets 0 : \RR$\;
  $\probabilities \gets [$\lForEachMap{$v=0,\dots,\vee_1$}{$\PP[X_1 = v]]$}
  \For{$\chunk \in \Chunks(\probabilities, b)$}{
    $\vleft \gets \seen$\;
    $\vright \gets \vleft + \frac{\card{\chunk}}{\vee_1}$\;
    $\seen \gets \vright$\;

    \If{$u_1 > \vleft$}{
      \tcp{$\alpha$: fraction of total tuples contained in the bucket}
      \tcp{$\beta$: queried fraction of the bucket's window}
      $$
        \alpha \, = \, \sum_{p \in \chunk} p,
        \quad \quad \quad
        \beta \, = \, \frac{\min\{\vright , u_1\} - \vleft}{\vright - \vleft}
      $$
      \BlankLine
      \BlankLine
      $\x \, \gets \, \x \, + \, \Predict \left(
        \begin{tabular}{rl}
          nTuples: &$\alpha \card{R}$ \\
          domains: &$\{\card{\chunk}, \vee_2, \dots, \vee_k\}$ \\
          selectivities: &$\{[0; \beta], [l_2, u_2], \dots, [l_k; u_k]\}$
        \end{tabular}
      \right)$\;
    }
  }
  \Return{\x}\;
  \BlankLine
  \BlankLine
  \SetKwData{a}{b}
  \SetKwData{xs}{xs}
  \SetKwProg{Fn}{Function}{ is}{end}
  \Fn{Chunks(\xs: [A], \a: int) : [[A]]}{
    partition \xs into \a chunks of approximately equal size\;
    \Return{the list of chunks}\;
  }
\end{algorithm}

\chapter{Results}\label{chp:results}
In this chapter, we describe the results
of our evaluation of the Elf cost model described in the previous chapter.
To this end, we try to answer the following questions.
\begin{enumerate}
  \item
    How closely does the number of visited nodes predict the execution time of the search?

  \item
    How closely does \autoref{alg:searchtime} predict
    the number of visited nodes during a search of the Elf?

  \item
    How accurately does \autoref{alg:nonuniform} compute the prediction
    for a non-uniform data distribution from the combination of several uniform distributions?

  \item
    Is every part of the model relevant?
    How significantly does the accuracy deteriorate when complex parts of the model are disregarded?
\end{enumerate}

The first question investigates that \autoref{hyp:linearvisit},
where we assumed that the execution time is linear to the number of visited nodes, is valid.
No predictions are made.
Instead, large deviations here for some scenarios indicate the inapplicability of our assumptions for that scenario.

The other questions investigate that our model for predicting the number of visited nodes is accurate.
Large deviations indicate systemic errors in our model.

In the remainder of this chapter, we describe our expected results,
our experimentation methods and implementation,
as well as observations and interpretation.

\section{Expected results}
For uniformly distributed and uncorrelated data sets,
we expected the visit count prediction to be exact up to
small (i.e.\ \SIrange{1}{2}{\percent}) errors caused by statistical \say{fringe} artefacts and rounding errors.

For non-uniform datasets, we expected the accuracy of the predicted visit count to decrease with the amount of \say{non-uniformity} introduced.
Using the histogram technique, we expected the visit count prediction to
again increase in accuracy as the number of histogram buckets increased.

With a correlation between two attributes of increasing strength,
we expected the actual visit count to deviate from the predicted count.
Furthermore, we expected this deviation to be corrected by applying
the technique described in \autoref{sec:correlation}.

Finally, we expected the predicted search time to be approximately
linear to the total visit count in all cases.

\section{Experiment setup}
We investigate the accuracy of our performance model in a multi-step process.
First, we select a configuration of parameters for the Elf.
We then build the Elf using this configuration, and measure the actual visit counts and search response times on the built Elf.
At the same time, we use our model to predict these metrics from the configuration parameters.
Finally, we compare the predicted and actual metrics using stochastic methods.
\Autoref{fig:dataflow} gives an overview of the data flow within our experiment setup,
while \autoref{tbl:evalparams} explains the used symbols and quantities.

\begin{table}[htb]
\begin{tabularx}{\textwidth}{lll}\toprule
Scope & Parameter & Explanation \\ \midrule
Global & $k$ & Number of dimensions in the data set \\
&$\card{R}$ & Number of rows in the data set \\
& $n$ & Number of evaluation queries \\
& $\operatorname{Cov}(X_i,X_j)$ & Covariance between different attributes \\ \addlinespace \addlinespace
For each dimension $i$ & $\mathit{max}_i$ & Source distribution cardinality \\
  & $\vee_i$ & Attribute cardinality \\
  & $X_i$ & Stochastic distribution of the row values \\ \addlinespace \addlinespace
For each query& &\\
and dimension $i$ & $[l_i, u_i]$ & Query window, determines $\sigma_i$ \\ \midrule
Metrics & $t$ & Measured query batch execution time \\
        & $\hat t$ & Predicted query batch execution time \\
        & $v$ & Measured visit count \\
        & $\hat v$ & Predicted visit count \\
\bottomrule
\end{tabularx}
\caption{Overview of the Elf configuration parameters and metrics}\label{tbl:evalparams}
\end{table}
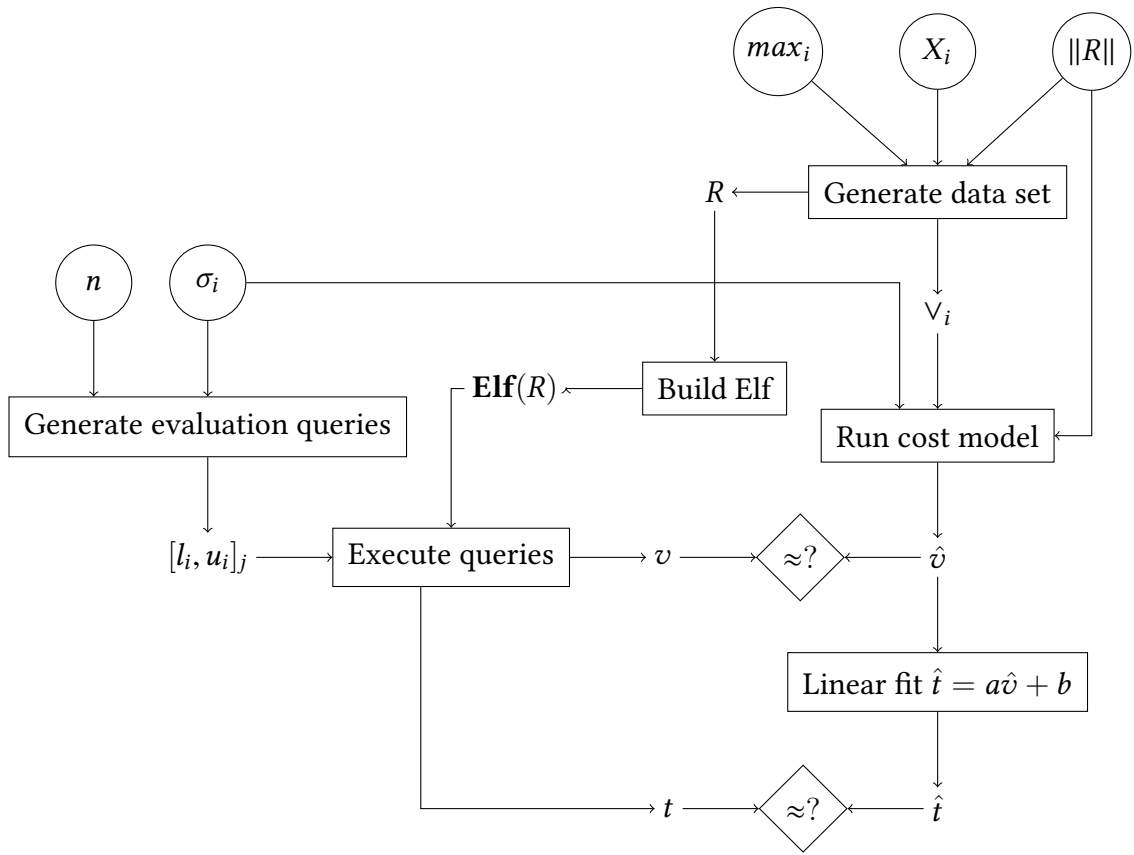
\begin{figure}[htb]
  \centering
  \centering
\pgfdeclarelayer{background}
\pgfdeclarelayer{foreground}
\pgfsetlayers{background,main,foreground}
\begin{tikzpicture}
[
  node distance = 1cm,
  inner sep=1mm,
  auto,
  data/.style={},
  param/.style={draw,circle,minimum size=1cm},
  op/.style={draw,inner sep=2mm},
  measure/.style={},
  eval/.style={draw,diamond},
]
\node (generate) at (376.87bp,450.0bp) [op] {Generate data set};
\node (vee_i) [below=of generate,data] {$\vee_i$};
\node (xi) [above=of generate,param] {$X_i$};
\node (predict) [below=of vee_i,op] {Run cost model};
\node (vhat) [below=of predict,measure] {$\hat v$};
\node (compv) [left=of vhat,eval] {$\approx?$};
\node (v) [left=of compv,measure] {$v$};
\node (exec) [left=of v,op] {Execute queries};
\node (r) [left=of generate,data] {$R$};
\node (build) [below=2cm of r,op] {Build Elf};
\node (elf) [left=of build,data] {$\Elf(R)$};
\node (cardr) [right=of xi] [param] {$\card{R}$};
\node (li_ui) [left=of exec,data] {$[l_i,u_i]_j$};
\node (genquery) [above=of li_ui, op] {Generate evaluation queries};
\node (sigmai) [above=of genquery,param] {$\sigma_i$};
\node (n) [xshift=-1.5cm,above=of genquery.north,param] {$n$};
\node (lm) [below=of vhat,op] {Linear fit $\hat t = a \hat v + b$};
\node (that) [below=of lm,measure] {$\hat t$};
\node (compt) [left=of that,eval] {$\approx?$};
\node (t) [left=of compt,measure] {$t$};
\node (maxi) [left=of xi,param] {$\mathit{max}_i$};
\draw [->] (cardr) -- node {$$} (generate);
\draw [->] (exec.south) ++(-.4cm,0) |- node {$$} (t);
\draw [->] (genquery) -- node {$$} (li_ui);
\draw [->] (sigmai.east) -| node {$$} ([xshift=-.5cm]predict.north);
\draw [->] (t) -- node {$$} (compt);
\draw [->] (r) -- node {$$} (build);
\draw [->] (build) -| node {$$} (elf);
\draw [->] (elf) -| node {$$} (exec);
\draw [->] (lm) -- node {$$} (that);
\draw [->] (maxi) -- node {$$} (generate);
\draw [->] (generate) -- node {$$} (vee_i);
\draw [->] (v) -- node {$$} (compv);
\draw [->] (sigmai) -- node {$$} (genquery);
\draw [->] (n) -- node {$$} ([xshift=-1.5cm]genquery.north);
\draw [->] (vee_i) -- node {$$} (predict);
\draw [->] (generate) -- node {$$} (r);
\draw [->] (exec) |- node {$$} (v);
\draw [->] (li_ui) |- node {$$} (exec);
\draw [->] (xi) -- node {$$} (generate);
\draw [->] (predict) -- node {$$} (vhat);
\draw [->] (cardr) |- node {$$} (predict);
\draw [->] (vhat) -- node {$$} (compv);
\draw [->] (that) -- node {$$} (compt);
\draw [->] (vhat) -- node {$$} (lm);
\end{tikzpicture}
  \caption[Overview of the data flow in our experiment setup]{
    Overview of the data flow in our experiment setup.
    Input parameters are shown circled, while actions are shown in rectangles.
  }
  \label{fig:dataflow}
\end{figure}

\subsection{Parameter selection}
As seen in the table above, a large number of parameters is available that the Elf behaviour depends on.
A holistic evaluation of the entire parameter space was therefore considered impossible.
Instead, we tried to intelligently sample from the large parameter space
to produce a wide range of interesting behaviour in edge cases.

We restricted our evaluation to data sets with a dimensionality of 3--15.
This aligns with our definition of multi-dimensional data
(in contrast to high-dimensional data),
and encompasses important benchmark data sets such as the TPC-H \model{lineitem} relation.
For robust measurements, query batch execution times are needed
that are significantly above the timing noise floor.
To achieve this,
we chose relation sizes $\card{R}$ ranging from \num{10000} to \num{1000000} rows,
and used \num{100} evaluation queries for each query batch.

With fixed bounds for the above parameters,
we derived extreme values for the properties of the data set.
Larger attribute cardinalities $\vee_i$ quickly lead to very small individual groups,
i.e.\ attribute $i$ resembles a primary key of the relation.
This leads to a \say{shallow} Elf that resembles a map from $i$ to invididual tuples,
not considering the hash map property of the Elf in the first dimension.
Therefore, only values $\vee_i \ll \card{R}$ produce meaningful results.
Together with the lower bound of $\vee_i=1$, we can now interpolate between the extremes.
For simplicity, we will only introduce correlations explicitly,
expressed through the covariance matrix $\operatorname{Cov}$.
In the case of no correlations, the resulting attribute cardinalities $\vee_i$ are,
for sufficiently large $\card{R}$, equal to
the cardinality $\mathit{max}_i$ of the uniform distributions we sample from.

Finally, the query windows $[l_i, u_i]$, and therefore the partial and total selectivities,
are always chosen at random.
This emphasizes the fact that arbitrary range queries can be executed on the Elf,
even if their execution might be expensive compared to a linear scan.
The number of evaluation queries $n$ is chosen between 100--500
to avoid long testing times while avoiding the \say{noise floor} with too few queries.

\subsection{Implementation}
\begin{listing}[t]
  \caption{Python implementation of the basic Elf size prediction (\autoref{alg:searchtime})}
  \label{lst:searchtime}
\begin{verbatim}
def algorithm2(cardR, doms, sel):
    visits = ['X', mpf(1)] + [None]*(len(doms)-2+1)
    mono = ['X', mpf(0)] + [None]*(len(doms)-2)
    avgsize = ['X', mpf(cardR)] + [None]*(len(doms)-2)
    gamma = ['X', buckets(doms[1], cardR)] + [None]*(len(doms)-2)

    for i in range(2, len(doms)+1):
        visits[i] = math.ceil(sel[i-1] * (visits[i-1] - mono[i-1]) * gamma[i-1])
        if i == len(doms):
            break
        mono_per_list = monobuckets(gamma[i-1], avgsize[i-1])
        mono[i] = math.ceil(sel[i-1] * (visits[i-1] - mono[i-1]) * mono_per_list)
        avgsize[i] = (avgsize[i-1] - mono_per_list) / (gamma[i-1] - mono_per_list)
        gamma[i] = buckets(doms[i], avgsize[i])

    return visits

def buckets(n, k):
    return n * (math.pow(n,k) - (n-1)**k) / math.pow(n,k)
def monobuckets(k, n):
    return k * ((k-1)/k)**(n-k)
\end{verbatim}
\end{listing}
\begin{listing}[t]
  \caption{Python implementation of the histogram prediction (\autoref{alg:nonuniform})}
  \label{lst:nonuniform}
\begin{verbatim}
def algorithm3(nbuk, doms, sels):
    n = int(doms[1])
    x = 0
    seen = 0.0

    for chunk in chunks([prob(n, k, bias) for k in range(0, n)], nbuk):
        left = seen
        right = left + float(len(chunk))/n
        seen = right

        if sels[1] > left:
            alpha = sum(chunk)
            beta = (min(right, sels[1]) - left) / (right-left)

            doms[1] = len(chunk)
            sels[1] = beta
            x += sum(algorithm2(alpha*cardR, doms, sels))

    return x
\end{verbatim}
\end{listing}

All tests were performed on an Intel~Core~i7-3820QM CPU of the \say{Ivy Bridge} architecture
clocked at \SI{2.3}{\giga\hertz},
with $2\times\SI{8}{\giga\byte}$ of DDR3 memory clocked at \SI{1333}{\mega\hertz},
running on a 2012 Apple~MacBook~Pro under Mac~OS~X~10.11.1.

Our implementations of algorithms~\ref{alg:searchtime} and~\ref{alg:nonuniform}
are shown in listings~\ref{lst:searchtime} and~\ref{lst:nonuniform}, respectively.
We implemented the visit prediction model in Python, using the \model{mpmath} library \cite{mpmath} for arbitrary-precision arithmetic.
This is especially important for the implementation of $\buckets$ and $\monobuckets$, since these functions deal with very large intermediate values.
For example, when computing Algorithm~\ref{alg:searchtime} for a small relation with $\card{R}=\num{100000}$, values around $2^{\num{1660964}}$ are encountered.
For scale, IEEE~754 double-precision floating-point arithmetic can only represent finite numbers up to $2^{\num{1024}}$.
The use of the comparatively slow arbitrary-precision arithmetic is not required, but was favoured for clarity and ease of implementation.

\subsection{Procedure}\label{sec:procedure}
Our experiment setup aims to mimic the setting of a commercial database system.
In particular, we divide the experiment into two phases: a \say{build} phase and a \say{search} phase.
In the build phase, we first generate a number of Elf configurations and evaluation query batches as described above.
We then build the Elf using the configuration, and store it in memory.
In the search phase, the queries are executed and the execution time is recorded.

These phases correspond to a background (or \say{off-line}) index build, and then a phase of on-line query processing.
In detail, we perform the following steps:

\paragraph{Build phase.}
We build a number of \emph{examples}, each comprising an Elf and a set of evaluation queries.
For each example, a data set is generated by sampling the random variable $X_i$ for each attribute $i$.
We then build the Elf for the data set, and linearise it.
The Elf is stored in memory, together with a generated batch of evaluation queries.
The parameters for the data set and queries are detailed in the specific result sections below.
This process is repeated for each example.

\paragraph{Search phase.}
The first batch of evaluation queries is executed against the first Elf, and the total response time for the batch measured.
This process is repeated for each example.
The entirety of this process is repeated five times, with the median value becoming the ground truth response time for the example.
Repeating the measurement ensures that any transient background CPU activity (i.e.\ caused by the operating system) does introduce noise.
Before repetitions of the same example run $i$, all other examples are evaluated,
to ensure that the CPU cache does not contain parts of the Elf of run $i$..

\subsection{Evaluation}
After completing the measurements of the actual values, we run our prediction algorithm
for the same scenario, using the known partial selectivities and attribute cardinalities.
For settings with uniform data distributions, \autoref{alg:searchtime} is used, while for
non-uniform data distributions, \autoref{alg:nonuniform} is used.

As shown in \autoref{fig:dataflow},
the prediction model only has access to the domain sizes and selectivities of the test scenario.
This is consistent with the information available in a real-life query planning scenario.
Since the prediction is unaware of the number of evaluation queries $n$ in the batch,
we expect the actual visit counts to be higher by the constant factor of $n$.

While the predicted and actual visit counts ($\hat v$ and $v$, respectively) can be compared directly,
the response time of the query batch is measured in seconds,
while the visit count is measured in the number of accessed memory cells.
Our hypothesis is that these two quantities vary in a linear fashion.
Therefore, we train a linear regression model \cite{SWB-445320133} on the relationship
\begin{equation}
  t \approx a\hat v + b + \epsilon,
\end{equation}
where $t$ is the response time of the query batch,
and $\hat v$ is the predicted number of visited nodes.
The error term $\epsilon$ indicates the influence of impact factors not contained in our cost model,
such as caching effects and differences between CPU architectures.

The linear fit results in the optimal coefficients $a$ and $b$,
that we use to predict new execution times $\hat t$ according to $\hat t = a\hat v + b$.
Now we can evaluate the prediction errors $\hat t - t$ and $\hat v - v$.
We expect $a$ to represent the mean access time for a single visited memory cell,
while $b$ represents the constant overhead of the Elf search and measurement,
which includes e.g.\ building the result set.

We measure the accuracy of the predicted visit counts
and execution times using a number of stochastic measures:
\begin{enumerate}
  \item
    The \emph{Coefficient of Determination} $R^2$ measures the fraction of variance in the actual values that is \emph{explained} by the variance in the predicted values.
    A high $R^2$ value suggests that the prediction is accurate, i.e.\
    it is unlikely that there is a large derivation between the predicted and actual values.
    We will write $R^2(a\sim b)=c$ to denote that a fraction $c$
    of the variance of $a$ is explained by the variance of $b$.

  \item
    The \emph{Coefficent of Variation} $c_v = \sigma / \mu$, where $\sigma$ and $\mu$ denote the standard deviation and mean of a sample, respectively.
    $c_v$ provides a relative measurement of the variance of the sample.
    Generally, we will first show a high $c_v$ to suggest that the sample covers an \say{interesting} part of the Elf parameter space, where the different parameter values contained in the sample lead to significant changes in the Elf performance.
    Then, we will show a high $R^2$ value to propose that our model explains
    a large fraction of this high variance.

  \item
    The mean absolute percentage error (MAPE) measures the relative deviation of the predicted values from the actual values \cite{armstrong1992error}.
    We assume that the absolute prediction error is linear to the actual search time, since the magnitude of the noise introduced by caching effects should be linear to the number of memory accesses.
    Therefore, we have chosen a relative error metric.


  \item The \emph{Error over magnitude} distribution visually indicates how the absolute prediction error varies with the magnitude of the prediction.
  If the resulting plot has a non-linear trend, the predicted and actual values likely do not have a linear relationship.
\end{enumerate}

\section{Qualitative accuracy}
\begin{figure}[htb]\centering
  \pgfplotstableread{data/eval_visits_skylines.dat}{\evalvisitcounts}
  \begin{tikzpicture}
    \begin{axis}[name=big,xmin=1,ymin=1,width=\textwidth,legend pos=north west,
    xlabel={Iteration},ylabel={$\max_i \visits_i$},height=6cm,xmax=40]
    \addplot table [x=i, y=predicted] {\evalvisitcounts};
    \addlegendentry{predicted}
    \addplot table [x=i, y=actual] {\evalvisitcounts};
    \addlegendentry{actual}
    \end{axis}

    \pgfplotstableread{data/eval_visits_predicted.dat}{\evalvisitpred}
    \begin{loglogaxis}[
    name=left,
    at=(big.below south west), anchor=above north west,
    xmin=1,ymin=1,width=0.5\textwidth,legend pos=north west,
    xlabel={$\vee_1$},ylabel={$\hat v$},
    height=6cm,
    enlarge x limits=false,
    cycle list name=my black white,
    ]
    \foreach \i in {1,2,...,7}
    {
      \addplot +[mark=none,unbounded coords=discard,very thick] table [x=gamma1, y=visited\i] {\evalvisitpred};
    }
    \end{loglogaxis}

    \pgfplotstableread{data/eval_visits_actual.dat}{\evalvisitact}
    \begin{loglogaxis}[
    name=right,
    at=(left.right of north east), anchor=left of north west,
    xmin=1,ymin=1,width=0.5\textwidth,legend pos=north west,
    xlabel={$\vee_1$},ylabel={$v$},
    height=6cm,
    enlarge x limits=false,
    cycle list name=my black white,
    ]
    \foreach \i in {1,2,...,7}
    {
      \addplot +[mark=none,unbounded coords=discard,very thick] table [x=gamma1, y=visited\i] {\evalvisitact};
    }
    \end{loglogaxis}
  \end{tikzpicture}
  \caption[Qualitative measurement of the visit count prediction accuracy]%
  {Qualitative measurement of the visit count prediction accuracy.
  Bottom left shows the predicted visit count for each dimension.
  Bottom right shows the actual visit counts.
  Above shows the \say{skyline}, i.e.\ the maximum of each graph, in a combined view.}\label{fig:evalqualvisits}
\end{figure}
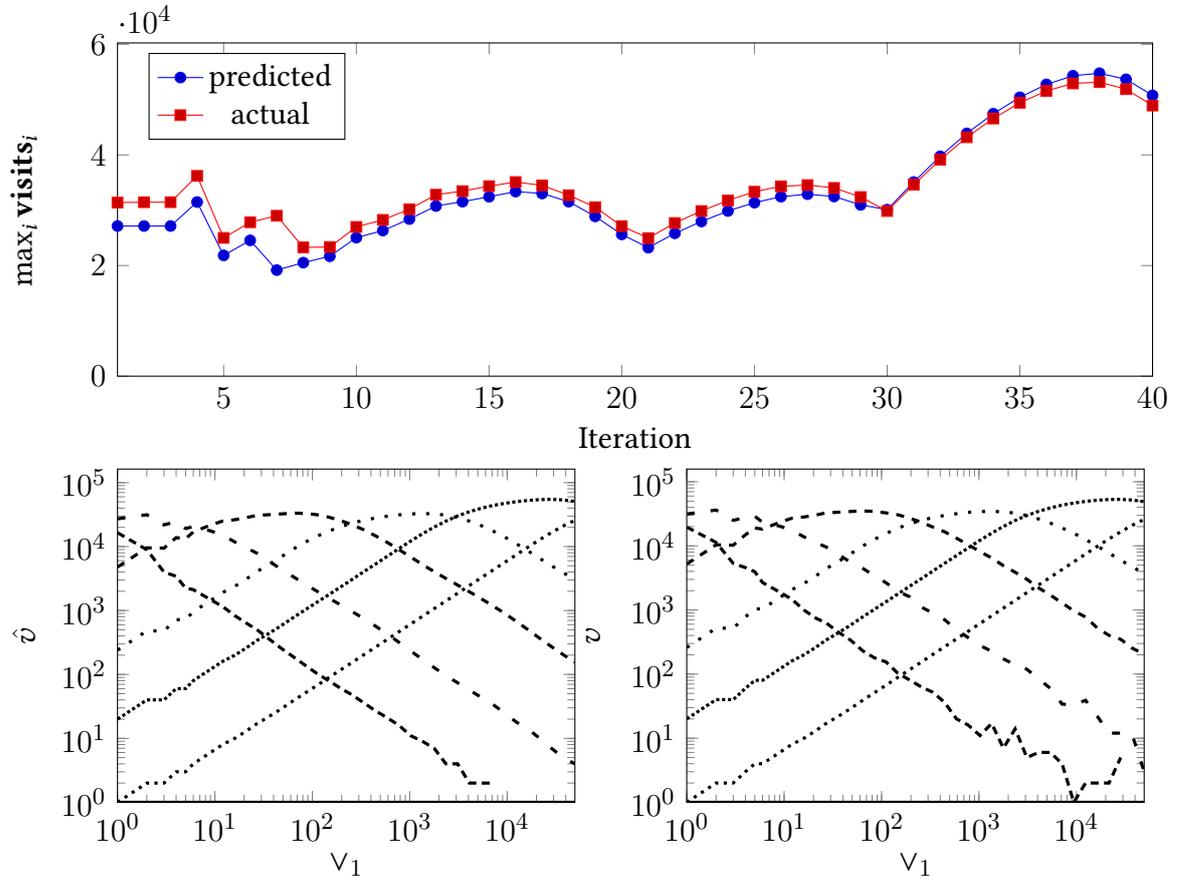
As a first litmus test for the accuracy of our prediction,
we varied the cardinality $\vee_1$ of the first dimension, while otherwise choosing a fixed set of parameters.
This is the same setup as the one we initially used to motivate our cost model
in \autoref{sec:visitmotivation}.
We also used this setup during development to first gain confidence of the accuracy while constructing the prediction model.

For this experiment, we chose $\sigma_1=\sigma_3=\sigma_5 = 0.6$, and $\sigma_2=\sigma_4=1$
to ensure a variety of selectivities while keeping visual clarity of the graphs.
Additionally, we chose $\card{R}=\num{50000}$ and $n=500$ as these values yield a non-shallow Elf and average query batch runtimes of \SIrange{100}{800}{\milli\second}.

\Autoref{fig:evalqualvisits} shows the results of this experiment as $\vee_1$ is varied.
For visual clarity, in addition to the predicted and actual visit counts in each dimension,
we show the \say{skyline} of the graph, i.e.\ the maximum value, in the magnified graph above.
While not statistically sound, we can observe a visual resemblance between the two graphs,
especially considering the position of local extrema.

\section{Accuracy under varying selectivities and cardinalities}
While useful during the development of the model, a qualitative analysis is obviously not sufficient
to confirm the accuracy of our prediction model.
Furthermore, the qualitative analysis did not include the predictive power of the visit count towards predicting the execution time.
Therefore, we conducted a set of experiments for measuring the quality of different aspects of our prediction model.
In particular, we gauge the accuracy of our visit count prediction,
and the influence of the visit count on the execution time.
In the following sections, we present our observations and results,
while we discuss the importance of these results in the following chapter.

The first experiment investigates the accuracy of
different parts of performance model on uniform, correlation-free data sets
For the first test, we chose $k=7$ and $\card{R}=\num{50000}$.
These choices are not arbitrary; our early tests indicated that under these conditions,
choosing attribute cardinalities between \num{1} and \num{1000} leads to both Elfs
that are very shallow and Elfs that are fully saturated, i.e.\ few dimension lists occur
before level $k+1$.
Therefore, these choices lead to a wide variance in execution time.
We randomised the other parameters for each example as follows:
\begin{itemize}
  \item Data distributions in each column: $X_i \sim \Ud{0}{\vee_i}$, where $\vee_i \sim \Ud{10}{65000}$, i.e.\ the cardinalities are randomly selected as well.
  \item For each of $n=500$ evaluation queries: $\sigma_i \sim \Uc{0.2}{1.0}$ to ensure we traverse the entire depth of the Elf for each query.
  Additionally, window positions are randomised to ensure uniform coverage of $[0,1]$ by query windows.
  \item No correlations.
\end{itemize}

\paragraph{Visit count prediction.}
Using these parameter values, we observe an $R^2$ value of
0.99, and a MAPE of \SI{5.3}{\percent}.
The actual visit counts exhibit a mean of \num{17509009} and a standard deviation of \num{12047885},
leading to a coefficient of variation of \num{0.60},
i.e.\ the average visit count for an example varies from the mean by \SI{60}{\percent}.
The maximum value is \num{17} times as large as the minimum value.

\begin{figure}[tbh]
  \begin{tikzpicture}
    \begin{axis}[xmin=0,ylabel={Relative error of $\hat v$},width=0.9\textwidth,
      height=6cm,
      xlabel={$\hat v$}]
      \addplot [only marks] table [x=x, y=r] {data/dim7_v_p_resids.dat};
      \draw[thick] (axis cs:\pgfkeysvalueof{/pgfplots/xmin},0) -- (axis cs:\pgfkeysvalueof{/pgfplots/xmax},0);
    \end{axis}
  \end{tikzpicture}
  \caption{Residuals for the visit prediction}\label{fig:visitsresid}
\end{figure}
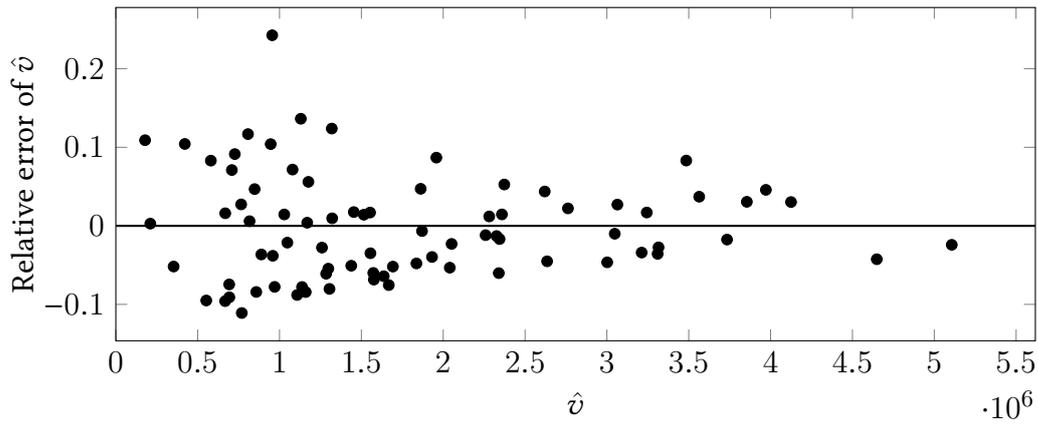

The distribution of the prediction errors is shown in \autoref{fig:visitsresid}.
We observe that the errors are distributed evenly around the value of zero error,
with a constant trend line.

\begin{figure}[tbh]\centering
  \begin{tikzpicture}
    \begin{axis}[name=a,xmin=0,
      width=0.45\textwidth,
      xlabel={$t$},ylabel={$t-\hat t$},
      y unit=ms, x unit=ms]
      \addplot [only marks] table [x=x, y=r] {data/dim7_t_v_resids.dat};
      \draw[thick] (axis cs:\pgfkeysvalueof{/pgfplots/xmin},0) -- (axis cs:\pgfkeysvalueof{/pgfplots/xmax},0);
    \end{axis}
    \begin{axis}[
      name=b,at=(a.right of north east),xshift=2cm, anchor=above north west,
      xmin=0,
      width=0.45\textwidth,
      xlabel={$t$},ylabel={$t-\hat t$},
      y unit=ms,x unit=ms]
      \addplot [only marks] table [x=x, y=r] {data/dim7_t_p_resids.dat};
      \draw[thick] (axis cs:\pgfkeysvalueof{/pgfplots/xmin},0) -- (axis cs:\pgfkeysvalueof{/pgfplots/xmax},0);
    \end{axis}
  \end{tikzpicture}
  \caption[Absolute error of the execution time prediction]{Absolute error of the execution time prediction.
  Left: $t$ is predicted from the ground truth visit count.
  Right: $t$ is predicted from the modelled visit count.
  }\label{fig:timeresid}
\end{figure}
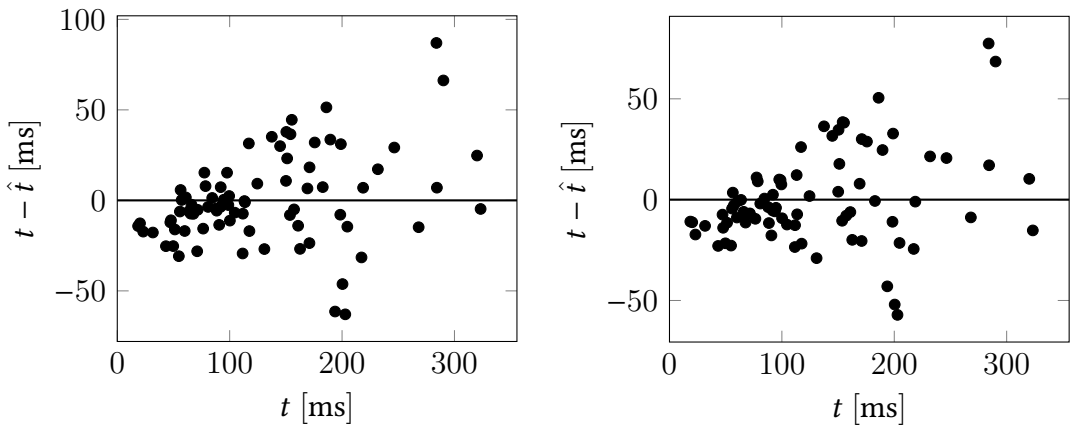

\paragraph{Response time prediction.}
Predicting the response time $\hat t$ from the \emph{ground truth} visit count $v$,
we obtain a corelation coefficient of $R^2 = 0.88$.
The estimates for the parameters $a$ and $b$ are \SI{1.22e-05}{\milli\second} and \SI{15.3}{\milli\second}, respectively.
The residual values are shown in \autoref{fig:timeresid}, left.

When predicting the response time $\hat t$ from the \emph{predicted} visit count $\hat v$,
we obtain a slightly higher corelation coefficient of $R^2 = 0.90$.
The estimates for the parameters $a$ and $b$ are \SI{6.27e-05}{\milli\second} and \SI{11.4}{\milli\second}, respectively.
The residual values are shown in \autoref{fig:timeresid}, right.

\section{Accuracy and adequacy under varying dimensionality}
One simple cost heuristic for the Elf could be the cardinality and/or selectivity
of the first attribute of the Elf,
since out of all parameters, we expected these to have the greatest influence on the Elf performance.
To compare the accuracy of these simple cost formulas to the accuracy of our model,
we tested the accuracy of the predictions for varying dimensionalities of the data set.

We chose to limit our range of dimensions from 3 to 15.
This choice can be explained as follows.
Three is the minimal number of dimensions that still lead to \say{interesting}
Elfs, since the first attribute contains the unique hash map value,
and the last attribute always contains a unique key for each tuple to ensure that
we have a well-defined data set.
The maximum is motivated by our definition of multi-dimensional (not high-dimensional) data sets,
and also through practical reasons:
To ensure a wide variety of Elf appearances, we need to increase the number of tuples
approximately exponentially with the number of dimensions,
since otherwise the Elf is too sparsely populated due to the curse of dimensionality,
and contains only monolists that simply increase in size.

To ensure a steady increase in Elf size while keeping build times manageable,
we chose the heuristic of choosing a number of
$ \card{R} := 3600e^{0.37k}$
tuples for dimensionality $k$.
This formula is approximately an exponential fit of
$(\num{7},\num{50000})$ and $(\num{15},\num{1000000})$.
The former point was motivated in the previous section
and the latter point is motivated by our tolerance in experiment run time,
where the build of a single Elf should not exceed five minutes.

As in the previous experiment, we also randomised all attribute cardinalities
within $\Ud{10}{65000}$, and all partial selectivities within $\Uc{0.2}{0.8}$.
We then measured the accuracy of the visit count prediction,
the influence of the visit count on the execution time,
and the overall accuracy of the time prediction.
For comparison, we also predicted the execution time using the {\naive} assumptions
that the time might be linear to either the selectivity or cardinality of the first column.
We use the $R^2$ value as the metric guiding the accuracy.

\begin{figure}[htb]
  \centering
  \begin{tikzpicture}
\begin{axis}[width=12cm,height=6cm,
  name=a,ymin=0,ymax=1,xlabel=Number of dimensions $k$,ylabel=Prediction accuracy,
  legend pos=outer north east]
  \addplot table[x=dim, y=vfromp] {data/multidim_accuracy.dat};
  \addlegendentry{$R^2(v\sim\hat v)$}
  \addplot table[x=dim, y=tfromp] {data/multidim_accuracy.dat};
  \addlegendentry{$R^2(t\sim\hat v)$}
  \addplot table[x=dim, y=tfromv] {data/multidim_accuracy.dat};
  \addlegendentry{$R^2(t \sim v)$}
  \addplot table[x=dim, y=tfromsimplep2] {data/multidim_accuracy.dat};
  \addlegendentry{$R^2(t\sim \sigma_1)$}
  \addplot table[x=dim, y=tfromsimplep] {data/multidim_accuracy.dat};
  \addlegendentry{$R^2(t\sim \vee_1)$}

\end{axis}
\end{tikzpicture}
  \caption{
    Prediction accuracy for varying dataset dimensionalities
  }
  \label{fig:multidimres}
\end{figure}
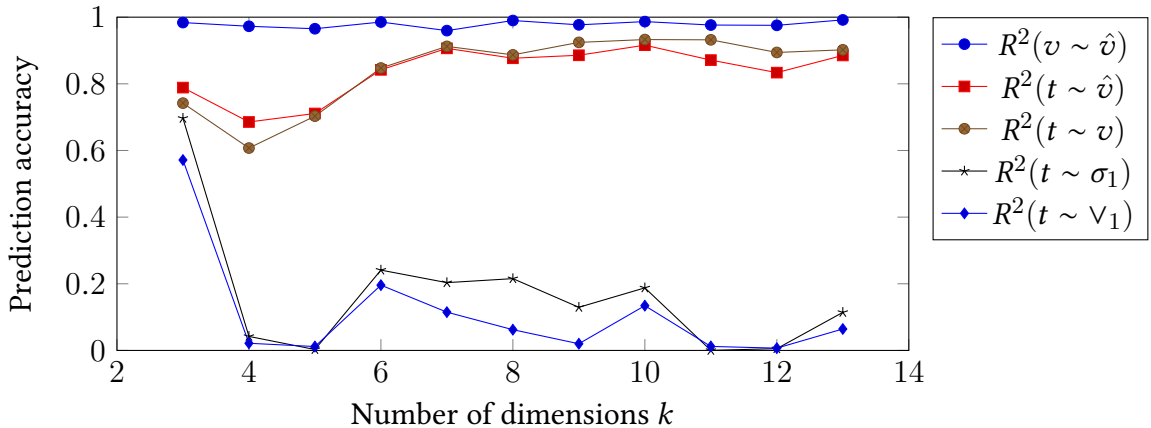

\Autoref{fig:multidimres} shows the measured $R^2$ values for all prediction scenarios described above.
Starting at low dimensionality, all predictions, including the {\naive} prediction,
show accurate results.
However, as dimensionality increases, accuracy of the latter drops off.
On the other hand, the visit count prediction consistently shows high $R^2$ values above \num{0.95}.
The accuracy of the overall execution time prediction follows the
fraction of execution time that is explained by the actual visit count.

\section{Accuracy on uniformly correlated data sets}
In \autoref{sec:nonuniform}, we proposed that correlations between different attributes of the Elf
cause the apparent cardinality of the deeper attribute to shrink.
To evaluate this theory, we generated a data set that exhibits this type of correlation.
Due to time constraints, we did not evaluate the effects of other types of correlations.

Our test data sets are constructed so that they exhibit
(2) a mean of $\bar\vee_1 \approx \num{6681}$ different values in $X_1$, and
(3) a mean of $\bar\vee_2 \approx \num{5410}$ different values in $X_2$.
However, the values of $X_2$ are not distributed uniformly over $[0;5409]$,
but instead are correlated such that $X_2 \sim X_1 + \Ud{0}{a}, a=21$.
Therefore, now $\mathit{max}_2 = 21 \neq \num{5410} \approx \vee_2$,
in contrast to the previous scenarios, where $\mathit{max}_i = \vee_i$.
Since the prediction is purely based on $\vee_2$,
we expect an overprediction of the dimension list lengths unless corrected using our method.

We created a set of $80$ examples, each with $n=500$ evaluation queries
and $\vee_1 \sim \Ud{1}{10000}$.
While $\sigma_1 \sim \Uc{0.2}{0.8}$, we always choose $\sigma_2=1$, since we assumed
that only one of the correlated dimensions is included in the query.

\begin{figure}[tbh]\centering
\begin{tikzpicture}
  \begin{axis}[
    ybar,
    height=5cm,
    enlarge y limits={abs=20pt},
    width=0.9\textwidth,
    ylabel={MAPE of $\visits_3-\mono_3$},
    symbolic x coords={16,20,21,22,26,vee2},
    xticklabels={16,20,21,22,26,$\vee_2$},
    xtick=data,
    nodes near coords={\pgfmathprintnumber\pgfplotspointmeta\%},
    nodes near coords align={vertical},
    yticklabels={,,},
    ]
    \addplot[
      y filter/.code={\pgfmathparse{#1*100}\pgfmathresult}
    ] coordinates {
      (16,0.268)
      (20,0.090)
      (21,0.058)
      (22,0.063)
      (26,0.167)
      (vee2,1.21)
    };
  \end{axis}
\end{tikzpicture}
  \caption[Prediction accuracy for a correlated data set]{
    Accuracy of the predicted number of non-monolist nodes in a correlated data set,
    for different assumptions about the cardinality of the correlated column.
  }
  \label{fig:corrmisprediction}
\end{figure}
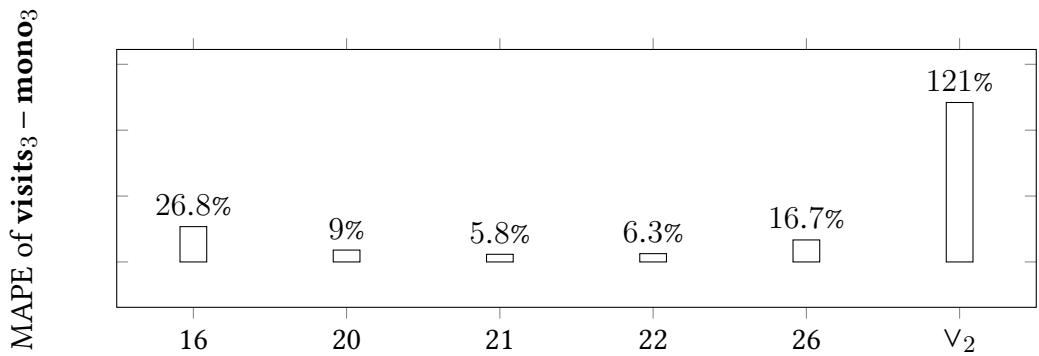

We then predicted the number of visited \emph{non-monolist nodes} at depth three of the Elf,
i.e.\ the number of interior nodes that remain after the two correlated columns.
We compared this prediction to the observed actual count of these visits.
The MAPE for different assumptions about the cardinality of $X_2$ is shown in \autoref{fig:corrmisprediction}.
The right-most bar shows the accuracy of the prediction that would be made if the correlation is not accounted for; then, the apparent cardinality of $X_2$ would be used.
In this case, the MAPE is \SI{121}{\percent}.
On the other hand, the error is much lower when the correct
\say{correlation cardinality} of $a=21$ is assumed, yielding a MAPE of \SI{5.8}{\percent}.

\section{Accuracy on non-uniform distributions}
Our basic method, as described in \autoref{sec:basicmodel} and evaluated above, does not account for non-uniform  distributions of the data contained in the Elf.
In \autoref{sec:nonuniform}, we described our approach that predicts the Elf visit count for non-uniform data sets using histograms.
In particular, this method relies on \autoref{thm:homomorph}, where we conjectured that the Elf
visit count is approximately the sum of the visit counts of a partition of the Elf.

We expect that, for non-uniform data sets, the histogram method has a significantly lower prediction error than our basic method.
In this section, we first demonstrate the validity of the approach based on a small example,
and then show how it scales to larger histograms.

\subsection{Simple piecewise uniform distribution}
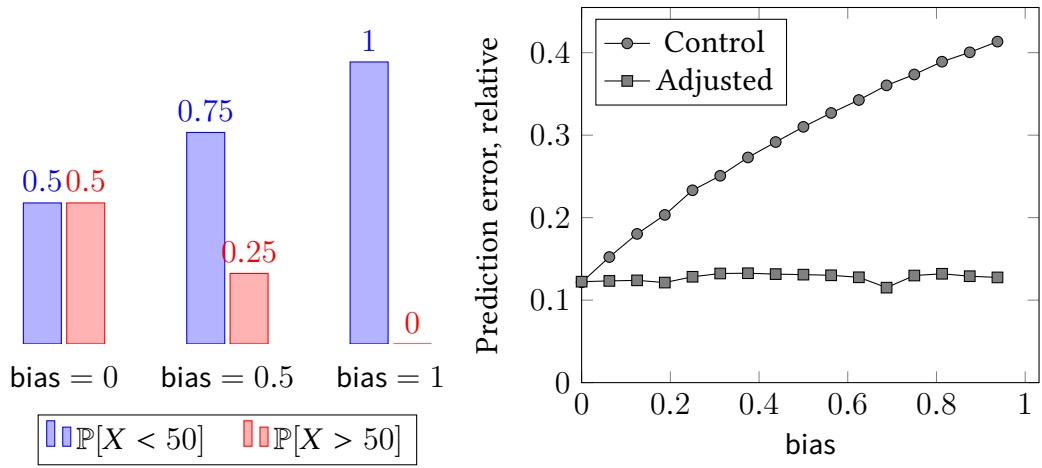
\begin{figure}[tb]
\begin{minipage}[b]{.5\linewidth}\centering
\begin{tikzpicture}
\begin{axis}[
    width=\linewidth,
    height=.7\linewidth,
    axis line style = { draw = none },
    bar width=0.5cm,
    ybar,
    ymin=0,ymax=1,
    enlarge x limits=0.2,
    legend style={at={(0.5,-0.25)},
      anchor=north,legend columns=-1},
    xtick style={draw=none},
    xticklabels={$\model{bias}=0$,$\model{bias}=0.5$,$\model{bias}=1$},
    xtick={1,2,3},
    ytick=\empty,
    nodes near coords,
    nodes near coords align={vertical},
    ]
\addplot coordinates {(1,0.5) (2,0.75) (3,1)};
\addplot coordinates {(1,0.5) (2,0.25) (3,0)};
\legend{$\PP[X < 50]\quad$,$\PP[X>50]$}
\end{axis}
\end{tikzpicture}
\end{minipage}%
\begin{minipage}[b]{.5\linewidth}\centering
\begin{tikzpicture}
\begin{axis}[
  width=\linewidth,
  xlabel={$\model{bias}$},
  ylabel={Prediction error, relative},
  cycle list name=black white,
  legend pos=north west,
  ymin=0,
  xmin=0
]
\addplot table[x=bias,y={relerr_uniform}] {data/twobucket_bias_corrector_error.dat}; \addlegendentry{Control}
\addplot table[x=bias,y={relerr_twobucket}] {data/twobucket_bias_corrector_error.dat}; \addlegendentry{Adjusted}
\end{axis}
\end{tikzpicture}
\end{minipage}
\caption[Prediction accuracy for a non-uniform data set]{Left: Distribution of the data set for different \model{bias} values.
      Right: Residuals of the basic and distribution-adjusted visit predictions}\label{fig:twobucketbias}
\end{figure}

We first investigate the general validity of our approach using a simple synthetic non-uniform data set that should provoke the non-uniformity prediction error.
To this end, we generated the Elf for a seven-dimensional data set containing \num{100000} tuples.
Each value in the first column was generated as follows: (1) with probability $0.5+\model{bias}$, sample from $\Ud{0}{49}$, (2) and with probability $0.5-\model{bias}$, sample from $\Ud{50}{99}$ (see \autoref{fig:twobucketbias}).
Here, \model{bias} is a parameter that represents the \say{skew} of the data set towards the lower values.

The values in the other columns are uniformly distributed over $\Ud{0}{15}$ to obtain a non-degenerate Elf.
In each dimension, the 500 test queries were assigned a partial selectivity $\alpha$ for each query and dimension, where $\alpha \sim \Uc{0.2}{1}$ to achieve a reasonable total selectivity.

We then measured the number of visited nodes.
This number was predicted using two different methods.
The control was computed using \autoref{alg:searchtime}, with $\vee_1 = 100$.
The adjusted prediction was computed using \autoref{alg:nonuniform}, with $l=2$ and the bucket probabilities as described above.
In this scenario, we expect the adjusted prediction to completely compensate the bias, since within each individual bucket, the data distribution is exactly uniform.

\Autoref{fig:twobucketbias} shows the relative prediction error for both models,
as functions of $\model{bias}$.
For $\model{bias}=0$, both methods perform similarly.
However, as \model{bias} increases, we observe the expected decrease in accuracy for the non-adjusted method.
In comparison, the relative error of the adjusted method stays at a constant level for all values of \model{bias}.

\subsection{Bucket-approximated binomial distribution}
To test the large-scale accuracy of the histogram technique with many buckets,
we used a synthetic data set that interpolates between a uniform distribution and a binomial distribution.

To this end, we generated $\Elf_7$ for a data set of \num{100000} tuples.
The values in the first column were generated as follows: with probability $\model{skew} \in [0,1]$, sample from $\Bin{n,0.5}, n=100$, and with probability $1-\model{skew}$, sample from $\Ud{0}{99}$.
Therefore, when $\model{skew}=0$, the data set is entirely uniform, and when $\model{skew}=1$, it is entirely binomial.

We therefore formulate the probability of a value $k$ appearing in $X_1$ as follows:
$$\PP_{\lvar{correct}} = \PP[X_1 = k] \, \, = \, \, (1-\model{skew})\cdot\frac{1}{n} \, + \, \model{skew} \cdot \binom{n}{k} 0.5^n$$
This probability, which we call $\PP_{\lvar{correct}}$, is one of the inputs to \autoref{alg:nonuniform}.
In order to evaluate the adequacy of the histogram technique, we compare it to itself in a
\say{placebo}-like variant; a variant that surely does not reveal any information about the data distribution at all.
As the placebo, we use the assumption $\PP_{\lvar{placebo}}$,
which states that $X_1$ is in fact entirely uniformly distributed for all values for \model{bias}.

\begin{figure}[tb]
  \centering
\pgfdeclarelayer{background}
\pgfdeclarelayer{foreground}
\pgfsetlayers{background,main,foreground}\begin{tikzpicture}
\begin{axis}[name=a,enlargelimits=false,ymin=1,ymax=20,xlabel=$\model{bias}$,
point meta min=.1,point meta max=0.4,width=7cm,ylabel={Number of buckets}]
\addplot[matrix plot*,
    point meta=explicit,mesh/cols=8,
    shader=faceted,
    y filter/.expression={y < 21 ? y : nan},
    ]
    table[meta=z] {data/binom_buckets.dat};
\end{axis}

\begin{axis}[name=b,anchor=west,width=7cm,at=(a.east),xshift=1cm,
enlargelimits=false,colorbar sampled,ymin=1,ymax=20,xlabel=$\model{bias}$,yticklabels={,,},
point meta min=.1,point meta max=0.4
]\addplot[matrix plot*,
    point meta=explicit,mesh/cols=8,
    shader=faceted,
    y filter/.expression={y < 21 ? y : nan},
    ]
    table[meta=z] {data/binom_buckets_uniformpred.dat};
\end{axis}

\node [below=1.5cm of a.south] {(a) $\PP_{\lvar{correct}}$};
\node [below=1.5cm of b.south] {(b) $\PP_{\lvar{placebo}}$};
\end{tikzpicture}
  \caption{Comparison of relative error of $t \sim \hat v$ for different distribution assumptions.}
  \label{fig:binombuckets}
\end{figure}
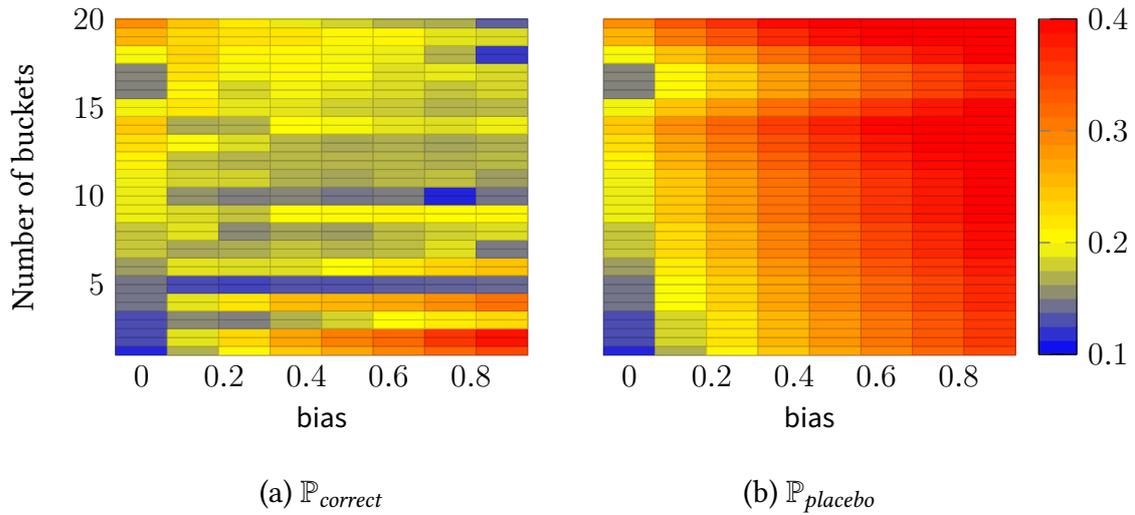

\Autoref{fig:binombuckets} shows the prediction accuracy for this experiment for both assumptions about the data distribution.
For low values of \model{bias} (left border), where $X_1$ indeed follows a uniform distribution,
both methods show small prediction errors.
As \model{bias} increases, both methods deteriorate at first.
However, the $\PP_{\lvar{correct}}$ prediction recovers as the number of buckets increases (towards top).
up to a point where the variance of the probability density of $X_1$ is explained through sufficiently many histogram buckets.
On the other hand, the results of the $\PP_{\lvar{placebo}}$ prediction do not improve;
they even deteriorate slightly as the number of histogram buckets increases.

\chapter{Discussion}\label{chp:discussion}
In this chapter, we discuss our experimental results that examined the
accuracy and adequacy of our cost model.
Furthermore, we discuss threats to validity of our results,
and describe how we modified our experimental technique to mitigate them.

\section{Interpretation}
In this section, we interpret our experimental results and discuss
their impact on the validity of our performance model.

First, the results of our qualitative analysis show a strong visual similarity
between the predicted and actual number of visited Elf nodes.
In particular, the predicted position and magnitude of local extrema accurately matches the
observed performance.
To recap \autoref{sec:visitmotivation}, we developed our model based on the proposition
of two major impact factors:
\begin{enumerate}
  \item Higher cardinality at the top leads to larger fanout, which leads to larger size.
  \item Higher cardinality at the bottom leads to earlier monolists, which leads to lower size.
\end{enumerate}
Therefore, we conclude that these two factors indeed dominate the size and shape of the Elf structure.
This finding corroborates the heuristic for selecting a dimension order
given by the original Elf authors~\cite{elf},
who suggested moving small-cardinality attributes closer to the top of the Elf.

The results of our statistical analysis show that for uniform data sets between five and twenty dimensions,
our model explains around \SI{99}{\percent} of the variance in visit counts observed.
Therefore, we conclude that in this scenario,
our model is a very accurate prediction of the Elf size at each level of the tree.

Our linear regression model that predicts the execution from the number of visited nodes
explains \SIrange{88}{90}{\percent} of the variance in execution time.
While having lower accuracy than the visit count prediction,
this accuracy is adequate for parameter tuning scenarios.
Especially considering the impact of suboptimal dimension orders of the Elf,
where the execution time differs by several orders of magnitude,
the accuracy is overwhelmingly sufficent.
We explain the lower accuracy through impact factors we disregarded for our model,
primarily the length of the dimension lists at each level of the tree.
See also the next subsection, where we discuss other possible impact factors.

\paragraph{Accuracy on data sets of varying dimensionality.}
When varying the dimensionality of the data set,
the visit count prediction retains more than \SI{95}{\percent} of accuracy throughout
dimensionalities of 2 to 15.
No trend is visible that indicates a deterioration in accuracy as the dimensionality increases further.
However, our results suggest that the visit count loses some of its predictive influence
on the execution time as dimensionality increases.
Therefore, the execution time is only predicted with at least \SI{80}{\percent} accuracy.
We hypothesise that this effect is caused by the increasing size of the Elf,
resulting in non-uniform access costs for the dimension lists in different dimensions.
Note that for the dimension experiment, we also used an approximately
exponential growth in the data set size to yield saturated Elfs.
Likely, the quality of cache alignment locality differs from dimension to dimension,
e.g.\ a memory access in a deeper dimension might be more expensive on average,
since nearby memory areas are less likely to have already been visited and put into cache.
Possibly, this can be mitigated by explicitly assigning weights to the memory accesses at each depth.
However, for our case of multi-dimensional (not high-dimensional) data sets,
the deterioration is still acceptable in magnitude.

As would be expected, for low-dimensional data sets, the selectivity and cardinality of the first attribute
are also good predictors for the search time.
However, their accuracy diminishes quickly as the dimensionality increases, rendering
these factors useless.
Similar experiments with the parameters for the deeper dimensions led to worse values.
This validates the need for a cost model like ours that specifically incorporates
the shape of the data structure at each level, instead of just at the top.

\paragraph{Accuracy on correlated data sets.}
For data sets with uniform correlations, we observe that the correlation impairs the accuracy
of the performance prediction, unless corrected.
Large errors already emerge in the intermediate visit counts around the correlated columns.
On the other hand, we find that if the correlated cardinality is known, and used
as the assumed cardinality of the correlated column, the prediction error vanishes.
Therefore, we conclude that using the correlated cardinality improves the prediction accuracy.
Note that one drawback of our technique is that the partial selectivity in the deeper column
must be 1.
Using a partial selectivity $\neq 1$ in the deeper column leads to non-uniform data sets in the deeper levels.
Due to time constraints, our model does not include the combination of a data set that is
\emph{both} correlated and non-uniformly distributed.

\paragraph{Accuracy on non-uniform data sets.}
For non-uniform data sets without correlations, we observed that the prediction worsens
as the data distribution deviates further from a uniform distribution.
If more accuracy is needed, data distributions with a coefficient of variation of less than
$c_v=\frac{\sigma}{\mu} = \frac{25}{50} = \num{0.5}$ can be efficiently and accurately
approximated through a five-bucket histogram.
However, the results suggest that the histogram technique itself introduces an inaccuracy
of the prediction.
This can be seen in the successively worsening accuracy for the $\PP_{\lvar{placebo}}$ prediction
as the number of buckets increases.
Also, we observe that accuracy does not increase monotonically with the number of buckets used.
Instead, some bucket counts yield consistently better results, especially the case of $k=5$ buckets.
We presume that this positive effect is a discretisation artefact,
which occurs for bucket counts that are divisors
of the total width of the attribute domain, which is $\vee_1=100$ here.
If the bucket count such a divisor, all buckets have exactly the same width,
instead of some buckets containing more values than others.

To summarise, we will answer the questions we posed in the beginning of our evaluation
(see \autoref{chp:results}).
We find that for data sets between 6--15 dimensions, the number of visited nodes
predicts the execution time with an accuracy of at least \SI{85}{\percent}.
Since we predict the number of visited nodes with an accuracy of over \SI{95}{\percent},
we predict the execution time with an accuracy of over \SI{80}{\percent}.
In comparison to simple cost formulas,
our model shows far superior accuracy.
In particular, as the dimensionality of the data set increases, the accuracy of
{\naive} predictions drops rapidly, while our method retains its accuracy.

\section{Threats to Validity}
During the development of our cost model and evaluation technique,
we considered a number of threats to validity.
In this section, we describe each threat and how we adapted our method to try to mitigate it.

\subsection{Internal validity}
As threats to interval validity, we considered
(1) the possibility of interfering errors,
(2) random noise from too few samples,
(3) cache interference between repetitions of the experiment,
(4) the performance impact of our measurement itself,
(5) and impact factors that we did not include in our model.

\begin{description}
  \item[Interfering errors]
Through the presence of \emph{multiple} errors  in our model,
different errors could possibly interfere to cause the appearance of no or low error error.
In particular, we considered the combination of an underprediction of the visit count in one of the dimensions
and an overprediction of the visit count in another dimension.
In this case, even though the model is inaccurate, the sum of visit counts might appear correct.
To decrease the likelihood of this threat,
we varied as many parameters as possible during the same experiment.
Furthermore, we also qualitatively compare the accuracy of the visit count of each dimension individually.
\item[Random sampling noise]
Timing noise can arise in our experiment setup from several sources.
First, the CPU cache might be in differing configurations at the start of each test run.
Second, operating system activity might interrupt and pause the search routine,
leading to spurious differences in execution time.
Finally, a too small number of \emph{examples} (see \autoref{sec:procedure})
can lead to too little variance,
which in turn leads to an overstatement of the prediction accuracy of the model.
To mitigate these threats, we repeated the search for each experiment five times
and chose a number of examples, \num{80}, that is considerably larger than the number of parameters
being fit (two: $a$ and $b$) to prevent this overfitting.

\item[Cache poisoning]
The performance of main-memory index structures can vary heavily due to varying
efficiency of the CPU cache.
Therefore, care must be taken to reset the cache to a known state after each experiment.
To this end, we always separate two repeated executions of the same query batch so they are not back-to-back, but instead have an evaluation of all other examples between them.
In this way, even if the first repetition fills the cache favourably for the query,
this cache configuration will be lost at the start of the second repetition, since the other examples
trigger the eviction policy of the cache to evict the data of the first example.
This means that a well-populated CPU cache from a previous repetition cannot leak
into the the next repetition and manipulate results.

\item[Instrumentation influences]
Our run-time instrumentation of the algorithm consisted of \say{tracking points}
that incremented a global visit counter for each visited node, and another counter
for the specific dimension that the access occured in.
Furthermore, the search itself never causes memory writes, except for adding result tuples
to the result set.
Therefore, even though the instrumentation looks somewhat harmless, it arguably has a \say{structural}
performance impact that could cause measurable differences in CPU behaviour, e.g.\
hardware memory barriers and instruction reordering.
However, since the counters are accessed so frequently and they are small in number,
we assume that they are stored in the CPU cache and are not written through to main memory.
Therefore, incrementing the counters on average only has a small constant performance impact per visited node, which is harmless.

\item[Unknown impact factors]
Finally, our experimental technique is not safe-guarded against impact factors that we did not consider in our cost model.
The most important example of such a threat is the presence of non-uniform memory access times
during the Elf search.
Non-uniform access times invalidate our hypothesis that the cost of accessing a single dimension list entry is constant.
Non-uniform access times could be caused by two influences, multiple entries per cache line and memory prefetching;
several small dimension lists or monolists could be loaded into a single cache line,
and future memory accesses could be predicted and pre-fetched by the CPU.
In both cases, access to the later elements appears to be faster than access to the first elements.
We argue that this threat does not have major influence,
because most dimension lists are short and therefore most dimension list scans are only
very small occurences of sequential memory accesses.
\end{description}

\subsection{External validity}
Our main threat to external validity is that our experiment scenario does not represent
an accurate environment for the production use of the Elf within a full-fledged
database system.
In the case of such an inaccuracy, our results would not be applicable to the use of the Elf in this important scenario.
We safeguard against this threat by taking measures to achieve \emph{procedural similarity}
with an on-line database system.
Our main concerns for procedural similarity lie in the careful selection of query sequencing rules
and the correct placement of caching barriers, which we will describe below.

Observe that designing an experiment that measures \say{the speed of an index structure} is non-trivial.
First, running only a single query on the structure likely has low signal-to-noise ratio,
since the single query is
executed rapidly.
Also, only running a single query inevitably leads to overfitting of the cost model.
Secondly, running the same query repeatedly does improve the signal-to-noise ratio, but instead
creates a new threat in the face of the CPU cache.
After the first few runs, the further runs are likely directly answered by the cache,
instead of the scenario that is actually simulated, which is queries that are
not known ahead-of-time to the database system.
Therefore, we conclude that a reasonable evaluation must always use a batch of queries at once.
This conclusion is shared by benchmark designers;
the TPC-H benchmark, for example, contains rules for sequencing the queries during test runs,
and its metrics are based on the combined performance over all types of queries \cite{council2008tpc}.
In our sequencing rules, the evaluation queries are always evaluated in random order, but deterministic between repeated runs of the same batch.

Concerning caching effects, care must be taken to neither eliminate too many nor too few caching effects.
Only desired caching effects should influence the measurements, while undesired effects should be eliminated.
One major undesired caching effect would be the locality of reference between the index build phase and the execution phase.
In a real-world scenario, the index is usually built well in advance of the actual query execution.
Therefore, the cache should not be populated with the results of the build phase at the start of the query phase.
We achieve this by first building all examples in-order, and then searching all examples in-order.
This means that when an example is first searched, its build is already separated in time from the search
by at least one search or build.
On the other hand, one desired effect is the effect of \say{hotness} of parts of the index structure.
Different queries that access the same relation might access intersecting parts of the index structure,
for example the top level of the Elf.
State-of-the art query schedulers even reorder queries to form batches, as they exhibit desirable performance improvements \cite{Ahmad:2011:ISR:2035111.2035115,Duggan:2011:PPC:1989323.1989359}.
We mimic this behaviour by executing multiple queries against the same Elf in sequence,
without interruptions.
This temporal locality, together with the spatial locality of the Elf search algorithm, allows CPU caching to affect the Elf performance within a batch,
as it would in a real-world database system.


\chapter{Conclusion}\label{chp:conclusion}

In this work, we developed an improved understanding of the Elf index structure,
and gained insights into its performance characteristics.
To conclude this thesis, we summarise the main results of our contributions,
and describe possibilities for future work.

\subsubsection*{Comparison to proposed multi-dimensional index structures}
We highlighted the importance of range queries with low total selectivity, but high partial selectivities,
and how this class of queries is problematic for {\naive} approaches to multi-dimensional indexing.
Based on a literature review of related work on index structures that support range queries,
we described how the Elf compares to them and what differentiates it from
e.g.\ other tree-based structures like the R-Tree and the {\kdtree}.
Furthermore, we described the adverse effects of the curse of dimensionality
and how the Elf is uniquely fit to tackle them.
Finally, we discussed the importance of an accurate cost model for
improving query processing performance.

\subsubsection*{Performance model of the Elf size and Query execution time}
Motivated by the advantages brought by accurate performance predictions,
we developed a cost model for the Elf.
First, we developed assumptions about the relationship between
(a) the size and shape of the Elf, and
(b) the execution time of range queries on the Elf.
Through an analysis of the Elf design and search algorithm, we derived a set of equations
that governs the shape of the Elf, which we expressed in a set of metrics that we introduced.
We provided an algorithm that iteratively computes these metrics.

\subsubsection*{Analysis of the impact of irregular data distributions}
Starting out from our basic performance model of the Elf,
we showed how the Elf behaves in situations that cause the performance of other
proposed index structures to degenerate.
In particular, we described how the Elf is impacted
when data distributions are skewed and when different attributes are correlated.
Interestingly, in these scenarios, the Elf does not degenerate,
but only requires simple and rather intuitive parameter adjustments to compensate for these effects.
This makes our cost model much more widely applicable
compared to cost models that are only accurate under the independence and uniformity assumptions.

\subsubsection*{Empirical evidence of model accuracy and adequacy}
Finally, we conducted an empirical evaluation using synthetic benchmarks,
which showed that the Elf search time is indeed
dominated by the factors included in our model.
Additionally, we demonstrated how better knowledge about the data distribution can be directly
incorporated to improve prediction accuracy,
using the techniques we introduced before.
In total, our results suggest that for the Elf data structure,
the traditional assumptions about the performance of disk-memory index structures still hold:
the execution time is closely related to the size of the accessed memory region.
The accessed memory region, in turn, is closely related to parameters of the query and data set.

To summarise, we have shown that the Elf data structure is uniquely positioned within
the growing field of multi-dimensional main-memory index structures.
We highlighted the importance of both efficiently using the resources of modern database systems,
and adapting to modern query workloads
demanded in large-scale data analytics.




\section{Future work}
In this section, we describe two avenues of future work.
First, we propose extensions to our cost model that could widen the range of
explained parameter influences.
Finally, we propose a variant of the Elf structure itself.

\paragraph{Automatic discovery and quantification of correlations.}
Our model only handles uniform correlations, but not arbitrary types of correlations.
Additionally, correlations have to be manually considered when collecting the prediction parameters.
However, automatic discovery and strength measurement techniques have been proposed to identify correlations.
These techniques could be incorporated to automate the cardinality correction for the Elf.

\paragraph{Simplication of the cost formula.}
Our performance model represents a cost formula for the Elf.
However, for simplicity, we only give a multi-step algorithm to compute the cost.
Presumably, the set of equations that govern this algorithm could be simplified, to yield
a more compact rendition of the Elf performance.
Especially the matrix of differentials
$\left(\frac{\partial \visits_i}{\partial \Box_j}\right)_{i,j}$
for $\Box \in \{\vee, \sigma\}$
could provide interesting insight into the relative importance of the parameters
on the search performance.
A sufficiently simple form of the cost formula could even allow automated
numerical optimisation of the Elf cost for the discovery of preferred Elf configurations.

\paragraph{Explicit incorporation of caching effects.}
In the first levels of the Elf, a larger fraction of dimension lists is accessed,
compared to the deeper levels of the Elf.
Due to the memory layout of the Elf and the in-order traversal,
this means that in the first levels of the Elf,
the memory accesses have a closer resemblance to a sequential scan than the memory accesses
in deeper levels.
Therefore, for very large numbers of dimensions,
it is possible that the average time for visiting a node near the top of the Elf
is shorter than the average time for visiting a node near the bottom.
Consequently, not all visit counts would be expected to have uniform access time.
Therefore, the total execution time would not be linear to the total number of visits;
instead, the visits for some of the levels of the Elf would have to be weighed differently from others.

\paragraph{Modelling the hash map property of the Elf.}
In our analysis, we have disregarded the unique property of the first level of the Elf, which
can be stored in an optimised way, since every attribute value is present.
In \autoref{sec:hashmap}, we described that this hash map is a unique kind of dimension list
which can be scanned with zero cost.
Therefore, the hash map could simply be treated as a list of smaller sub-Elfs, whose search time can again be predicted using our model.
For our model, we disregarded the hash map and simply always treated it as an attribute with a cardinality of one.

\paragraph{Predicting from query bounds instead of selectivities.}
Our prediction model does not use the actual query bounds,
but rather the partial selectivities, to predict the fraction of data requested by the query.
One possible problem here stems from the discretisation of the partial selectivity
to the number of visited dimension list entries.
If a dimension list has two entries, and the query has a partial selectivity of \num{0.5},
we expect one entry of the dimension list to be visited;
however, we are not \emph{confident} that the second entry is not visited as well.
This does not impair the accuracy of the prediction, but rather the robustness as the selectivity estimate varies.
Therefore, predicting from the query bounds instead of
the partial selectivity could lead to increased robustness.




\paragraph{Elf as a framework for multi-dimensional index structures.}
Finally, we note an interesting observation made during our studies of the Elf.
While in its original form, a dimension list is simply an array of values and pointers in memory,
any type of map from values to pointers could conceivably be used in place of the array.
This train of thought leads to a new role of the Elf as a \emph{composition scheme} for
building a multi-dimensional index structure out of single-dimensional index structures.
For example, to mitigate the worsened performance of large dimension list, a search tree
structure like the {\bplustree} could again be used to improve performance.




\printbibliography[heading=bibintoc]


\end{document}